\documentclass[prb,twocolumn,aps,showpacs,amsmath,nofootinbib,longbibliography]{revtex4-1}
\usepackage{graphicx}
\usepackage{multirow}
\usepackage{dcolumn}
\usepackage{amssymb,amscd,xypic,bm,dsfont,wasysym,amsmath,mathrsfs,latexsym,psfrag,accents,mathtools}
\usepackage{float}
\usepackage{color}
\usepackage{xcolor}

		\newcommand{\e}[1]{\begin{align}{#1}\end{align}}	
		
		\newcommand{\la}[1]{\label{#1}}
		\newcommand{\q}[1]{Eq.\ (\ref{#1})}
		\newcommand{\s}[1]{Sec.\ \ref{#1}}
		\newcommand{\fig}[1]{Fig.\ \ref{#1}}		

\newcolumntype{L}[1]{>{\raggedright\arraybackslash}p{#1}}
\newcolumntype{C}[1]{>{\centering\arraybackslash}p{#1}}
\newcolumntype{R}[1]{>{\raggedleft\arraybackslash}p{#1}}

\newcommand{\hatgone}{\hat{g}_1}
\newcommand{\hatgtwo}{\hat{g}_2}
\newcommand{\hatgthree}{\hat{g}_3}
\newcommand{\hatgonetwo}{\hat{g}_{12}}
\newcommand{\hatgi}{\hat{g}_i}
\newcommand{\hatgj}{\hat{g}_j}
\newcommand{\hatgk}{\hat{g}_k}
\newcommand{\hatgij}{\hat{g}_{ij}}

\newcommand{\gij}{{g}_{ij}}
\newcommand{\di}{{D}_i}
\newcommand{\cij}{{C}_{i,j}}

\newcommand\gkpar{G_{\sma{\pi,k_z}}}

\newcommand\gwz{G_{\sma{\tilde{\Gamma}}}}
\newcommand\gwp{G_{\sma{\tilde{X}}}}
\newcommand\gn{\caln}
\newcommand\gs{G_{\circ}}
\newcommand\gx{G_{\sma{\tilde{X}}}}

\newcommand\pper{P_{\sma{\perp}}}

\newcommand\myspace{\;\;\;\;}

\newcommand\tilx{\tilde{X}}
\newcommand\tilz{\tilde{Z}}
\newcommand\tilg{\tilde{\Gamma}}
\newcommand\tilu{\tilde{U}}

\newcommand{\bk}{\boldsymbol{k}}
\newcommand{\bkone}{\boldsymbol{k_1}}
\newcommand{\bktwo}{\boldsymbol{k_2}}
\newcommand{\br}{\boldsymbol{r}}
\newcommand{\bG}{\boldsymbol{G}}
\newcommand{\bR}{\boldsymbol{R}}
\newcommand{\bdelta}{\boldsymbol{\delta}}
\newcommand{\bdeltapar}{\boldsymbol{\delta}_{\sma{\parallel}}}

\newcommand\kzeq{\;\stackrel{\mathclap{\normalfont\mbox{$\sma{\bar{k}_z}$}}}{=}\;}

\newcommand\glidez{\calp_{\sma{\tilde{Z}}}^{\sma{\eta}}}

\newcommand\glide{\calp_{\sma{\tilde{\Gamma}}}^{\sma{\eta}}}
\newcommand\glidep{\calp_{\sma{\tilde{\Gamma}}}^{\sma{+}}}
\newcommand\glidem{\calp_{\sma{\tilde{\Gamma}}}^{\sma{-}}}
\newcommand{\calti}{{\cal T}_{\sma{\cal I}}}

\newcommand{\pgdel}{\pdg{g}_{\boldsymbol{\delta}}}
\newcommand{\bmz}{\bar{M}_z}
\newcommand{\calbmx}{{\cal \bar{M}}_x}
\newcommand{\calgdel}{\pdg{\breve{g}}_{\sma{\bdelta}}}
\newcommand{\brevegdel}{\pdg{\breve{g}}_{\sma{\bdelta}}}
\newcommand{\brevegdelpar}{\pdg{\breve{g}}_{\sma{\bdelta}\shortparallel}}
\newcommand{\brevebmx}{\breve{\bar{M}}_x}
\newcommand{\bmx}{\bar{M}_x}
\newcommand{\uit}{U_{\sma{\cali T}}}
\newcommand{\utg}{U_{\sma{Tg \bdelta}}}

\newcommand{\ubmx}{U_{\sma{\bmx}}}
\newcommand{\utbmz}{U_{\sma{T\bmz}}}

\newcommand{\caltgdel}{\breve{T}_{\sma{g \bdelta}}}

\newcommand{\brevetbmx}{\breve{T}_{\sma{\bmx}}}
\newcommand{\hattbmx}{\hat{T}_{\sma{\bmx}}}
\newcommand{\brevetbmz}{\breve{T}_{\sma{\bmz}}}
\newcommand{\breveti}{\breve{T}_{\sma{\cali}}}
\newcommand{\caltbmz}{\calt_{\sma{\bmz}}}

\newcommand{\hatgdel}{\pdg{\hat{g}}_{\sma{\bdelta}}}
\newcommand{\hatgdelpar}{\pdg{\hat{g}}_{\sma{\bdelta} \shortparallel}}
\newcommand{\hatTg}{\pdg{\hat{T}}_{\sma{g\bdelta}}}

\newcommand{\sx}{\sigma_{\sma{1}}}
\newcommand{\sy}{\sigma_{\sma{2}}}
\newcommand{\sz}{\sigma_{\sma{3}}}

\newcommand{\tx}{\tau_{\sma{1}}}

\newcommand{\tz}{\tau_{\sma{3}}}

\newcommand{\ins}[1]{\;\;\;\;\text{#1}\;\;\;\;}

\newcommand{\kpar}{\boldsymbol{k}_{{\shortparallel}}}

\newcommand{\Dpar}{D_{g{\shortparallel}}}

\newcommand{\pgdelpar}{\pdg{g}_{\boldsymbol{\delta}\shortparallel}}

\newcommand{\calb}{{\cal B}}
\newcommand{\calc}{{\cal C}}
\newcommand{\cald}{{\cal D}}

\newcommand{\calh}{{\cal H}}
\newcommand{\cali}{{\cal I}}

\newcommand{\calm}{{\cal M}}
\newcommand{\caln}{{\cal N}}

\newcommand{\calp}{{\cal P}}
\newcommand{\calq}{{\cal Q}}

\newcommand{\calt}{{\cal T}}
\newcommand{\calu}{{\cal U}}

\newcommand{\caly}{{\cal Y}}
\newcommand{\calz}{{\cal Z}}

\newcommand{\noi}[1]{\noindent (#1)}
\newcommand{\imp}{\;\;\Rightarrow\;\;}
\newcommand{\mo}{\sma{-1}}

\newcommand{\braket}[2]{\big\langle #1 \big| #2 \big\rangle}
\newcommand{\ketbra}[2]{\big|  #1  \big\rangle \big\langle #2 \big| }

\newcommand{\bra}[1]{\big\langle#1\big|}
\newcommand{\ket}[1]{\big|#1\big\rangle}

\newcommand{\bea}{\begin{eqnarray}}
\newcommand{\enea}{\end{eqnarray}}
\newcommand{\beq}{\begin{equation}}
\newcommand{\eneq}{\end{equation}}

\newcommand{\pdg}[1]{{#1}^{\phantom{\dagger}}}

\newcommand{\lin}{\notag \\}

\newcommand{\eq}{=&\;}
\newcommand{\ab}{\alpha\beta}

\newcommand{\low}{L$\ddot{\text{o}}$wdin\;}

\newcommand{\W}{{\cal W}}

\newcommand{\bpm}{\begin{pmatrix}}
\newcommand{\epm}{\end{pmatrix}}

\newcommand{\bal}{\begin{align}}
\newcommand{\eal}{\end{align}}

\newcommand{\R}{\mathbb{R}}

\newcommand{\dg}[1]{#1^{\scriptstyle{\dagger}}}

\newcommand{\sma}[1]{\scriptscriptstyle{#1}}

\newcommand{\noc}{n_{\sma{{occ}}}}

\newcommand{\Z}{\mathbb{Z}}

\newcommand{\qed}{\nobreak \ifvmode \relax \else
      \ifdim\lastskip<1.5em \hskip-\lastskip
      \hskip1.5em plus0em minus0.5em \fi \nobreak
      \vrule height0.75em width0.5em depth0.25em\fi}

\begin{document}

\title{Topological Insulators from Group Cohomology}

\author{A. Alexandradinata$^{1,2}$, Zhijun {Wang}$^1$, and B. Andrei Bernevig$^1$}
\affiliation{${^1}$Department of Physics, Princeton University, Princeton, NJ 08544, USA}
\affiliation{${^2}$Department of Physics, Yale University, New Haven, CT 06520, USA}

\date{\today}

\begin{abstract}
We classify insulators by generalized symmetries that combine space-time transformations with quasimomentum translations.\ Our group-cohomological classification generalizes the nonsymmorphic space groups, which extend point groups by real-space translations, i.e., nonsymmorphic symmetries unavoidably translate the spatial origin by a fraction of the lattice period.\ Here, we \emph{further} extend nonsymmorphic groups by reciprocal translations, thus placing real and quasimomentum space on equal footing.\ We propose that group cohomology provides a symmetry-based classification of quasimomentum manifolds, which in turn determines the band topology.\ In this sense, cohomology underlies band topology.\ Our claim is exemplified by the first theory of time-reversal-invariant insulators with nonsymmorphic spatial symmetries.\ These insulators may be described as `piecewise topological', in the sense that subtopologies describe the different high-symmetry submanifolds of the Brillouin zone, and the various subtopologies must be pieced together to form a globally consistent topology.\ The subtopologies that we discovered include: a glide-symmetric analog of the quantum spin Hall effect, an hourglass-flow topology (exemplified by our recently-proposed KHgSb material class), and quantized non-Abelian polarizations.\ Our cohomological classification results in an atypical bulk-boundary correspondence for our topological insulators.
\end{abstract}




\maketitle


Spatial symmetries have enriched the topological classification of insulators and superconductors.\cite{fu2011,Classification_Chiu,AZ_mirror,ChaoxingNonsymm,chen2013,Shiozaki2014,Nonsymm_Shiozaki,AAchen,Varjas,Varjas2}
A basic geometric property that distinguishes spatial symmetries regards their transformation of the spatial origin: symmorphic symmetries preserve the origin, while nonsymmorphic symmetries unavoidably translate the origin by a fraction of the lattice period.\cite{Lax} This fractional translation is responsible for band topologies that have no analog in symmorphic crystals. Thus far, all experimentally-tested topological insulators have relied on symmorphic space groups.\cite{xia2009,hsieh2009a,Hsieh2012,Xu2012,tanaka2012,CeX} Here, we propose the first nonsymmorphic theory of time-reversal-invariant insulators, which complements previous theoretical proposals with magnetic, nonsymmorphic space groups.\cite{moore2010,ChaoxingNonsymm,chen2013,Shiozaki2015,singlediraccone} Motivated by our recently-proposed KHg$X$ material class ($X{=}$Sb,Bi,As),\cite{Hourglass} we present here a complete classification of spin-orbit-coupled insulators with the space group ($D_{6h}^4$) of KHg$X$.

\begin{figure}[ht]
\centering
\includegraphics[width=8.6 cm]{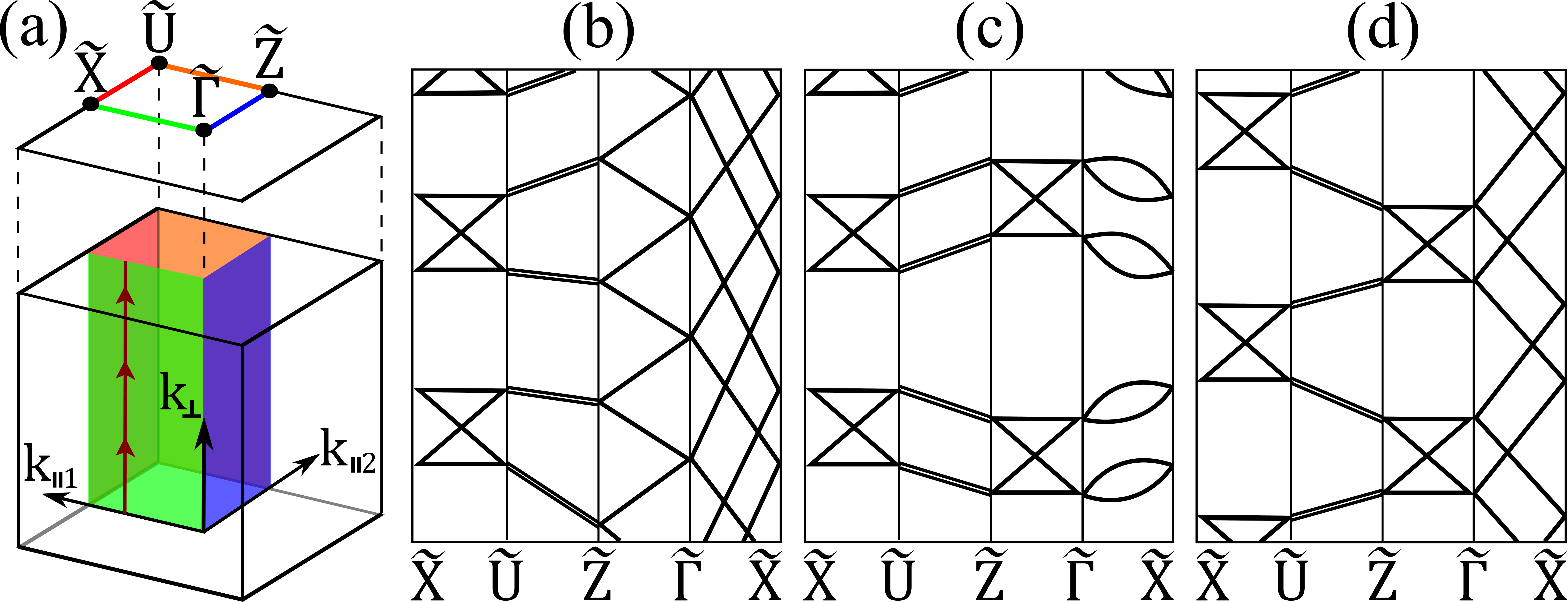}
    \caption{(a) Mirror (red, blue) and glide (green, brown) planes in the 3D Brillouin torus of KHg$X$. These planes project to the high-symmetry line $\tilx \tilu \tilz \tilg \tilx$ in the 2D Brillouin torus of the 010 surface. (b-d) Examples of possible piecewise topologies in the space group of KHg$X$, as illustrated by their surface bandstructures along $\tilx \tilu \tilz \tilg \tilx$. (b) describes a `quantum glide Hall effect' (along $\tilz \tilg$), and an odd mirror Chern number (along $\tilg \tilx$); these two subtopologies must be pieced together at their intersection point $\tilg$.   (c) describes an hourglass-flow topology ($\tilx \tilu \tilz \tilg$), and an even mirror Chern number ($\tilg \tilx$). (d) A trivial topology is shown for comparison.} \label{fig:representatives}
\end{figure}

The point group ($D_{6h}$) of KHg$X$, defined as the quotient of its space group by translations, is generated by four spatial transformations -- this typifies the complexity of most space groups. This work describes a systematic method to topologically classify space groups with similar complexity; in contrast, previous classifications\cite{fu2011,Hsieh2012,Classification_Chiu,ChaoxingNonsymm,chen2013,Shiozaki2014,Nonsymm_Shiozaki} (with one exception by us\cite{AAchen}) have expanded the Altland-Zirnbauer symmetry classes\cite{kitaev2009,schnyder2009} to include only a single point-group generator. For point groups with multiple generators, different submanifolds of the Brillouin torus are invariant under different symmetries, e.g., mirror and glide planes are respectively mapped to themselves by a symmorphic reflection and a glide reflection, as illustrated in \fig{fig:representatives}(a) for $D_{6h}^4$. Wavefunctions in each submanifold are characterized by a lower-dimensional topological invariant which depends on the symmetries of that submanifold, e.g., mirror planes are characterized by a mirror Chern number\cite{teo2008} and glide planes by a glide-symmetric analog\cite{Hourglass,Nonsymm_Shiozaki} of the quantum spin Hall effect\cite{kane2005B} (in short, a quantum glide Hall effect). The various invariants are dependent because wavefunctions must be continuous where the submanifolds overlap, e.g., the intersection of planes in \fig{fig:representatives}(a) are lines that project to $\tilg,\tilx,\tilu,$ and $\tilz$. We refer to such insulators as `piecewise topological', in the sense that various subtopologies (topologies defined on different submanifolds) must be pieced together consistently to form a 3D topology.

This work addresses two related themes: (i) a group-cohomological classification of quasimomentum submanifolds, and (ii) the connection between this cohomological classification and the topological classfication of band insulators. In (i), we ask how a mirror plane differs from a glide plane. Are two glide planes in the same Brillouin torus always equal? This equality does \emph{not} hold for $D_{6h}^4$: in one glide plane, the symmetries are represented \emph{ordinarily}, while in the other we encounter generalized `symmetries' that combine space-time transformations with quasimomentum translations ($\W$). Specifically, $\W$ denotes a discrete quasimomentum translation in the reciprocal lattice. These `symmetries' then generate an extension of the point group by $\W$, i.e., $\W$ becomes an element in a \emph{projective} representation of the point group. The various representations (corresponding to different glide planes) are classified by group cohomology, and they result in different subtopologies (e.g., one glide plane in $D_{6h}^4$ may manifest a quantum glide Hall effect, while the other cannot). In this sense, cohomology underlies band topology.

To determine the possible subtopologies within each submanifold and then combine them into a 3D topology, we propose a general methodology through Wilson loops of the Berry gauge field;\cite{AA2014,Maryam2014} these loops represent quasimomentum transport in the space of filled bands.\cite{zak1989} As exemplified for the space group $D_{6h}^4$, our method is shown to be efficient and geometrically intuitive --  piecing together subtopologies reduces to a problem of interpolating and matching curves. The novel subtopologies that we discover include: (i) the quantum glide Hall effect in Fig.\ \ref{fig:representatives}(a), (ii) an hourglass-flow topology, as illustrated in Fig.\ \ref{fig:representatives}(b) and exemplified\cite{Hourglass} by KHg$X$, and (iii) quantized, non-abelian polarizations that generalize the abelian theory of polarization.\cite{kingsmith1993}

Our topological classification of $D_{6h}^4$ is the first physical application of group extensions by quasimomentum translations. It generalizes the construction of nonsymmorphic space groups, which extend point groups by real-space translations.\cite{ascher1,ascher2,cohomologyhiller,mermin_fourierspace,rabsonbenji} Here, we \emph{further} extend nonsymmorphic groups by reciprocal translations, thus placing real and quasimomentum space on equal footing. A consequence of this projective representation is an atypical bulk-boundary correspondence for our topological insulators. This correspondence describes a mapping between topological numbers that describe bulk wavefunctions and surface topological numbers\cite{fidk2011}  -- such a mapping exists if the bulk and surface have in common certain `edge symmetries' which form a subgroup of the full bulk symmetry; this edge subgroup is responsible for quantizing both bulk and surface topological numbers, i.e., these numbers are robust against gap- and edge-symmetry-preserving deformations of the Hamiltonian. In our case study, the edge symmetry is projectively represented in the bulk, where quasimomentum provides the parameter space for parallel transport; on a surface with reduced translational symmetry, the \emph{same} symmetry is represented ordinarily. In contrast, all known symmetry-protected correspondences\cite{Maryam2014} are one-to-one and rely on the identity between bulk and surface representations; our work explains how a \emph{partial} correspondence arises where such identity is absent.

The outline of our paper: we first summarize our main results in \s{sec:summary}, which also serves as a guide to the whole paper.
We then preliminarily review the tight-binding method in Sec.\ \ref{sec:general}, as well as introduce the spatial symmetries of our case study. Next in Sec.\ \ref{sec:reviewWilson}, we review the Wilson loop and the bulk-boundary correspondence of topological insulators; the notion of a partial correspondence is introduced, and exemplified with our case study of $D_{6h}^4$. The method of Wilson loops is then used to construct and classify a piecewise topological insulator in Sec.\ \ref{sec:curvematching}; here, we also introduce the quantum glide Hall effect. Our topological classification relies on extending the symmetry group by quasimomentum translations, as we elaborate in Sec.\ \ref{sec:wilsoniansymm}; the application of group cohomology in band theory is introduced here. We offer an alternative perspective of our main results in Sec.\ \ref{sec:outlook}, and end with an outlook.


\section{Summary of results} \la{sec:summary}


A topological insulator in $d$ spatial dimensions may manifest robust edge states on a $(d-1)$-dimensional boundary. Letting $\bk$ parametrize the $d$-dimensional Brillouin torus, we then split the quasimomentum coordinate  as $\bk {=} (k_{\perp},\kpar)$, such that $k_{\perp}$ corresponds to the coordinate orthogonal to the surface, and $\kpar$ is a wavevector in a $(d-1)$-dimensional surface-Brillouin torus. We then consider a family of noncontractible circles $c(\kpar)$, where for each circle, $\kpar$ is fixed, while $k_{\perp}$ is varied over a reciprocal period, e.g., consider the brown line in \fig{fig:representatives}(a). We propose to classify each quasimomentum circle by the symmetries which leave that circle invariant. For example, in centrosymmetric crystals, spatial inversion is a symmetry of $c(\kpar)$ for inversion-invariant $\kpar$ satisfying $\kpar{=}{-}\kpar$ modulo a surface reciprocal vector. The symmetries of the circle are classified by the second group cohomology 
\e{H^2(\gs,\Z_2\times \Z^{d} \times \Z). \la{classifymani}} 
As further elaborated in \s{sec:wilsoniansymm} and App.\ \ref{app:wilsonproj}, $H^2$ classifies the possible group extensions of $\gs$ by $\Z_2{\times} \Z^{d} {\times} \Z$, and each extension describes how the symmetries of the circle are represented. The arguments in $H^2$ are defined as:\\
 
\noi{a} The first argument, $\gs$, is a magnetic point group\cite{magnetic_groups} consisting of those space-time symmetries that (i) preserve a spatial point, and (ii) map the circle $c(\kpar)$ to itself. For $d=3$, the possible magnetic point groups comprise the 32 classical point groups\cite{tinkhambook} without time reversal ($T$), 32 classic point groups with $T$, and 58 groups in which $T$ occurs only in combination with other operations and not by itself. However, we would only consider subgroups of the 3D  magnetic point groups (numbering $32{+}32{+}58{=}122$) which satisfy (ii); these subgroups might also include spatial symmetries which are spoilt by the surface, with the just-mentioned spatial inversion a case in point.\\

\noi{b} The second argument of $H^2$ is the direct product of three abelian groups that we explain in turn. The $\Z_2$ group is generated by a $2\pi$ spin rotation; its inclusion in the second argument implies that we also consider half-integer-spin representations, e.g., at inversion-invariant $\kpar$ of fermionic insulators, time reversal is represented by $T^2{=}{-}I$. \\

\noi{c} The second abelian group ($\Z^{d}$) is generated by discrete real-space translations in $d$ dimensions; by extending a magnetic point group ($\gs$) by $\Z^d$, we obtain a magnetic space group; nontrivial extensions are referred to as nonsymmorphic.\\ 

\noi{d} The final abelian group ($\Z$) is generated by the discrete quasimomentum translation in the surface-normal direction, i.e., a translation along $c(\kpar)$ and covering $c(\kpar)$ once. A nontrivial extension by quasimomentum translations is exemplified by one of two glide planes in the space group $D_{6h}^4$ [cf.\ \s{sec:wilsoniansymm}]. \\

Having classified quasimomentum circles through \q{classifymani}, we outline a systematic methodology to topologically classify band insulators. The key observation is that quasimomentum translations in the space of filled bands is represented by Wilson loops of the Berry gauge field; the various group extensions, as classified by \q{classifymani}, correspond to the various ways in which symmetry may constrain the Wilson loop; studying the Wilson-loop spectrum then determines the topological classification. A more detailed summary is as follows:\\

\noi{i} We consider translations along $c({\kpar})$ with a certain orientation that we might arbitrarily choose, e.g., the triple arrows in \fig{fig:representatives}(a). These translations are represented by the Wilson loop $\W(\kpar)$, and the phase  ($\theta$) of each Wilson-loop eigenvalue traces out a `curve' over $\kpar$. In analogy with Hamiltonian-energy bands, we refer to each `curve' as the energy of a Wilson band in a surface-Brillouin torus. The advantage of this analogy is that the Wilson bands may be interpolated\cite{fidk2011,ZhoushenHofstadter} to Hamiltonian-energy bands in a semi-infinite geometry with a surface orthogonal to $k_{\perp}$. Some topological properties of the Hamiltonian and Wilson bands are preserved in this interpolation, resulting in a bulk-boundary correspondence that we describe in \s{subsec:bb}. There, we also introduce two complementary notions of a total and a partial correspondence; the latter is exemplified by the space group $D_{6h}^4$.\\

\noi{ii} The symmetries of $c(\kpar)$ are formally defined as the group of the Wilson loop in \s{sec:wilsoniansymm}; any group of the Wilson loop corresponds to a group extension classified by \q{classifymani}. That is, our cohomological classification of quasimomentum circles determines the representation of point-group symmetries that constrain the Wilson loop, whether linear or projective. The particular representation determines the rules that govern the connectivity of Wilson energies (`curves'), as we elaborate in \s{sec:rules}; we then connect the `curves' in all possible legal ways, as in \s{sec:connect} -- distinct connectivities of the Wilson energies correspond to topologically inequivalent groundstates. This program of interpolating and matching curves, when carried out for the space group $D_{6h}^4$, produces the classification summarized in Tab.\ \ref{classification}. \\


\begin{table}[ht]
	\centering
		\begin{tabular} {C{1.5cm} C{0.001cm} |C{1.5cm}|C{1.5cm}|C{1.5cm}|} \cline{3-5}
			 & & $\calq_{\sma{\tilg \tilz}}$ & $\calq_{\sma{\tilx \tilu}}$& $\calc_e$\\ \cline{3-5} 
			\cline{1-1} \multicolumn{1}{|C{2cm}|}{$\glide=0$} &   & $\Z_2$ & $\Z_2$ & $2\Z$   \\	\cline{3-5}	
			\cline{1-1} \multicolumn{1}{|C{2cm}|}{$\glide=e/2$} &  & - & $\Z_2$  &  $2\Z+1$ \\ \cline{1-1}	 \cline{3-5}
		\end{tabular}
		\caption{Classification of time-reversal-invariant insulators with the space group $D_{6h}^4$. A quantized polarization invariant ($\glide$) distinguishes between two families of insulators: modulo the electron charge $e$, $\glide{=}e/2$ (${=}0$) characterizes the `quantum glide Hall effect' (resp., its absence). Specifically, $\glide$ is the polarization of one of two glide subspaces (as labelled by $\eta {=}{\pm}1$), but time reversal symmetry ensures there is only one independent polarization: $\glidep{=}\glidem$ modulo $e$. The $\glide{=}0$ family is further sub-classified by two non-Abelian polarizations: $\calq_{\sma{\tilg \tilz}} {\in}\, \Z_2$ and $\calq_{\sma{\tilx \tilu}} {\in}\, \Z_2$, and a mirror Chern number ($\calc_e$) that is constrained to be even; where $\calq_{\sma{\tilg \tilz}} {\neq} \calq_{\sma{\tilx \tilu}}$, the insulator manifests an hourglass-flow topology. The $\glide{=}e/2$ family is sub-classified by $\calq_{\sma{\tilx \tilu}}{\in}\, \Z_2$ and odd $\calc_e$.
		 \label{classification}}
\end{table}

Beyond $D_{6h}^4$, we note that \q{classifymani} and the Wilson-loop method provide a unifying framework to classify chiral topological insulators,\cite{Haldane1988} and all topological insulators with robust edge states protected by space-time symmetries. Here, we refer to topological insulators with either symmorphic\cite{Classification_Chiu,fu2011,AAchen} or nonsymmorphic spatial symmetries\cite{ChaoxingNonsymm,unpinned,Shiozaki2015,Nonsymm_Shiozaki}, the time-reversal-invariant quantum spin Hall phase,\cite{kane2005B} and magnetic topological insulators.\cite{moore2010,fang2013,liu2013,magnetic_ti} These case studies are characterized by extensions of $\gs$ by $\Z_2{\times} \Z^{d}$; on the other hand, extensions by quasimomentum translations are necessary to describe the space group $D_{6h}^4$, but have not been considered in the literature. In particular, $D_{6h}^4$ falls outside the K-theoretic classification of nonsymmorphic topological insulators in Ref.\ [\onlinecite{Nonsymm_Shiozaki}]. \\

Finally, we remark that the method of Wilson loops (synonymous\cite{AA2014} with the method of Wannier centers\cite{Maryam2014}) is actively being used in topologically classifying band insulators.\cite{AA2014,yu2011,alexey2011,Maryam2014,berryphaseTCI} The present work advances the Wilson-loop methodology by: (i) relating it to group cohomology through \q{classifymani}, (ii) providing a systematic summary of the method (in this Section), and (ii) demonstrating how to classify a piecewise-topological insulator for the case study $D_{6h}^4$ (cf.\ \s{sec:curvematching}).

\section{Preliminaries}

\subsection{Review of the tight-binding method} \label{sec:general}

In the tight-binding method, the Hilbert space is reduced to a finite number of \low orbitals $\varphi_{\boldsymbol{R},\alpha}$, for each unit cell labelled by the Bravais lattice (BL) vector $\boldsymbol{R}$.\cite{slater1954,goringe1997,lowdin1950} In Hamiltonians with discrete translational symmetry, our basis vectors are
\begin{align} \label{basisvec}
\phi_{\boldsymbol{k}, \alpha}(\boldsymbol{r}) = \tfrac{1}{\sqrt{N}} \sum_{\boldsymbol{R}} e^{i\boldsymbol{k} \cdot (\boldsymbol{R}+\boldsymbol{r_{\alpha}})} \pdg{\varphi}_{\boldsymbol{R},\alpha}(\boldsymbol{r}-\boldsymbol{R}-\boldsymbol{r_{\alpha}}),
\end{align}
where $\alpha=1,\ldots, n_{tot}$, $\boldsymbol{k}$ is a crystal momentum, $N$ is the number of unit cells, $\alpha$ labels the \low orbital, and $\boldsymbol{r_{\alpha}}$ denotes the position of the orbital $\alpha$ relative to the origin in each unit cell. The tight-binding Hamiltonian is defined as 
\begin{align}
H(\boldsymbol{k})_{\alpha \beta} = \int d^dr\,\phi^*_{\boldsymbol{k},\alpha}(\boldsymbol{r}) \,\hat{H} \,\phi_{\boldsymbol{k},\beta}(\boldsymbol{r}), 
\end{align}
where $\hat{H}$ is the single-particle Hamiltonian. The energy eigenstates are labelled by a band index $n$, and defined as $\psi_{n,\boldsymbol{k}}(\boldsymbol{r}) = \sum_{\alpha=1}^{n_{tot}}  \,u_{n,\boldsymbol{k}}(\alpha)\,\phi_{\boldsymbol{k}, \alpha}(\boldsymbol{r})$, where  
\begin{align}
\sum_{\beta=1}^{n_{tot}} H(\boldsymbol{k})_{\ab} \,u_{n,\boldsymbol{k}}(\beta)  = \varepsilon_{n,\boldsymbol{k}}\,u_{n,\boldsymbol{k}}(\alpha).
\end{align}
We employ the braket notation:
\begin{align} \label{energyeigen}
H(\boldsymbol{k})\,\ket{u_{n,\boldsymbol{k}}} = \varepsilon_{n,\boldsymbol{k}}\,\ket{u_{n,\boldsymbol{k}}}.
\end{align}
Due to the spatial embedding of the orbitals, the basis vectors $\phi_{\boldsymbol{k},\alpha}$ are generally not periodic under $\boldsymbol{k} \rightarrow \boldsymbol{k}+\boldsymbol{G}$ for a reciprocal vector $\boldsymbol{G}$. This implies that the tight-binding Hamiltonian satisfies:
\begin{align} \label{aperiodic}
H(\boldsymbol{k}+\boldsymbol{G}) = V(\boldsymbol{G})^{\mo}\,H(\boldsymbol{k})\,V(\boldsymbol{G}),
\end{align}
where $V(\boldsymbol{G})$ is a unitary matrix with elements: $[V(\boldsymbol{G})]_{\ab} = \delta_{\ab}\,e^{i\boldsymbol{G}\cdot \boldsymbol{r_{\alpha}}}$. We are interested in Hamiltonians with a spectral gap that is finite for all $\bk$, such that we can distinguish occupied from empty bands;  the former are projected by
\begin{align} \label{periodP}
P(&\bk) = \sum_{n =1}^{\noc} \ket{u_{n,\bk}}\bra{u_{n,\bk}} \lin
\eq V(\bG)\,P(\bk+\bG)\,V(\bG)^{\mo},
\end{align}
where the last equality follows directly from Eq.\ (\ref{aperiodic}).


\subsection{Crystal structure and spatial symmetries}

\begin{figure}[H]
\centering
\includegraphics[width=8 cm]{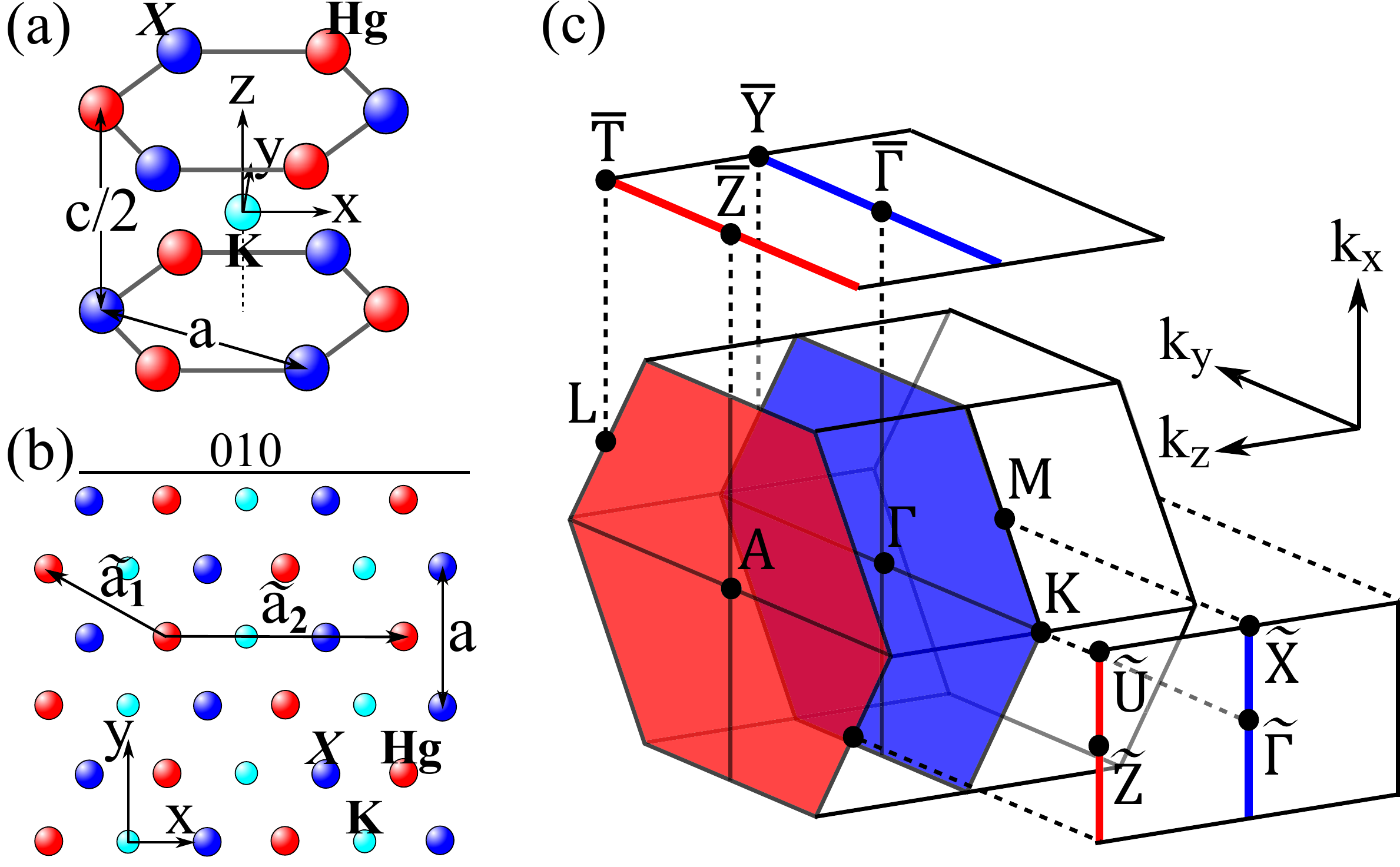}
    \caption{(a) 3D view of atomic structure. The Hg (red) and $X$ (blue) ions form a honeycomb layers with AB stacking. The K ion (cyan) is located at an inversion center, which we also choose to be our spatial origin. (b) Top-down view of atomic structure that is truncated in the 010 direction; two of three Bravais lattice vectors are indicated by $\tilde{\boldsymbol{a}}_1$ and $\tilde{\boldsymbol{a}}_2$. (c) Center: bulk Brillouin zone (BZ) of KHg$X$, with two mirror planes of $\bmz$ colored red and blue. Top: 100-surface BZ. Right: 010-surface BZ.  } \label{fig:structure}
\end{figure}

The crystal structure KHg$X$ is chosen to exemplify the spatial symmetries we study. As illustrated in Fig.\ \ref{fig:structure}, the Hg and $X$ ions form honeycomb layers with AB stacking along $\vec{z}$; here, $\vec{x},\vec{y},\vec{z}$ denote unit basis vectors for the Cartesian coordinate system drawn in the same figure. Between each AB bilayer sits a triangular lattice of K ions. The space group ($D_{\sma{6h}}^{4} {\equiv} P6_3/mmc$) of KHg$X$ includes the following symmetries: (i)  an inversion ($\cali$) centered around a K ion (which we henceforth take as our spatial origin),  the reflections (ii) $\bar{M}_z=t(c\vec{z}/2)M_z$, and (iii) $\bar{M}_x=t(c\vec{z}/2)M_x$, where $M_j$ inverts the coordinate $j \in \{x,y,z\}$. In (ii-iii) and the remainder of the paper, we denote, for any transformation $g$, $\bar{g} =  t(c\vec{z}/2) \, g$ as a product of $g$ with a translation ($t$) by half a lattice vector ($c\vec{z}/2$). Among (ii-iii), only $\bar{M}_x$ is a glide reflection, wherefor the fractional translation is unremovable\cite{Lax} by a different choice of origin. While we primarily focus on the symmetries (i-iii), they do not generate the full group of $D_{\sma{6h}}^{4}$, e.g., there exists also a six-fold screw symmetry whose implications have been explored in our companion paper.\cite{Hourglass}\\

We are interested in symmetry-protected topologies that manifest on surfaces. Given a surface termination, we refer to the subset of bulk symmetries which are preserved by that surface as edge symmetries.  The edge symmetries of the 100 and 001 surfaces are symmorphic, and they have been previously addressed in the context of KHg$X$.\cite{Hourglass}  Our paper instead focuses on the 010 surface, whose edge group (nonsymmorphic $Pma2$) is generated by two reflections: glideless $\bmz$ and glide $\bmx$.  \\


\section{Wilson loops and the bulk-boundary correspondence} \label{sec:reviewWilson}

We review the Wilson loop in Sec.\ \ref{subsec:reviewwilson}, as well as introduce the loop geometry that is assumed throughout this paper. The relation between Wilson loops and the geometric theory of polarization is summarized in Sec.\ \ref{subsec:bb}. There, we also introduce the notion of a partial bulk-boundary correspondence, which our nonsymmorphic insulator exemplifies.

\subsection{Review of Wilson loops} \label{subsec:reviewwilson}

\begin{figure}[h]
\centering
\includegraphics[width=8.6 cm]{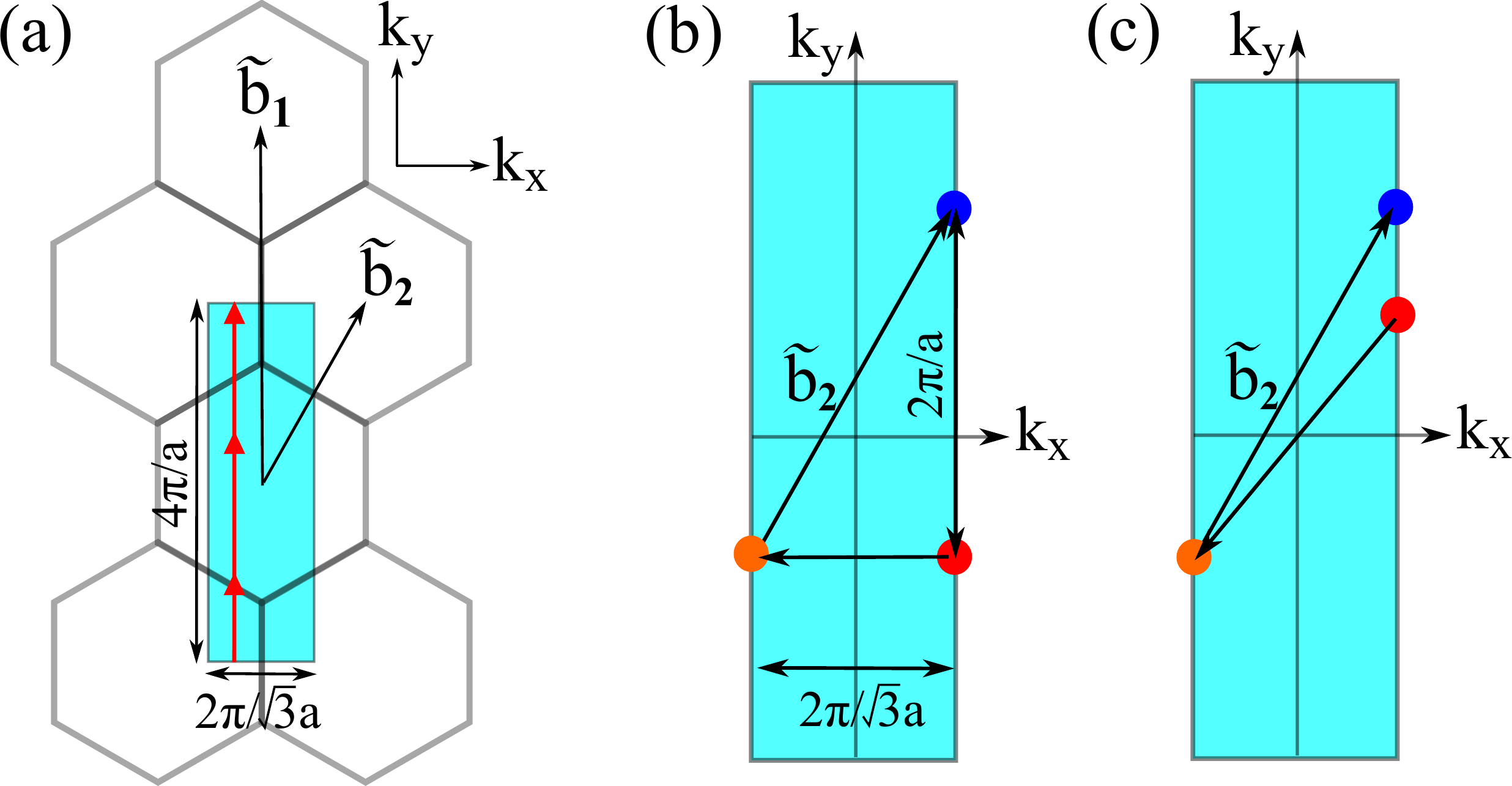}
    \caption{(a) A constant-$k_z$ slice of quasimomentum space, with two of three reciprocal lattice vectors indicated by $\tilde{\boldsymbol{b}}_1$ and $\tilde{\boldsymbol{b}}_2$. While each hexagon corresponds to a Wigner-Seitz primitive cell, it is convenient to pick the rectangular primitive cell that is shaded in cyan. A close-up of this cell is shown in (b). Here, we illustrate how the glide reflection ($\bmx$) maps $(k_y,\pi/\sqrt{3}a,k_z) \rightarrow (k_y,-\pi/\sqrt{3}a,k_z)$ (red dot to brown)  which connects to $(2\pi/a+k_y,\pi/\sqrt{3}a,k_z)$ (blue) through $\tilde{\boldsymbol{b}}_2$. Figure (c) serves two interpretation. In the first, $T\bmz$ maps  $(k_y,\pi/\sqrt{3}a,k_z) \rightarrow (-k_y,-\pi/\sqrt{3}a,k_z)$ (red dot to brown)  which connects to $(2\pi/a-k_y,\pi/\sqrt{3}a,k_z)$ (blue) through $\tilde{\boldsymbol{b}}_2$. If we interpret Figure (c) as the $k_z=0$ cross-section, the same vectors illustrate the effect of time reversal.} \label{fig:wherewilson}
\end{figure}

The matrix representation of parallel transport along Brillouin-zone loops is known as the Wilson loop of the Berry gauge field. It may be expressed as the path-ordered exponential (denoted by $ \bar{\text{exp}}$) of the Berry-Wilczek-Zee connection\cite{wilczek1984,berry1984} $\boldsymbol{A}(\boldsymbol{k})_{ij} = \braket{u_{i,\boldsymbol{k}}}{\nabla_{\boldsymbol{k}}\,u_{j,\boldsymbol{k}}}$: 
\begin{align} \label{wloopdifferentiable}
\W[l] = \bar{\text{exp}}\,\big[{-\int_l} dl \cdot \boldsymbol{A}(\boldsymbol{k})\,\big].
\end{align}
Here, recall from Eq.\ (\ref{energyeigen}) that $\ket{u_{j,\boldsymbol{k}}}$ is an occupied eigenstate of the tight-binding Hamiltonian; $l$ denotes a loop and $\boldsymbol{A}$ is a matrix with dimension equal to the number ($\noc$) of occupied bands. The gauge-invariant spectrum of $\W[l]$ is the non-abelian generalization of the Berry phase factors (Zak phase factors\cite{zak1989}) if $l$ is contractible (resp. non-contractible).\cite{AA2014,Maryam2014} In this paper, we consider only a family of loops parametrized by $\kpar {=}(k_x {\in} [-\pi/\sqrt{3}a,+\pi/\sqrt{3}a],k_z{\in} [-\pi/c,+\pi/c])$, where for each loop $\kpar$ is fixed while $k_y$ is varied over a non-contractible circle $[{-}2\pi/a,{+}2\pi/a]$ (oriented line with three arrowheads in Fig.\ \ref{fig:wherewilson}(a)). We then label each Wilson loop as $\W(\kpar)$ and denote its eigenvalues by exp$[i\theta_{\sma{n,\kpar}}]$ with $n{=}1,\ldots, \noc$. Note that $\kpar$ also parametrizes the 010-surface bands, hence we refer to $\kpar$ as a surface wavevector; here and henceforth, we take the unconventional ordering $\bk {=} (k_y,k_x,k_z) {=} (k_y,\kpar)$. To simplify notation in the rest of the paper, we reparametrize the rectangular primitive cell of Fig.\ \ref{fig:wherewilson} as a cube of dimension $2\pi$, i.e., $k_x{=}{\pm} \pi/\sqrt{3}a \rightarrow k_x{=} {\pm} \pi$, $k_y{=}{\pm} 2\pi/a \rightarrow k_y{=}{\pm} \pi$, and $k_z{=} {\pm} \pi/c \rightarrow k_z{=}{\pm} \pi$. The time-reversal-invariant $\kpar$ are then labelled as: $\tilde{\Gamma}{=}(0,0)$, $\tilde{X} {=} (\pi,0)$, $\tilde{Z} {=} (0,\pi)$ and $\tilde{U} {=} (\pi,\pi)$. For example, $\W(\tilde{\Gamma})$ would correspond to a loop parametrized by $(k_y,0,0)$.

\subsection{Bulk-boundary correspondence of topological insulators} \label{subsec:bb}

The bulk-boundary correspondence describes topological similarities between the Wilson loop and the surface bandstructure. To sharpen this analogy, we refer to the eigenvectors of $\W(\kpar)$ as forming Wilson bands with energies $\theta_{n,\kpar}$. The correspondence may be understood in two steps: \\

\noi{i} The first is a spectral equivalence between $(-i/2\pi)$log$\W(\kpar)$ and the projected-position operator $\pper(\kpar)\hat{y}\pper(\kpar)$, where
\begin{align}
\pper(\kpar) = \sum_{n=1}^{\noc}\int_{-\pi}^{\pi} \frac{dk_y}{2\pi} \ketbra{\psi_{n,k_y,\kpar}}{\psi_{n,k_y,\kpar}}
\end{align}
projects to all occupied bands with surface wavevector $\kpar$, and $\psi_{\sma{n,\bk}}(\br){=}\text{exp}(i\bk\cdot \br)u_{\sma{n,\bk}}(\br)$ are the Bloch-wave eigenfunctions of $\hat{H}$. For the position operator $\hat{y}$, we have chosen natural units of the lattice where $1 {\equiv} a/2 {=}\tilde{\boldsymbol{a}}_1{\cdot} \vec{y}$, and $\tilde{\boldsymbol{a}}_1/a{=}{-}\sqrt{3}\vec{x}/2{+}\vec{y}/2$ is the lattice vector indicated in Fig.\ \ref{fig:structure}(b). Denoting the eigenvalues of $\pper(\kpar)\hat{y}\pper(\kpar)$ as ${y_{\sma{n,\kpar}}}$, the two spectra are related as $y_{\sma{n,\kpar}} = \theta_{\sma{n,\kpar}}/2\pi$ modulo one.\cite{AA2014} Some intuition about the projected-position operator may be gained from studying its eigenfunctions; they form a set of hybrid functions $\{|\kpar,n \rangle| n {\in} \{1,2,\ldots, \noc\} \}$ which maximally localize in $\vec{y}$ (as a Wannier function) but extend in $\vec{x}$ and $\vec{z}$ (as a Bloch wave with momentum $\kpar{=}\,(k_x,k_z)$). In this Bloch-Wannier (BW) representation,\cite{Maryam2014}  the eigenvalue ($y_{\sma{n,\kpar}}$) under $\pper \hat{y} \pper$ is merely the center-of-mass coordinate of the BW function ($\ket{n,\kpar}$).\cite{alexey2011,AA2014} Since $\pper$ is symmetric under translation $(\,t(\tilde{\boldsymbol{a}}_1)\,)$ by $\tilde{\boldsymbol{a}}_1$, while $t(\tilde{\boldsymbol{a}}_1)\,\hat{y}\,t(\tilde{\boldsymbol{a}}_1)^{\sma{-1}}{=} \hat{y}-I$, each of $\{y_{\sma{n,\kpar}}| n {\in} \{1,2,\ldots, \noc\}\}$ represents a family of BW functions related by integer translations. The Abelian polarization ($\calp/e$) is defined as the net displacement of BW functions:\cite{kingsmith1993,vanderbilt1993,resta1994} 
\begin{align} \label{abelianpol}
\frac{\calp_{\kpar}}{e} \eq \frac1{2\pi}\,\sum_{j=1}^{\noc} \theta_{j,\kpar} = \frac{i}{2\pi}\int_{-\pi}^{\pi} \text{Tr}  {A}_y(k_y,\kpar) dk_y,
\end{align}
where all equalities are defined modulo integers, and Tr$A_y(\bk){=}\sum_{i=1}^{\noc}\braket{u_{i,\boldsymbol{k}}}{\nabla_y u_{i,\boldsymbol{k}}}$ is the Abelian Berry connection.\\

\noi{ii} The next step is an interpolation\cite{fidk2011,ZhoushenHofstadter}  between $\pper\hat{y}\pper$ and an open-boundary Hamiltonian ($H_s$) with a boundary termination. Presently, we assume for simplicity that each of $\{\pper, \hat{y},H_s\}$ is invariant under space-time transformations of the edge group. A simple example is the 2D quantum spin Hall (QSH) insulator, where time reversal ($T$) is the sole edge symmetry: by assumption $T$ is a symmetry of the periodic-boundary Hamiltonian (hence also of $\pper$); furthermore, since $T$ acts locally in space, it is also a symmetry of $\hat{y}$ and $H_s$. It has been shown in Ref.\ \onlinecite{fidk2011} that the discrete subset of the $H_s$-spectrum (corresponding to edge-localized states) is deformable into a subset of the fully-discrete $\pper\hat{y}\pper$-spectrum. More physically, a subset of the BW functions mutually and continuously hybridize into edge-localized states when a boundary is slowly introduced, and the edge symmetry is preserved throughout this hybridization. Consequently, $\pper\hat{y}\pper$ (equivalently, log[$\W$]) and $H_s$ share certain traits which are only well-defined in the discrete part of the spectrum, and moreover these traits are robust in the continued presence of said symmetries. The trait that identifies the QSH phase (in both the Zak phases and the edge-mode dispersion) is a zigzag connectivity where the spectrum is discrete; here, eigenvalues are well-defined, and they are Kramers-degenerate at time-reversal-invariant momenta but otherwise singly-degenerate, and furthermore all Kramers subspaces are connected in a zigzag pattern.\cite{yu2011,alexey2011,AA2014}  In the QSH example, it might be taken for granted that the representation ($T^2{=}{-}I$) of the edge symmetry is identical for both $H_s$ and $\W$; the invariance of $T^2{=}{-}I$ throughout the interpolation accounts for the persistence of Kramers degeneracies, and consequently for the entire zigzag topology. The QSH phase thus exemplifies a \emph{total} bulk-boundary correspondence, where the entire set of boundary topologies (i.e., topologies that are consistent with the edge symmetries of $H_s$) is in one-to-one correspondence with the entire set of $\W$-topologies (i.e., topologies which are consistent with symmetries of $\W$, of which the edge symmetries form a subset). One is then justified in inferring the topological classification purely from the representation theory of surface wavefunctions -- this surface-centric methodology has successfully been applied to many space groups.\cite{AAchen,ChaoxingNonsymm,TCIbyrepresentation} \\

While this surface-centric approach is technically easier than the representation theory of Wilson loops, it ignores the bulk symmetries that are spoilt by the boundary. On the other hand, $\W$-topologies encode these bulk symmetries, and are therefore more reliable in a topological classication. In some cases,\cite{AA2014,hughes2011,turner2012,berryphaseTCI} these bulk symmetries \emph{enable} $\W$-topologies that have no boundary analog. Simply put, some topological phases do not have robust boundary states, a case in point being the $\Z$ topology of 2D inversion-symmetric insulators.\cite{AA2014} In our nonsymmorphic case study, it is an out-of-surface translational symmetry ($t(\tilde{\boldsymbol{a}}_1)$) that \emph{disables} a $\W$-topology, and consequently a naive surface-centric approach would over-predict the topological classification -- this exemplifies a \emph{partial} bulk-boundary correspondence. As we will clarify, the $t(\tilde{\boldsymbol{a}}_1)$ symmetry distinguishes between two representations of the same edge symmetries: an ordinary representation with the open-boundary Hamiltonian ($H_s$), and a projective one with the Wilson loop ($\W$). To state the conclusion upfront, the projective representation rules out a quantum glide Hall topology that would otherwise be allowed in the ordinary representation. This discussion motivates a careful determination of the $\W$-topologies in Sec.\ \ref{sec:curvematching}.

\section{Constructing a piecewise-topological insulator by Wilson loops} \label{sec:curvematching}

We would like to classify time-reversal-invariant insulators with the space group $D_{6h}^4$; our result should more broadly apply to hexagonal crystal systems with the edge symmetry $Pma2$ (generated by glide $\bmx$ and glideless $\bmz$) and a bulk spatial-inversion symmetry. Our final result in Tab.\ \ref{classification} relies on  topological invariants that we briefly introduce here, deferring a detailed explanation to the sub-sections below. The invariants are: (i) the mirror Chern number ($\calc_e$) in the $k_z{=}0$ plane, (ii) the quadruplet polarization $\calq_{\sma{\tilg \tilz}}$ ($\calq_{\sma{\tilx \tilu}}$) in the $k_x{=}0$ glide plane (resp.\ $k_x{=}\pi$), wherefor $\calq_{\sma{\tilg \tilz}}{\neq}\calq_{\sma{\tilx \tilu}}$ implies an hourglass flow, and (iii) the glide polarization $\glide{=}0$ ($e/2$) indicates the absence (resp.\ presence) of the quantum glide Hall effect in the $k_x{=}0$ plane. \\

Our strategy for classification is simple: we first derive the symmetry constraints on the Wilson-loop spectrum, then enumerate all topologically distinct spectra that are consistent with these constraints. Pictorially, this amounts to understanding the rules obeyed by curves (the Wilson bands), and connecting curves in all possible legal ways; we do these in Sec.\ \ref{sec:rules} and \ref{sec:connect} respectively.

\subsection{Local rules of the curves} \label{sec:rules}

We consider how the bulk symmetries  constrain the Wilson loop $\W(\kpar)$, with $\kpar$ lying on the high-symmetry line $\tilde{\Gamma}\tilde{X}\tilde{U}\tilde{Z}\tilde{\Gamma}$; note that $\tilde{\Gamma}\tilde{Z}$ and $\tilde{X}\tilde{U}$ are glide lines which are invariant under $\bmx$, while $\tilde{\Gamma}\tilde{X}$ and $\tilde{Z}\tilde{U}$ are mirror lines invariant under $\bmz$. The relevant symmetries that constrain $\W(\kpar)$ necessarily preserve the circle $(k_y{\in}[{-}\pi,\pi],\kpar)$, modulo translation by a reciprocal vector; such symmetries comprise the little group of the circle.\cite{AAchen} For example, (a) $T\cali$ would constrain $\W(\kpar)$ for all $\kpar$, (b) $T\bmx$ and $\bmz$ is constraining only for $\kpar$ along $\tilde{\Gamma}\tilde{X}$, and (c) $T$ matters only at the time-reversal-invariant $\kpar$. Along $\tilde{\Gamma}\tilde{X}$, we omit discussion of other symmetries (e.g., $TC_{2y}$) in the group of the circle, because they do not additionally constrain the Wilson-loop spectrum. For each symmetry, only three properties influence the connectivity of curves, which we first state succinctly: \\

\noi{i} Does the symmetry map each Wilson energy as $\theta {\rightarrow} \theta$ or $\theta {\rightarrow}{-}\theta$? Note here we have omitted the constant argument of $\theta_{\sma{\kpar}}$.  \\

\noi{ii} If the symmetry maps $\theta {\rightarrow} \theta$, does it also result in Kramers-like degeneracy? By `Kramers-like', we mean a doublet degeneracy arising from an antiunitary symmetry which representatively squares to $-1$, much like time-reversal symmetry in half-integer-spin representations.\\

\noi{iii} How does the symmetry transform the mirror eigenvalues of the Wilson bands? Here, we refer to the eigenvalues of mirror $\bmz$ and glide $\bmx$ along their respective invariant lines.\\

\noindent To elaborate, (i) and (ii) are determined by how the symmetry constrains the Wilson loop. We say that a symmetry represented by $\calt_{\pm}$ is time-reversal-like at $\kpar$, if for that $\kpar$
\begin{align} \label{Ttype}
\calt_{\pm}\,\W(\kpar)\,\calt_{\pm}^{\mo}&  =  {\W(\kpar)}^{\mo}, \lin
\ins{with}\calt_{\pm} \,i \,\calt_{\pm}^{\mo} = -i,& \ins{and} \calt_{\pm}^2 = \pm I.
\end{align}
Both $\calt_{\pm}$ map the Wilson energy as $\theta {\rightarrow} \theta$, but only $\calt_-$ symmetries guarantee a Kramers-like degeneracy. Similarly, a symmetry represented by $\calu$ is particle-hole-like at $\kpar$, if for that $\kpar$
\begin{align} \label{Ptype}
\calu\,\W(\kpar)\,\calu^{\mo}  =  {\W(\kpar)},\ins{with} \calu \,i \,\calu^{\mo} = -i,
\end{align}
i.e., $\calu$ maps the Wilson energy as $\theta {\rightarrow} {-}\theta$. Here, we caution that $\calt$ and $\calu$ are symmetries of the circle $(k_y{\in}[{-}\pi,\pi],\kpar)$ and preserve the momentum parameter $\kpar$; this differs from the conventional\cite{schnyder2009} time-reversal and particle-hole symmetries which typically invert momentum. \\

To precisely state (iii), we first elaborate on how Wilson bands may be labelled by mirror eigenvalues, which we define as $\lambda_j$ for the reflection $\bar{M}_j$ ($j \in \{x,z\}$). First consider the glideless $\bmz$, which is a symmetry of any \emph{bulk} wavevector which projects to $\tilde{\Gamma}\tilde{X}$ ($k_z{=}0$) and $\tilde{Z}\tilde{U}$ ($k_z{=}\pi$) in $\vec{y}$. Being glideless, $\bmz^2{=}\bar{E}$ ($2\pi$ rotation of a half-integer spin) implies two momentum-independent branches for the eigenvalues of $\bmz$: $\lambda_z {=} {\pm} i$; this eigenvalue is an invariant of any parallel transport within either $\bmz$-invariant plane. That is, if $\psi_1$ is a mirror eigenstate, any state related to $\psi_1$ by parallel transport must have the same mirror eigenvalue. Consequently, the Wilson loop block-diagonalizes with respect to $\lambda_z{=}{\pm}i$, and any Wilson band may be labelled by $\lambda_z$.\\

A similar story occurs for the glide $\bmx$, which is a symmetry of any bulk wavevector that projects to $\tilde{\Gamma}\tilde{Z}$ ($k_x{=}0$). The only difference from $\bmz$ is that the two branches of $\lambda_x$ are momentum-dependent, which follows from $\bmx^2 =t(\vec{z})\,\bar{E} $, with $t$ denoting a lattice translation. Explicitly, the Bloch representation of $\bmx$ squares to ${-}$exp$({-}ik_z)$, which implies for the glide eigenvalues: $\lambda_x(k_z){=}{\pm} i$exp$({-}ik_z/2)$.\\

To wrap up our discussion of the mirror eigenvalues, we consider the subtler effect of $\bmx$ along $\tilde{X}\tilde{U}$. Despite being a symmetry of any surface wavevector along $\tilde{X}\tilde{U}$:
\begin{align}
\bmx: (\pi,k_z) \longrightarrow ({-}\pi,k_z) = (\pi,k_z)-2\pi\vec{x},
\end{align}
with $2\pi\vec{x}$ a surface reciprocal vector, $\bmx$ is not a symmetry of any bulk wavevector that projects to $\tilde{X}\tilde{U}$, but instead relates two bulk momenta which are separated by half a bulk reciprocal vector, i.e., 
\begin{align}
\bmx: (k_y,\pi,k_z) \longrightarrow (k_y,{-}\pi,k_z) = (k_y+\pi,\pi,k_z)-\tilde{\boldsymbol{b}}_2, \notag
\end{align}
as illustrated in Fig.\ \ref{fig:wherewilson}(b). This reference to Fig.\ \ref{fig:wherewilson}(b) must be made with our reparametrization ($k_x{=}{\pm} \pi/\sqrt{3}a {\rightarrow} k_x{=} {\pm} \pi$, $k_y{=}{\pm} 2\pi/a {\rightarrow} k_y{=}{\pm} \pi$) in mind. We refer to such a glide plane as a \emph{projective} glide plane, to distinguish it from the \emph{ordinary} glide plane at $k_x{=}0$. The absence of $\bmx$ symmetry at each bulk wavevector implies that the Wilson loop cannot be block-diagonalized with respect to the eigenvalues of $\bmx$. However, quantum numbers exist for a generalized symmetry ($\calbmx$) that combines the glide reflection with parallel transport over half a reciprocal period. To be precise, let us define the Wilson line $\W_{\sma{-\pi \leftarrow 0}}$ to represent the parallel transport from $(0,\pi,k_z)$ to $(-\pi,\pi,k_z)$. We demonstrate in Sec.\ \ref{sec:wilsoniansymm} that all Wilson bands may be labelled by quantum numbers under $\calbmx {\equiv} \W_{\sma{-\pi \leftarrow 0}}\,\bmx$, and that these quantum numbers fall into two energy-dependent branches as: 
\begin{align} \label{relativistic}
\lambda_x(\theta+k_z) = \eta\, i \,\text{exp}\,[-i(\theta+k_z)/2] \;\;\text{with}\;\; \eta = \pm 1.
\end{align}
That is, $\lambda_x(\theta{+}k_z)$ is the $\calbmx$-eigenvalue of a Wilson band at surface momentum $(\pi,k_z)$ and Wilson energy $\theta$.\\

For the purpose of topological classification, all we need are the existence of these symmetry eigenvalues (ordinary and generalized) that fall into two branches (recall $\lambda_z{=}{\pm} i, \lambda_x{=}{\pm} i \text{exp}(-ik_z/2)$ along $\tilg \tilz$, and also Eq.\ (\ref{relativistic})\,), and (iii) asks whether the $\calt_{\pm}$- and $\calu$-type symmetries preserve ($\lambda_j {\rightarrow} \lambda_j$) or interchange ($\lambda_j {\rightarrow} {-}\lambda_j$) the branch. To clarify a possible confusion, both $\calt_{\pm}$- and $\calu$-type symmetries are antiunitary and therefore have no eigenvalues, while the reflections ($\bmz,\bmx$ (along $\tilg\tilz$), and $\calbmx$ (along $\tilx\tilu$)\,) are unitary. The answer to (iii) is determined by the commutation relation between the symmetry in question (whether $\calt_{\pm}$- or $\calu$-type) and the relevant reflection. To exemplify (i-iii), let us evaluate the effect of $T\cali$ symmetry along $\tilde{\Gamma}\tilde{Z}$. This may be derived in the polarization perspective, due to the spectral equivalence of $({-}i/2\pi)$log$\W(\kpar)$ and $\pper(\kpar)\hat{y}\pper(\kpar)$. Since $T \cali$ inverts all spatial coordinates but transforms any momentum to itself ($y_{\sma{n,\kpar}} {=} \theta_{\sma{n,\kpar}}/2\pi {\rightarrow} {-}y_{\sma{n,\kpar}}$), we identify $T\cali$ as a $\calu$-type symmetry (cf. Eq.\ (\ref{Ptype})). Indeed, while $T\cali$ is known to produce Kramers degeneracy in the Hamiltonian spectrum, $T\cali$ emerges as an unconventional particle-hole-type symmetry in the Wilson loop. Since $\bmx$ commutes individually with $\pper(0,k_z)$ and $\hat{y}$, all eigenstates of $\pper(0,k_z)\hat{y}\pper(0,k_z)$ may simultaneously be labelled by $\lambda_x$. That $T\cali$ maps $\lambda_x {\rightarrow} {-}\lambda_x$  then follows from $\bmx T \cali {=} t(\vec{z}) T \cali \bmx$, where $t(\vec{z})$ originates simply from the noncommutivity of $\cali$ with the fractional translation ($t(\vec{z}/2)$) in $\bmx$:
\begin{align} \label{algebrabmxi}
\bmx\, \cali = t(\vec{z})\, \cali\, \bmx.
\end{align}
To show $T\cali: \lambda_x {\rightarrow} {-}\lambda_x$ in more detail, suppose for a Bloch-Wannier function $|n,k_z\rangle$ that
\begin{align}
\pper \hat{y}\pper \,\ket{n,k_z} \eq y_{\sma{n,k_z}}\,\ket{n,k_z} \ins{and} \lin
 \bmx \,\ket{n,k_z} \eq \lambda_x(k_z)\,\ket{n,k_z}, 
\end{align}
with $\lambda_x(k_z) {=} {\pm} i\,$exp($-ik_z/2)$ and suppression of the label $k_x{=}0$. $[T\cali,\pper]{=}\{T\cali,\hat{y}\}{=}0$ then leads to 
\begin{align}
\pper \hat{y}\pper \,T\cali\ket{n,k_z} \eq  -y_{\sma{n,k_z}}\,\ket{n,k_z}, \ins{and} \lin
 \bmx \,T\cali\ket{n,k_z} \eq t(\vec{z}) T \cali \bmx\,\ket{n,k_z} \lin
\eq e^{-ik_z}\lambda_x^*\,T\cali\ket{n,k_z}, 
\end{align}
with exp$({-ik_z})\lambda_x^*{=}{-}\lambda_x$ following from $\lambda_x^2{=}{-}$exp$(-ik_z)$. To recapitulate, (a) $T\cali$ imposes a particle-hole-symmetric spectrum, and (b) two states related by $T\cali$ have opposite eigenvalues under $\bmx$. (a-b) is summarized by the notation $\calu:\lambda_x {\rightarrow} {-}\lambda_x$ in the top left entry of Tab.\ \ref{rulesgeneric}. The complete symmetry analysis is derived in Sec.\ \ref{sec:wilsoniansymm} and App.\ \ref{app:symmPyP}, and tabulated in Tab.\ \ref{rulesgeneric} and \ref{rulestrim}. These relations constrain the possible topologies of the Wilson bands, as we show in the next section.




 \begin{widetext}

\begin{table}[H]
	\centering
\begin{tabular}{C{1cm}|C{3.7cm}|C{4.9cm}|C{3.3cm}|C{3.3cm}|}
\cline{2-5}
& {$k_x=0 \;(\tilde{\Gamma}\tilde{Z})$} & {$k_x=\pi\;(\tilde{X}\tilde{U})$} & {$k_z=0\;(\tilde{\Gamma}\tilde{X})$} & {$k_z=\pi\;(\tilde{Z}\tilde{U})$} \\ \hline
\multicolumn{1}{ |c|  }{$T \cali$} & {$\calu: \lambda_{x}(k_z) \rightarrow -\lambda_x(k_z)$} & {$\calu:\lambda_x(k_z+\theta) \rightarrow -\lambda_x(k_z-\theta)$}& {$\calu:\lambda_z \rightarrow -\lambda_z$}& {$\calu: \lambda_z \rightarrow +\lambda_z$} \\ \hline  
\multicolumn{1}{ |c|  }{$T \bmz$} & {$\calt_+: \lambda_x(k_z) \rightarrow +\lambda_x(k_z)$} & {$\calt_+: \lambda_x(k_z+\theta) \rightarrow +\lambda_x(k_z+\theta)$}& {-}& {-} \\ \hline  
\multicolumn{1}{ |c|  }{$T\bmx$} & {-} & {-}& {$\calt_+: \lambda_z \rightarrow +\lambda_z$}& {$\calt_-: \lambda_z \rightarrow -\lambda_z$} \\ \hline  
\end{tabular}
		\caption{ Symmetry constraints of the Wilson bands at generic points along the mirror lines. $\calt_{\pm}$ and $\calu$ are possible characterizations of the symmetries ($T\cali,T\bmz,T\bmx$) in the left-most column: a $\calt_{\pm}$-type ($\calu$-type) symmetry is  time-reversal-like (resp.\ particle-hole-like) symmetry, as defined in Eq.\ (\ref{Ttype}) and (\ref{Ptype}). For $j \in \{x,z\}$, $\lambda_j$ is a symmetry eigenvalue which falls into one of two branches: $\lambda_z {=} {\pm} i$, and $\lambda_x(\alpha){=} {\pm} i$exp$({-}i\alpha/2)$. Along $k_x{=}0$, $\lambda_x$ is momentum-dependent; along $k_x{=}\pi$, it is energy-dependent as well, that is, $\lambda_x(k_z+\theta)$ is the glide eigenvalue of a Wilson band at momentum $k_z$ and energy $\theta$. 
		 \label{rulesgeneric}}
\end{table}

\begin{table}[H]

	\centering
\begin{tabular}{C{1cm}|C{4cm}|C{4cm}|C{4cm}|C{4cm}|}
\cline{2-5}
& $\tilde{\Gamma}=(0,0)$ & $\tilde{X}=(\pi,0)$  & $\tilde{Z}=(0,\pi)$ & $\tilde{U}=(\pi,\pi)$   \\ \hline
\multicolumn{1}{ |c|  }{$T$} & $\calt_-: \lambda_z \rightarrow -\lambda_z,$ & $\calt_-: \lambda_z \rightarrow -\lambda_z,$ & $\calt_-: \lambda_z \rightarrow -\lambda_z,$& $\calt_-: \lambda_z \rightarrow -\lambda_z,$  \\ 
\multicolumn{1}{ |c|  }{}& $\lambda_x(k_z) \rightarrow -\lambda_x(k_z)$ & $\lambda_x(k_z+\theta) \rightarrow -\lambda_x(k_z+\theta)$ & $\lambda_x(k_z) \rightarrow +\lambda_x(k_z)$& $\lambda_x(k_z+\theta) \rightarrow +\lambda_x(k_z+\theta)$  \\ \hline
\end{tabular}
		\caption{ Time-reversal constraint of the Wilson bands at surface wavevectors satisfying $\kpar {=} (k_x,k_z) {=} {-}\kpar$ modulo a reciprocal vector. To clarify a possible source of confusion, the actual time-reversal symmetry ($T$) is, by our definition of $\calt_{\pm}$ in Eq.\ (\ref{Ttype}), only `time-reversal-like' at $\kpar {=} {-}\kpar$, since $T:\kpar {\rightarrow}{-}\kpar$ is not a symmetry of the circle $(k_y{\in}[{-}\pi,\pi],\kpar)$ for generic $\kpar$. \label{rulestrim}}  
\end{table} 


\end{widetext}



\subsection{Connecting curves in all possible legal ways} \label{sec:connect}

Our goal here is to determine the possible topologies of curves (Wilson bands), which are piecewise smooth on the high-symmetry line $\tilde{\Gamma}\tilde{X}\tilde{U}\tilde{Z}\tilde{\Gamma}$. We first analyze each momentum interval separately, by evaluating the available subtopologies within each of $\tilde{\Gamma}\tilde{Z}$, $\tilde{Z}\tilde{U}$, etc. The various subtopologies are then combined to a full topology, by a program of matching curves at the intersection points (e.g.,$\tilde{Z}$) between momentum intervals.\\

Since our program here is to interpolate and match curves (Wilson bands), it is important to establish just how many Wilson bands must be connected. A combination of symmetry, band continuity and topology dictates this answer to be a multiple of four. Since the number ($\noc$) of occupied Hamiltonian bands is also the dimension of the Wilson loop, it suffices to show that $\noc$ is a multiple of four. Indeed, this follows from our assumption that the groundstate is insulating, and a property of connectedness between sets of Hamiltonian bands. For spin systems with \emph{minimally} time-reversal and glide-reflection symmetries, we prove in App.\ \ref{app:connectivity} that Hamiltonian bands divide into sets of four which are individually connected, i.e., in each set there are enough contact points to travel continuously through all four branches. The lack of gapless excitations in an insulator then implies that a connected quadruplet is either completely occupied, or unoccupied.




\subsubsection{Interpolating curves along the glide line $\tilde{\Gamma}\tilde{Z}$} \label{sec:interpolateGZ}

\noindent Along $k_x{=}0$ ($\tilg \tilz$), the rules are:\\

\begin{figure}[h]
\centering
\includegraphics[width=7 cm]{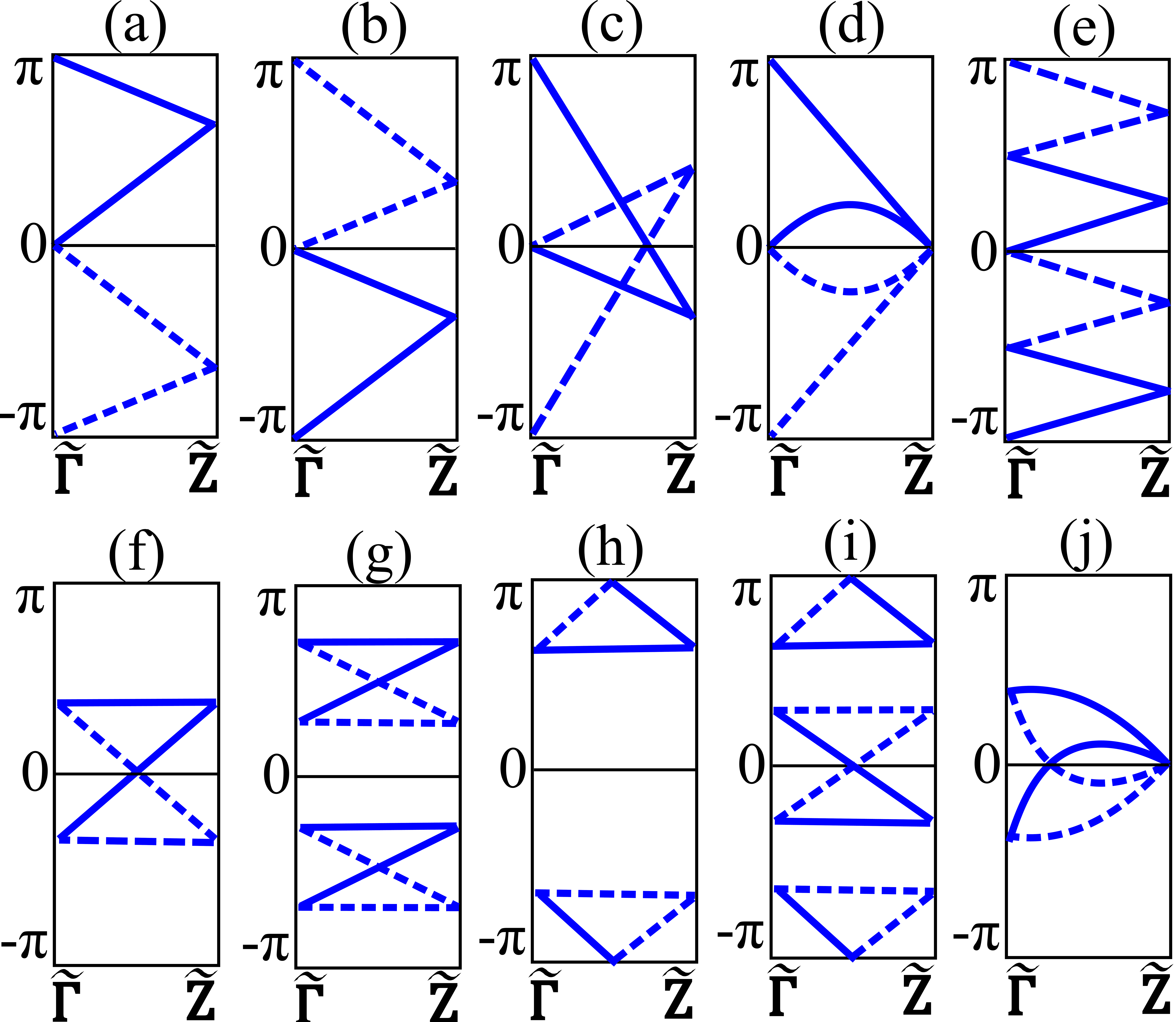}
    \caption{Possible Wilson spectra along $\tilde{\Gamma}\tilde{Z}$.} \label{fig:wilsonGZ}
\end{figure}

\noi{a} There are two flavors of curves (illustrated as blue solid and blue dashed lines in Fig.\ \ref{fig:wilsonGZ}), corresponding to two branches of the glide eigenvalue $\lambda_x{=}{\pm} i\text{exp}({-}ik_z/2)$. Only crossings between solid and dashed curves are robust, in the sense of being movable but unremovable.\\

\noi{b} At any point along $\tilde{\Gamma}\tilde{Z}$, there is an uncoventional particle-hole symmetry (due to $T\cali$) with conjugate bands (related by $\theta {\rightarrow}{-}\theta$) belonging in opposite glide branches; cf.\ first column of Tab.\ \ref{rulesgeneric}. Pictorially, [$\theta$, blue solid] $\leftrightarrow$ [$-\theta$, blue dashed].\\

\noi{c} At $\tilde{\Gamma}$, each solid curve is degenerate with a dashed curve, while at $\tilde{Z}$ the degeneracies are solid-solid and dashed-dashed; cf.\ Tab.\ \ref{rulestrim}. These end-point constraints are boundary conditions for the interpolation along $\tilde{\Gamma}\tilde{Z}$.\\

\noindent Given these rules, there are three distinct connectivities along $\tilg \tilz$, which we describe in turn: (i) a zigzag connectivity (Fig.\ \ref{fig:wilsonGZ}(a-e)) defines the quantum glide Hall effect (QGHE), and (ii) two configurations of hourglasses (e.g., Fig.\ \ref{fig:wilsonGZ}(f) vs \ref{fig:wilsonGZ}(h), and also \ref{fig:wilsonGZ}(g) vs \ref{fig:wilsonGZ}(i)\,) are distinguished by a connected-quadruplet polarization.\\

\noi{i} As illustrated in Fig.\ \ref{fig:wilsonGZ}(a-e), the QGHE describes a zigzag connectivity over $\tilde{\Gamma}\tilde{Z}$, where each cusp of the zigzag corresponds to a Kramers-degenerate subspace. While Fig.\ \ref{fig:wilsonGZ}(c-d) is not obviously zigzag, they are smoothly deformable to Fig.\ \ref{fig:wilsonGZ}(a) which clearly is. A unifying property of all five figures (a-e) is spectral flow: the QGHE is characterized by Wilson bands which robustly interpolate across the maximal energy range of $2\pi$. What distinguishes the QGHE from the usual quantum spin Hall effect:\cite{kane2005B} despite describing the band topology over all of $\tilde{\Gamma}\tilde{Z}$, the QGHE is solely determined by a polarization invariant ($\glide$) at a single point ($\tilde{\Gamma}$), which we now describe.   \\



\noindent \emph{Definition of} $\glide$: Consider the $(k_y{\in}[{-}\pi,\pi),\kpar{=}0)$ circle in the 3D Brillouin zone. Each point here has the glide symmetry $\bmx$, and the Bloch waves divide into two glide subspaces labelled by $\lambda_x/i {\equiv}\eta {=}{\pm} 1$.  This allows us to define an Abelian polarization ($\glide/e$) as the net displacement of Bloch-Wannier functions in either $\eta$ subspace:
\begin{align}
\frac{\glide}{e} = \frac1{2\pi}\,\sum_{j=1}^{\noc/2} \theta_{j,\tilde{\Gamma}}^{{\eta}} \;\;\text{mod}\;\; 1.
\end{align}
Here, the superscript $\eta$ indicates a restriction to the $\lambda_x{=}\eta i$, occupied subspace;  $\{\text{exp}(i\theta^{\sma{\eta}})\}$ are the eigenvalues of the Wilson loop $\W^{\sma{\eta}}(\tilde{\Gamma})$, and the second equality follows from the spectral equivalence introduced in Sec.\ \ref{sec:reviewWilson}. We have previously determined in this Section that  $\noc$ is a multiple of four, and therefore there is always an even number ($\noc/2$) of Wilson bands in either $\eta$ subspace. Furthemore, $\glidep{=}\glidem$ modulo $e$ follows from time reversal relating $\theta_{\sma{\tilg}}^{\sma{\eta}} {\rightarrow} \theta_{\sma{\tilg}}^{\sma{{-}\eta}}$; cf. Tab.\ \ref{rulestrim}. \\

\noindent We claim that the effect of spatial inversion ($\cali$) symmetry is to quantize $\glide$ to $0$ and $e/2$, which respectively corresponds to the absence and presence of the QGHE. Restated, the set of occupied Bloch states along a high-symmetry line (projecting to $\tilg$) holographically determines the topology in a high-symmetry plane (projecting to $\tilg \tilz$). To demonstrate this, (i-a) we first relate the Wilson spectrum at $\tilg$ to the invariant $\glide$, then (i-b) determine the possible Wilson spectra at $\tilz$. (i-c) These end-point spectra may be interpreted as boundary conditions for curves interpolating across $\tilg \tilz$ -- we find there are only two classes of interpolating curves which are distinguished by spectral flow. \\ 

\noi{i-a} To prove the quantization of $\glide$, consider how each glide subspace is individually invariant under $\cali$. This invariance follows from Eq.\ (\ref{algebrabmxi}), leading to the representative commutivity of $\cali$ and $\bmx$ where $k_z{=}0$. We will need further that  $\cali$ maps $\theta^{\sma{\eta}}_{\sma{\tilg}} {\rightarrow} {-}\theta^{\sma{\eta}}_{\sma{\tilg}}$ mod $2\pi$. This may be deduced from the polarization perspective, where $\theta^{\sma{\eta}}_{\sma{\tilg}}/2\pi$ is an eigenvalue of the position operator $\hat{y}$ (projected to the occupied subspace at surface wavevector $\tilg$ and with $\lambda_x{=}\eta i$); our claim then follows simply from $\cali$ inverting the position operator $\hat{y}$. $\theta_{\sma{\tilg}}^{\sma{\eta}}$ and ${-}\theta_{\sma{\tilg}}^{\sma{\eta}}$ may correspond either to two distinct Wilson bands (an inversion doublet), or to the same Wilson band (an inversion singlet at $\theta_{\sma{\tilg}}^{\sma{\eta}}{=}0$ or $\pi$). Since there are an even number of Wilson bands in each $\eta$ subspace, a $0$-singlet is always accompanied by a $\pi$-singlet -- such a singlet pair produces the only non-integral contribution to $\glide ({=} e/2)$; the absence of singlets corresponds to $\glide{=}0$. These two cases correspond to two classes of boundary conditions at $\tilg$. We remark briefly on fine-tuned scenarios where an inversion doublet may accidentally lie at $0$ (or $\pi$) without affecting the value of $\glide$. In complete generality, $\glide{=}e/2$ ($0$) corresponds to an odd (resp. even) number of bands at both $0$ and $\pi$, in one $\eta$ subspace.\\ 

\noi{i-b} What is left is to determine the possible boundary conditions at $\tilz$. We find here only one class of boundary conditions, i.e., any one boundary condition may be smoothly deformed into another, indicating the absence of a nontrivial topological invariant at $\tilz$. Indeed, the same nonsymmorphic algebra (Eq.\ (\ref{algebrabmxi})) has different implications where $k_z{=}\pi$: now $\cali$ relates Wilson bands in opposite glide subspaces, i.e., $\cali: \theta_{\sma{\tilde{Z}}}^{\sma{\eta}} {\rightarrow} \theta_{\sma{\tilde{Z}}}^{\sma{{-}\eta}}{=} {-}\theta_{\sma{\tilde{Z}}}^{\sma{\eta}}$. Consequently, the total polarization ($\calp_{\sma{\tilz}}$) vanishes modulo $e$, and the analogous $\glidez$ is well-defined but not quantized. With the additional constraint by $T$ (see Tab.\ \ref{rulestrim}), any Kramers pair belongs to the same glide subspace due to the reality of the glide eigenvalues; on the other hand, each Kramers pairs at $\theta$ is mapped by $\cali$ to another Kramers pair at ${-}\theta$, and $\cali$-related pairs belong to different glide subspaces.\\

\noi{i-c} Having determined all boundary conditions, we proceed to the interpolation. For simplicity, this is first performed for the minimal number (four) of Wilson bands; the two Wilson-energy functions in each $\eta$ subspace are defined as $\theta^{\sma{\eta}}_{\sma{ 1,\kpar}}$ and $\theta^{\sma{\eta}}_{\sma{2,\kpar}}$.   If $\glide{=}e/2$, the boundary conditions are
\begin{align*}
\theta^{\sma{\eta}}_{\sma{1,\tilg}}=0, \;\; \theta^{\sma{\eta}}_{\sma{2,\tilg}}=\pm \pi, \;\;\text{and}\;\;
\theta^{\sma{\eta}}_{\sma{1,\tilz}}=\theta^{\sma{\eta}}_{\sma{2,\tilz}} = -\theta^{\sma{-\eta}}_{\sma{1,\tilz}} = -\theta^{\sma{-\eta}}_{\sma{2,\tilz}}.
\end{align*}
In one of the glide subspaces (say, ${\eta}$), the two Wilson-energy functions are degenerate at $\tilz$, but are at everywhere else along $\tilg \tilz$ nondegenerate; particularly at $\tilg$, one Wilson energies is fixed to $0$, and other to $\pm \pi$. Consequently, the two energy functions sweep out an energy interval that contains at least $[0,\pi]$ (e.g., Fig.\ \ref{fig:wilsonGZ}(a)), but may contain more (e.g., Fig.\ \ref{fig:wilsonGZ}(c)). The particle-hole symmetry (due to $T \cali$; cf.\ Tab.\ \ref{rulesgeneric}) further imposes that the other two energy functions (in ${-}\eta$) sweep out at least $[{-}\pi,0]$ -- the net result is that the entire energy range is swept; this spectral flow is identified with the QGHE. \\

If $\glide{=}0$, the boundary conditions at $\tilg$ are instead
\begin{align}
\theta^{\sma{\eta}}_{\sma{1,\tilg}}=\theta^{\sma{-\eta}}_{\sma{1,\tilg}}=-\theta^{\sma{\eta}}_{\sma{2,\tilg}}=-\theta^{\sma{-\eta}}_{\sma{2,\tilg}},
\end{align}
leading to spectrally-isolated quadruplets, e.g., in Fig.\ \ref{fig:wilsonGZ}(f), (h) and (i). Since Kramers partners at $\tilg$ ($\tilz$) belong in opposite $\eta$ subspaces (resp.\ the same $\eta$ subspace), the interpolation describes an internal partner-switching within each quadruplet, resulting in an hourglass-like dispersion. The center of the hourglass is an unavoidable crossing\cite{elementaryenergybands} between opposite-$\eta$ bands -- this degeneracy is movable but unremovable. Finally, we remark that the interpolations distinguished by $\glide$ easily generalize beyond the minimal number of Wilson bands, e.g., compare Fig.\ \ref{fig:wilsonGZ}(e) with (g). $\blacksquare$\\

Given that the Abelian polarization depends on the choice of spatial origin,\cite{AA2014} it may seem surprising that a single polarization invariant ($\glide$) sufficiently indicates the QGHE; indeed, each of the inequivalent inversion centers is a reasonable choice for the spatial origin. In contrast, many other topologies are diagnosed by gauge-invariant {differences} in polarizations of different 1D submanifolds in the same Brillouin zone.\cite{fu2006,AA2014} Unlike generic polarizations, $\glide$ is invariant when a different inversion center is picked as origin, i.e., this globally shifts all $\theta {\rightarrow} \theta+\pi$ (e.g., Fig.\ \ref{fig:wilsonGZ}(a) to (b)), which leads to $\glide/e {\rightarrow} \glide/e $ modulo $\Z$, since each glide subspace is even-dimensional. We caution that $\glide$ will not remain quantized if the spatial origin lies away from an inversion center, due to the well-known $U(1)$ ambiguity of the Wilson loop.\cite{AA2014}\\

That the QGHE is determined solely by $\glide$ makes diagnosis especially easy: we propose to multiply the spatial-inversion ($\cali$) eigenvalues of occupied bands in a single $\eta$ subspace, at the two inversion-invariant $\bk$ which project to $\tilg$. This product being $+1$ ($-1$) then implies that $\glide{=}0$ (resp.\ ${=}e/2$). In contrast, the usual quantum spin Hall effect (without glide symmetry) cannot\cite{yu2011,alexey2011,AA2014} be formulated as an Abelian polarization, and a diagnosis would require the $\cali$ eigenvalues at all four inversion-invariant momenta of a two-torus.\cite{Fu1}   \\

\begin{figure}[H]
	\centering
	\includegraphics[width=4.4 cm]{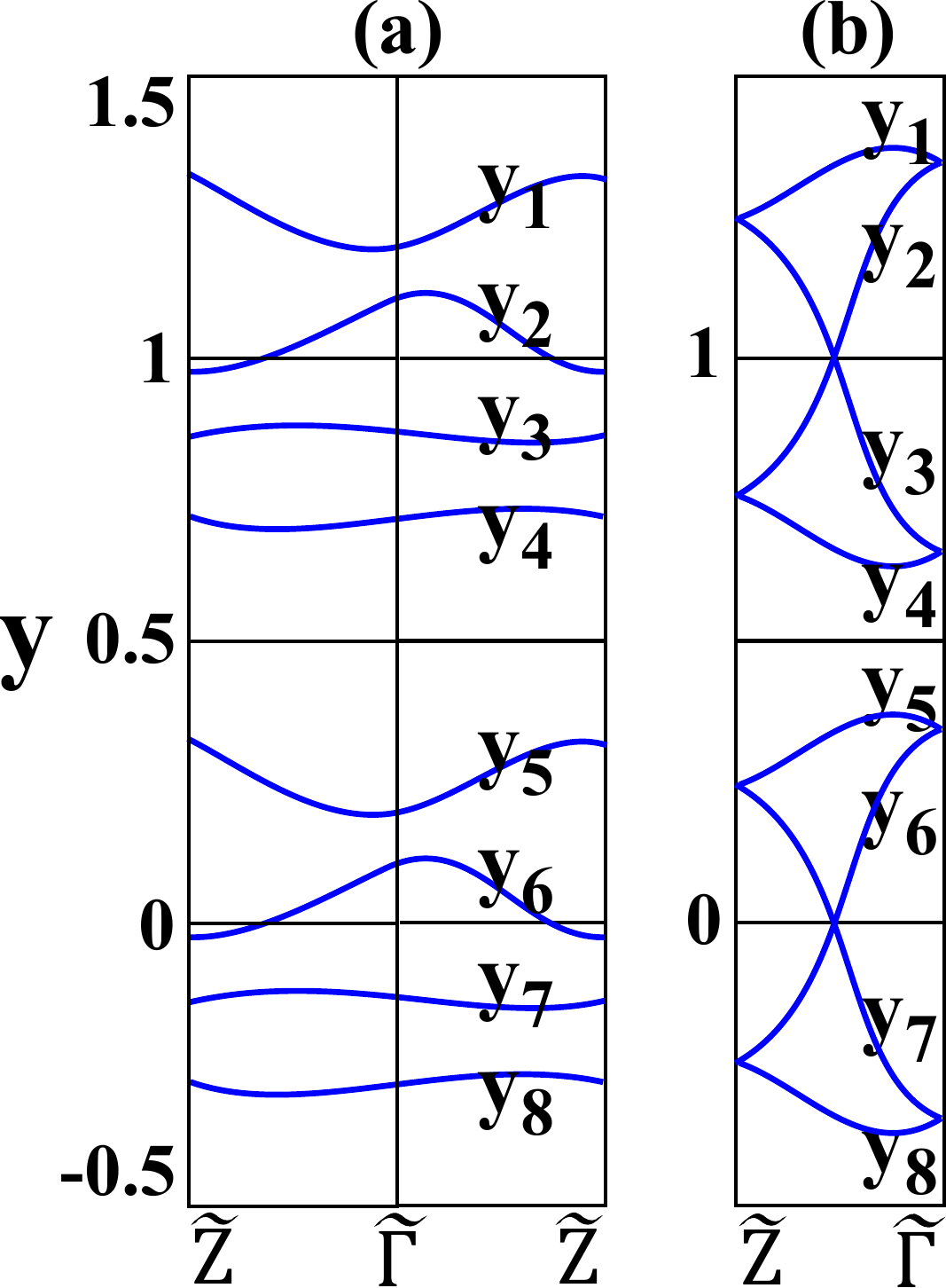}
			\caption{Comparison of the spectrum of $\pper \hat{y}\pper$ for a system without any symmetry (a), and one with time-reversal, spatial-inversion and glide symmetries (b). Only the spectrum for two spatial unit cells are shown.} \label{fig:quadrupletpol}
	\end{figure}

	\noi{ii} With trivial $\glide$, the spectrally-isolated interpolations further subdivides into two distinct classes, which are distinguished by an hourglass centered at $\theta{=}\pi$, e.g., contrast Fig.\ \ref{fig:wilsonGZ}(h-i) with (f-g),(j). This difference may be formalized by a $\Z_2$ topological invariant ($\calq_{\sma{\tilg \tilz}}$) which we introduced in our companion work,\cite{Hourglass} and will presently describe in the polarization perspective. $\calq_{\sma{\tilg \tilz}}$ characterizes a coarse-grained polarization  of quadruplets along $\tilg \tilz$, as we illustrate with Fig.\ \ref{fig:quadrupletpol}(b). Here, the center-of-mass position of this quadruplet may tentatively be defined by averaging four Bloch-Wannier positions: $\caly_{\sma{1}}({\kpar}) {=} (1/4)\sum_{n=1}^4y_{\sma{n,\kpar}}$, with $\kpar {\in}\tilg \tilz$. Any polarization quantity should be well-defined modulo $1$, which reflects the discrete translational symmetry of the crystal. However, we caution that $\caly$ is only well-defined mod $1/4$ for quadruplet bands without symmetry, due to the integer ambiguity of each of $\{y_n |n \in \Z \}$. To illustrate this ambiguity, consider in Fig.\ \ref{fig:quadrupletpol}(a) the spectrum of $\pper\hat{y}\pper$ for an asymmetric insulator with four occupied bands. Only the spectrum for two spatial unit cells (with unit period) are shown, and the discrete translational symmetry ensures $y_{\sma{j,\kpar}} = y_{\sma{j+4l,\kpar}}-l$ for $j,l \in \Z$. Clearly the centers of mass of $\{y_1,y_2,y_3,y_4\}$ and $\{y_2,y_3,y_4,y_5\}$ differ by $1/4$ at each $\kpar$, but both choices are equally natural given level repulsion across $\tilz \tilg \tilz$.
	
	However, a natural choice presents itself if the Bloch-Wannier bands may be grouped in sets of four, such that within each set there are enough contact points along $\tilg \tilz$ to continuously travel between the four bands. Such a property, which we call four-fold connectivity, is illustrated in Fig.\ \ref{fig:quadrupletpol}(b) over two spatial unit cells. Here, both quadrupets $\{y_1,y_2,y_3,y_4\}$ and $\{y_5,y_6,y_7,y_8\}$ are connected, and their centers of mass differ by unity; on the other hand, $\{y_2,y_3,y_4,y_5\}$ is not connected. The following discussion then hinges on this four-fold connectivity, which characterizes insulators with glide and time-reversal symmetries. Having defined a mod-one center-of-mass coordinate for one connected quadruplet, we extend our discussion to insulators with multiple quadruplets per unit cell, i.e., since there are $\noc$ number of Bloch-Wannier bands, where $\noc$ is the number of occupied bands, we now discuss the most general case where integral $\noc/4{\geq}1$. Let us define the net displacement of all $\noc/4$ number of connected-quadruplet centers: $\calq(\kpar)/e {=} \sum_{\sma{j{=}1}}^{\sma{\noc/4}} \caly_j(\kpar)$ mod $1$. A combination of time-reversal ($T$) and spatial-inversion ($\cali$) symmetry quantizes $\calq(\kpar)$ to $0$ or $e/2$, as we now show. We have previously described how $T\cali$ inverts the spatial coordinate but leaves momentum untouched, i.e., we have an unconventional particle-hole symmetry at each $\kpar$: $T\cali |\kpar,n \rangle {=}|\kpar,m\rangle$ with $m {\neq} n$ and $y_{\sma{n,\kpar}} {=} {-}y_{\sma{m,\kpar}}$ mod $1$. Consequently, $T\cali : \caly_j(\kpar) {\rightarrow} \caly_{j'}(\kpar){=}{-}\caly_j(\kpar)$ mod $1$, and the only non-integer contribution to $\calq/e$ (${=}1/2$) arises if there exists a particle-hole-invariant quadruplet ($\bar{j}$) that is centered at $\caly_{\bar{j}}{=}1/2{=}{-}\caly_{\bar{j}}$ mod $1$, as we exemplify in Fig.\ \ref{fig:wilsonGZ}(h-i); moreover, since each $y_{\sma{n,\kpar}}$ is a continuous function of $\kpar$, $\calq_{\sma{\kpar}}$ is constant (${\equiv}\calq_{\sma{\tilg \tilz}}$) over $\tilg \tilz$. Alternatively stated, $\calq_{\sma{\tilg \tilz}}$ is a quantized polarization invariant that characterizes the entire glide plane that projects to $\tilg \tilz$.

\subsubsection{Connecting curves along the glide line $\tilde{X}\tilde{U}$} \label{app:connectkxpi}

\begin{figure}[h]
\centering
\includegraphics[width=7 cm]{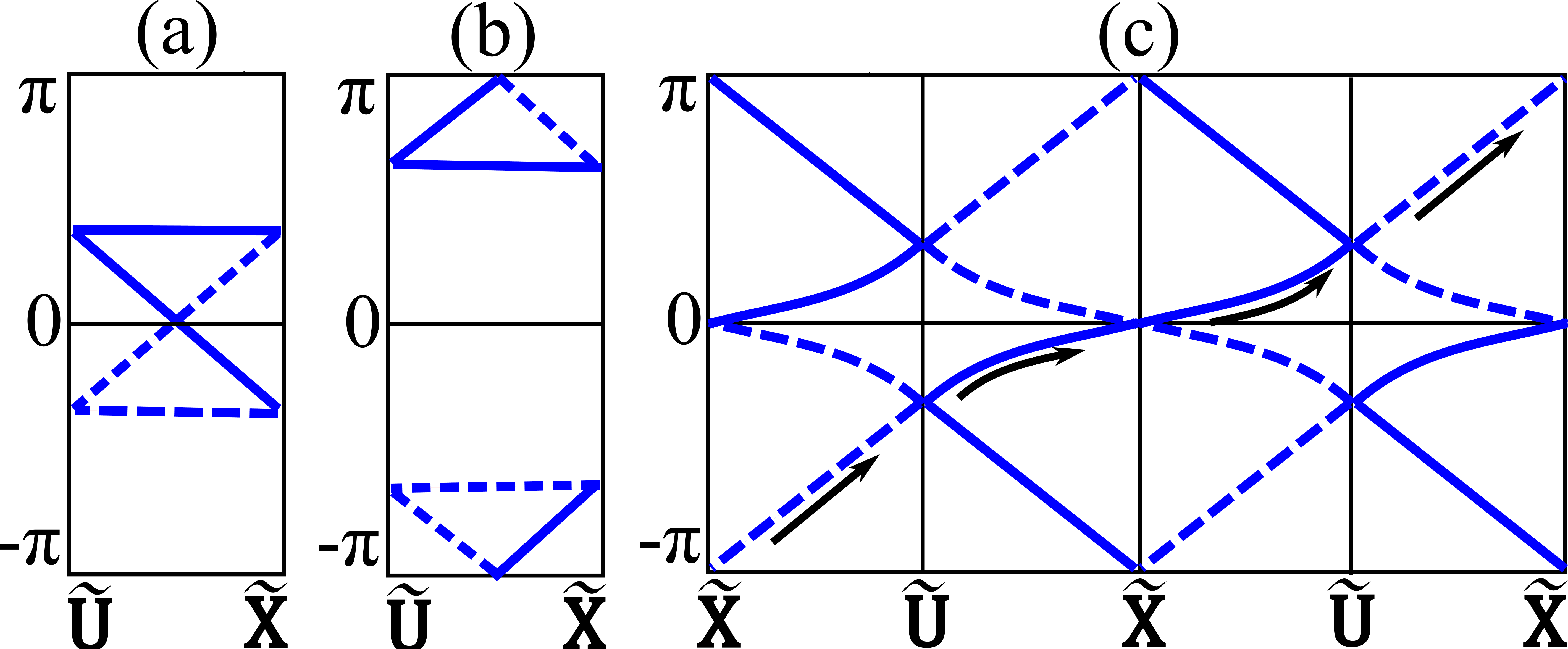}
    \caption{(a-b) Possible Wilson spectra along $\tilde{X}\tilde{U}$. (c) A hypothetical spectrum that is ruled out by a continuity argument.} \label{fig:wilsonXU}
\end{figure}

As discussed in Sec.\ \ref{sec:rules}, the relevant symmetries that constrain $\W(\kpar)$ comprise the little group of the circle $(k_y{\in}[{-}\pi,\pi],\kpar)$.\cite{AAchen} For any $\kpar{\in}\tilx \tilu$, the corresponding group has the symmetries $\bmx$, $T\cali$ and $T\bmz$; these are exactly the same symmetries of the group for $\kpar {\in} \tilg \tilz$. In spite of this similarity, the available subtopologies on $\tilx\tilu$ and $\tilg\tilz$ differ: while the two hourglass configurations (distinguished by a connected-quadruplet polarization) are available subtopologies on each line, the QGHE is only available along $\tilg \tilz$. This difference arises because the same symmetries are represented differently on each line -- the different projective representations are classified by the second cohomology group, as discussed in Sec.\ \ref{sec:wilsoniansymm}. For the purpose of topological classification, we need only extract one salient result from that Section: any Wilson band at $\kpar{=}(\pi,k_z)$ and Wilson-energy $\theta$ has simultaneously a `glide' eigenvalue: $\lambda_x(k_z+\theta)$ in Eq.\ (\ref{relativistic}); here, `glide' refers to the generalized symmetry $\calbmx {\equiv} \W_{\sma{-\pi \leftarrow 0}}\,\bmx$, which combines the ordinary glide reflection ($\bmx$) with parallel transport ($\W_{\sma{-\pi \leftarrow 0}}$). We might ask if $\eta {=} {\pm} 1$ in $\lambda_x$ labels a meaningful division of the Wilson bands, i.e., do we once again have two non-interacting flavors of curves, as we had for $\tilg \tilz$? The answer is affirmative if the Wilson bands are spectrally isolated, i.e., if all $\noc$ Wilson bands lie \emph{strictly} within an energy interval $[\theta_i,\theta_f]$ with $|\theta_f-\theta_i|<2\pi$, for all $k_z \in [0,2\pi)$. For example, the isolated bands of Fig.\ \ref{fig:wilsonXU}(a) lie within a window of $[-\pi/2,\pi/2]$, whereas no similar window exists in the hypothetical scenario of Fig.\ \ref{fig:wilsonXU}(c). 
If isolated, then at each $k_z$ the energy difference ($\theta_i-\theta_j$) between any two bands is strictly less than $2\pi$ -- therefore there is no ambiguity in labelling each band by $\eta$ from Eq.\ (\ref{relativistic}). Conversely, this potential ambiguity is sufficient to rule out bands with spectral flow, as we now demonstrate.\\

\noindent Let us consider a hypothetical scenario with spectral flow (Fig.\ \ref{fig:wilsonXU}(c)), as would describe a QGHE. There is then a smooth interpolation between bands in one energy period to any band in the next, as illustrated by connecting black arrows in Fig.\ \ref{fig:wilsonXU}(c). As we interpolate $\theta {\rightarrow} \theta {+}2\pi$ and $k_z {\rightarrow} k_z {+} 4\pi$, we of course return to the same eigenvector of $\W$, and therefore the `glide' eigenvalue must also return to itself. However, the energy-dependence leads to $\lambda_x {\rightarrow} {-}\lambda_x$. More generally for $4u$ number of occupied bands, $\theta {\rightarrow} \theta {+}2\pi$ while $k_z {\rightarrow}  k_z{+}4u \pi$, leading to the same contradiction. We remark that the essential properties that jointly lead to a contradiction are that: (i) Wilson bands come in multiples of four, as we discussed in the introduction to Sec.\ \ref{sec:connect}, (ii) the Kramers partners at $\tilx$ ($\tilu$) belong to opposite flavors (resp. the same flavor); cf. Tab.\ \ref{rulestrim}, and (iii) that bands connect in a zigzag. Certain details of Fig.\ \ref{fig:wilsonXU}(c) (e.g., that $\theta$ is quantized to special values at $\tilx$) are superfluous to our argument. Besides this argument, we furnish an alternative proof to rule out the QGHE in our companion paper.\cite{Hourglass}  We remark that the QGHE is perfectly consistent with the surface symmetries,\cite{Hourglass} and it is only ruled out by a proper account of the bulk symmetries. $\blacksquare$\\


Returning to our classification, Tab.\ \ref{rulesgeneric} and \ref{rulestrim} inform us of the constraints due to time-reversal and spatial-inversion symmetries. The summary of this symmetry analysis is that our rules for the curves along $\tilx \tilu$ are completely identical to that along $\tilde{\Gamma}\tilde{Z}$, \emph{assuming} that bands are spectrally isolated. We thus conclude that the only subtopologies are two hourglass-type interpolations (Fig.\ \ref{fig:wilsonXU}(a-b)), which are distinguished by a second connected-quadruplet polarization ($\calq_{\sma{\tilx \tilu}}$).

\subsubsection{Connecting curves along the mirror line $\tilde{\Gamma}\tilde{X}$}

\begin{figure}[H]
\centering
\includegraphics[width=5 cm]{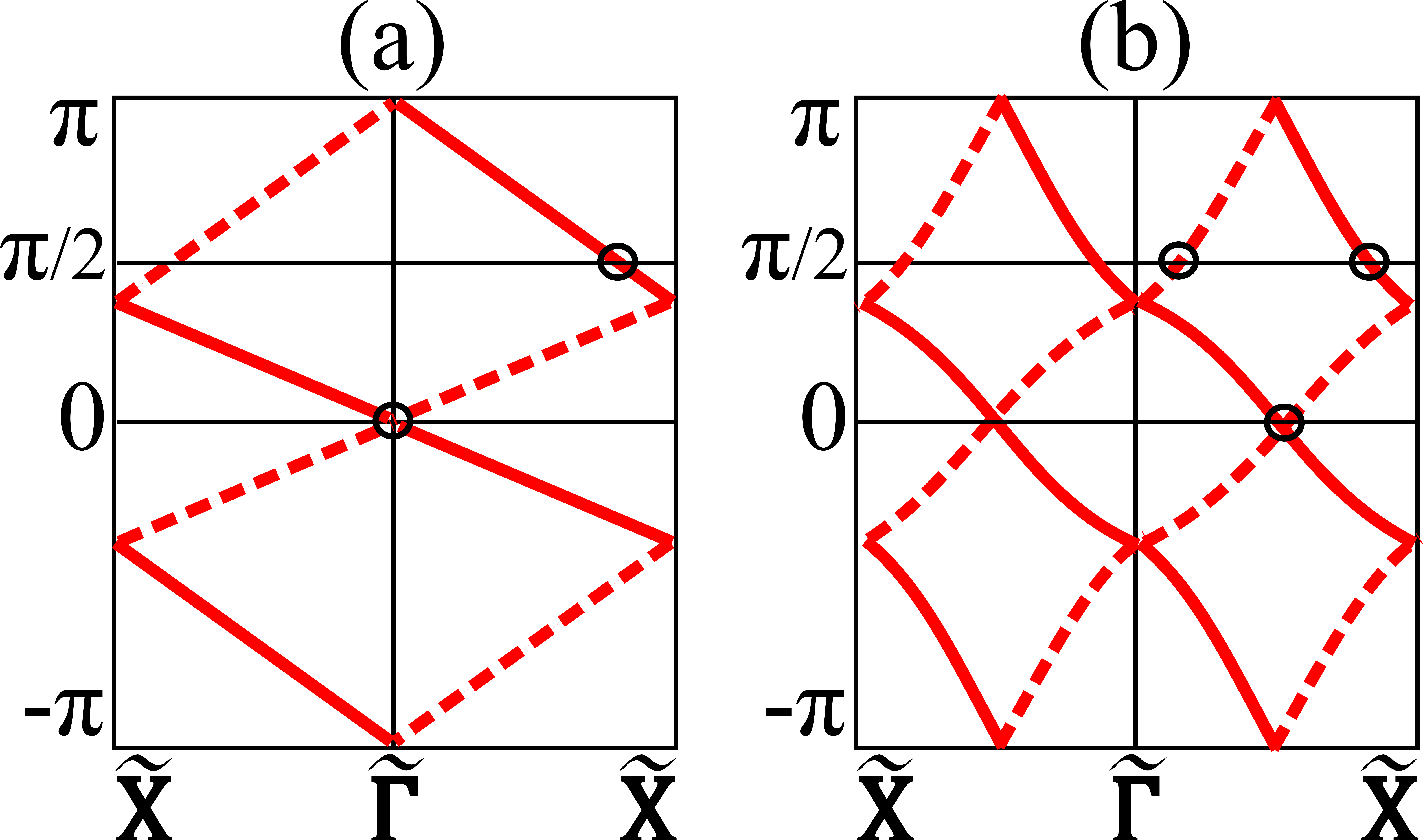}
    \caption{Illustrating the single-energy criterion to determine the parity of the mirror-Chern number ($\calc_e$); $\calc_e{=}{-}1$ in (a) and ${-}2$ in (b). Red solid (dashed) lines correspond to a Wilson band with mirror eigenvalue $\lambda_z{=}{+}i$ (${-}i$).} \label{fig:fermicriterion}
\end{figure}

\noi{a} Curves divide into two non-interacting flavors (red solid and red dashed lines in Fig.\ \ref{fig:fermicriterion}), corresponding to $\lambda_z={\pm} i$ subspaces. \\

\noi{b} At both boundaries ($\tilg$ and $\tilx$), each red solid curve is degenerate with a red dashed curve; cf. Tab.\ \ref{rulestrim}.\\

\noi{c} At any point along $\tilg \tilx$, [$\theta$, red solid] $\leftrightarrow$ [$-\theta$, red dashed], due to the $T\cali$ symmetry of Tab.\ \ref{rulesgeneric}.\\

\noindent These rules allow for mirror-Chern\cite{teo2008} sub-topologies in the torus that projects to $\tilx \tilg \tilx$, where $\lambda_z{=}{{\pm}} i$ subspaces have opposite chirality due to time-reversal symmetry; Fig.\ \ref{fig:fermicriterion}(a) exemplifies a Chern number ($\calc_e$) of $-1$ in the $\lambda_z{=}{+}i$ subspace, and Fig.\ \ref{fig:fermicriterion}(b) exemplifies $\calc_e{=}{-}2$. The allowed mirror Chern numbers ($\calc_e$) depend on our last rule:\\

\noi{d} Curves must match continuously at $\tilde{\Gamma}$ and $\tilde{X}$. \\

\begin{figure}[H]
\centering
\includegraphics[width=8.6 cm]{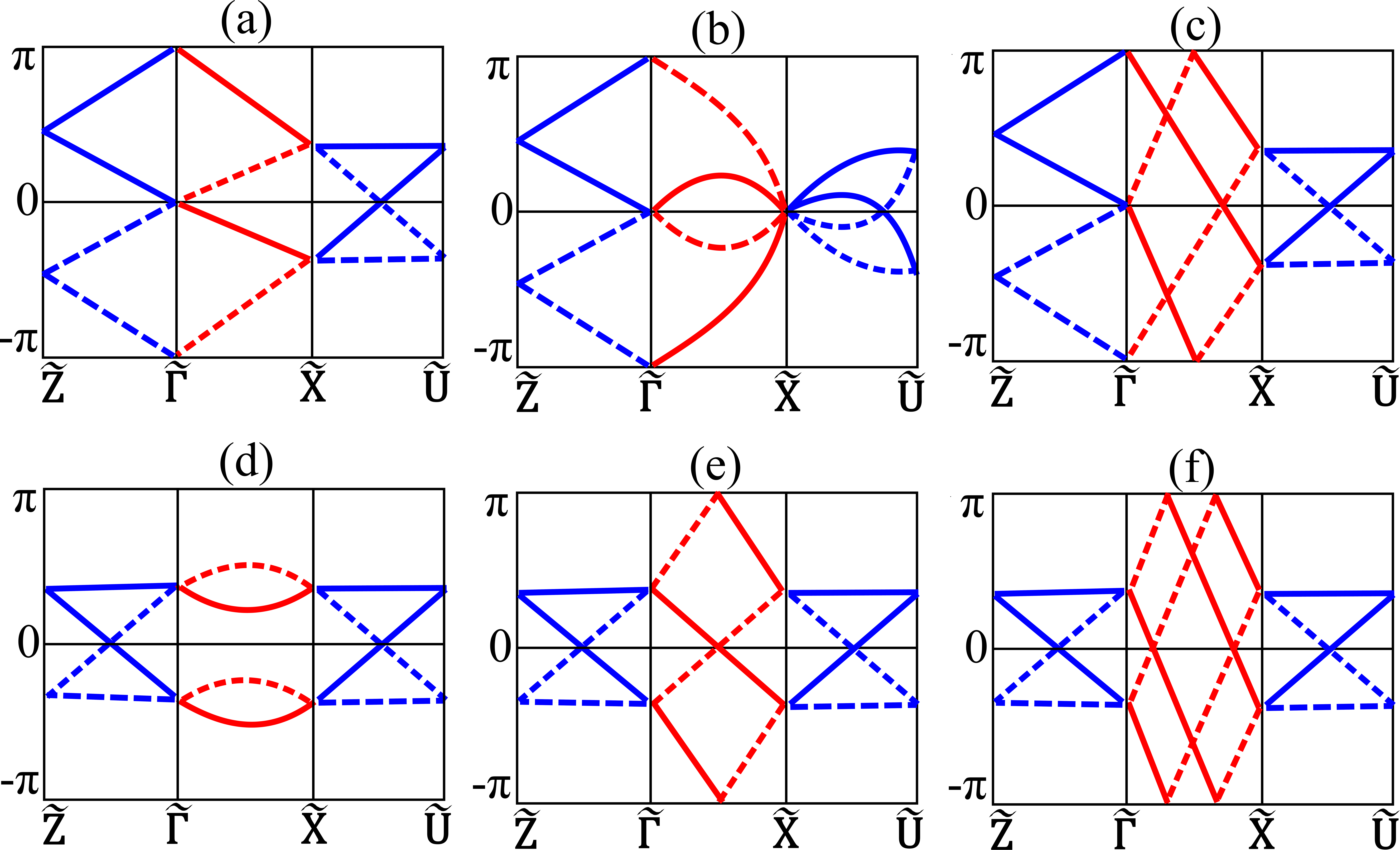}
    \caption{Possible Wilson spectra along $\tilde{\Gamma}\tilde{X}$ (a-c) With nontrivial glide polarization ($\glide$), the mirror Chern numbers ($\calc_e$) are respectively $-1,+1$ and $-3$ . (d-f) With trivial $\glide$, $\calc_e$ equals $0$,$-2$ and $-4$ respectively.} \label{fig:wilsonGX}
\end{figure}

\noindent This last rule imposes a consistency condition with the subtopologies at $\tilde{\Gamma}\tilde{Z}$ and $\tilx \tilu$: $\calc_e$ is odd (even) if and only if $\glide{=}e/2$ (resp.\ $\glide{=}0$), as illustrated in Fig.\ \ref{fig:wilsonGX}(a-c) (resp.\ \ref{fig:wilsonGX}(d-f)). \\

To demonstrate our claim, we rely on a single-energy criterion to determine the parity of $\calc_e$: count the number ($N_{+i}$) of $\lambda_z=+i$ states at an arbitrarily-chosen energy and along the full circle $\tilx \tilde{\Gamma}\tilde{X}$, then apply $N_{+i}$ mod $2{=}\calc_e$ mod $2$. Supposing we chose $\theta{=}\pi/2$ in Fig.\ \ref{fig:fermicriterion}(a), there is a single intersection (encircled in the figure) with a $\lambda_z{=}{+}i$ band, as is consistent with $\calc_e{=}{-}1$ being odd. For the purpose of connecting $\calc_e$ with $\glide$, we will need a slightly-modified counting rule that applies to the half-circle $\tilg \tilx$ instead of the full circle. Since time reversal relates $\lambda_z{=}{\pm} i$ bands at opposite momentum, we would instead count the total number of bands in both $\bmz$ subspaces, at our chosen energy and along the half-circle; one additional rule regards the counting of Kramers doublet, which comprise time-reversed partners at either $\tilg$ or $\tilx$. If such a doublet lies at our chosen energy, it counts not as two but as one; every other singlet state counts as one. With these rules, the parity of this weighted count ($\tilde{N}$) equals that of $\calc_e$. Returning to  Fig.\ \ref{fig:fermicriterion}(a) for illustration, we would still count the single $\lambda_z{=}{+}i$ crossing (encircled) at $\theta{=}\pi/2$, but if we instead pick $\theta{=}0$, we could count the Kramers doublet at $\tilg$ as unity; for both choices of $\theta$, $\tilde{N}{=}{+}1$. In comparison, the two $\theta{=}0$ states in Fig.\ \ref{fig:fermicriterion}(b) are both singlets and count collectively as two, which is consistent with this figure describing a $\calc_e{=}{-}2$ phase. \\

Though our single-energy criterion applies at any $\theta$, it is useful to particularize to $\theta{=}0$, where in counting $\tilde{N}$ we would have to determine the number of zero-energy Kramers doublets at $\tilg$, e.g., this would be one in Fig.\ \ref{fig:fermicriterion}(a), and zero in Fig.\ \ref{fig:fermicriterion}(b). This number may be identified, mod $2$, with the number of zero-energy inversion singlets, which we established in Sec.\ \ref{sec:interpolateGZ} to be unity if $\glide{=}e/2$, and zero if $\glide{=}0$. Moreover, the parity of zero-energy doublets at $\tilg$ may immediately be identified with the parity of $\tilde{N}$ (and thus also that of $\calc_e$), because every other contribution to $\tilde{N}$ has even parity, as we now show. We first consider the contribution at $\tilx$. Given that the only subtopologies at $\tilx \tilu$ are hourglasses, there are generically no zero-energy Kramers doublets at $\tilx$ (Fig.\ \ref{fig:wilsonGX}(a) and (c-f)), though in fine-tuned situations (Fig.\ \ref{fig:wilsonGX}(b)) there might be an even number. Away from the end points, any intersection comes in particle-hole-symmetric pairs (e.g., Fig.\ \ref{fig:wilsonGX}(c), (e-f)).


\subsubsection{Connecting curves along the mirror line $\tilde{Z}\tilde{U}$}

\begin{figure}[h]
\centering
\includegraphics[width=8.6 cm]{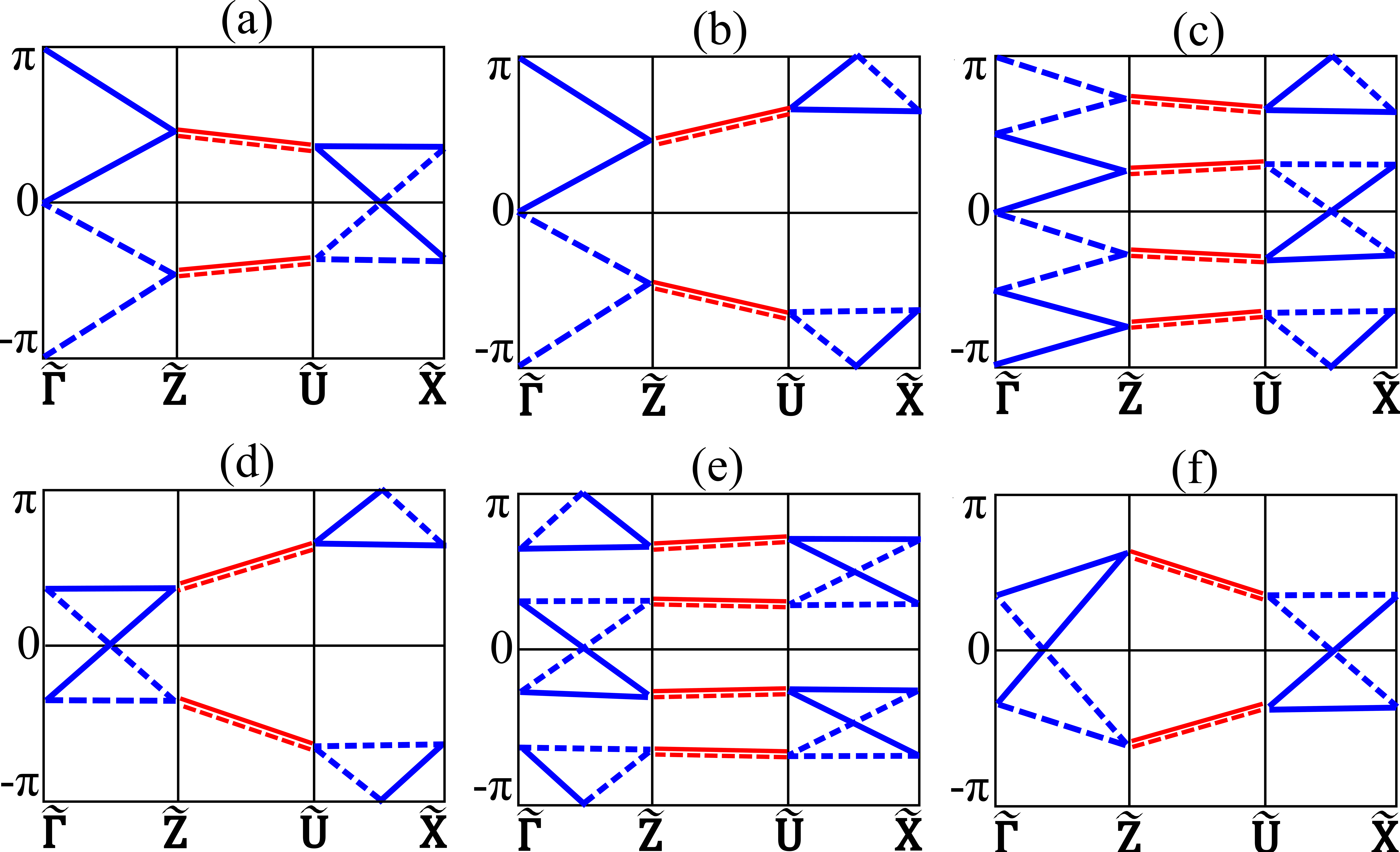}
    \caption{Possible Wilson spectra along $\tilde{Z}\tilde{U}$.} \label{fig:wilsonZU}
\end{figure}

\noi{a} As illustrated in Fig.\ \ref{fig:wilsonZU}, Each red, solid curve ($\bmz{=}+i$) is degenerate with a red, dashed curve ($\bmz{=}-i$). Doublet curves cannot cross due to level repulsion, and must be symmetric under $\theta\rightarrow -\theta$.\\

\noi{b} The curve-matching conditions at $\tilde{Z}$ and $\tilde{U}$ again impose consistency requirements. \\

These rules are stringent enough to uniquely specify the interpolation along $\tilde{Z}\tilde{U}$, given the subtopologies at $\tilg \tilz$ (specified by $\glide,\calq_{\sma{\tilg \tilz}}$) and at $\tilx \tilu$ ($\calq_{\sma{\tilx \tilu}}$). Alternatively stated, there are no additional invariants in this already-complete classification. To justify our claim, first consider $\glide=e/2$, such that doublets at $\tilu$ are matched with cusps of hourglasses (along $\tilde{U}\tilde{X}$), while doublets at $\tilde{Z}$ connect to cusps of a zigzag (along $\tilde{Z}\tilde{\Gamma}$). There is then only one type of interpolation illustrated in Fig.\ \ref{fig:wilsonZU}(a-c). If $\glide=0$, we have hourglasses on both glide lines $\tilg \tilz$ and $\tilx \tilu$. If on one glide line an hourglass is centered at $\theta=\pi$, while on the other line there is no $\pi$-hourglass (i.e., $\calq_{\sma{\tilg \tilz}} \neq \calq_{\sma{\tilx \tilu}}$), the unique interpolation is shown in Fig.\ \ref{fig:wilsonZU}(d-e): red doublets connect the upper cusp of one hourglass to the lower cusp of another, in a generalized zigzag pattern with spectral flow. A brief remark here is in order: when viewed individually along any straight line (e.g., $\tilg \tilz$ or $\tilz\tilu$), bands are clearly spectrally isolated; however, when viewed along a bent line ($\tilg \tilz \tilu \tilx$), the bands exhibit spectral flow. In all other cases for $\glide,\calq_{\sma{\tilg \tilz}}$ and $\calq_{\sma{\tilx \tilu}}$, bands along $\tilde{\Gamma}\tilde{Z}\tilde{U}\tilde{X}$ separate into spectrally-isolated quadruplets, as in Fig.\ \ref{fig:wilsonZU}(f).  

\section{Quasimomentum extensions and group cohomology in band insulators} \label{sec:wilsoniansymm}

Symmetry operations normally describe space-time transformations; such symmetries and their groups are referred to as \emph{ordinary}. Here, we encounter certain `symmetries' of the Wilson loop which additionally induce quasimomentum transport in the space of filled bands; we call them W-symmetries to distinguish them from the ordinary symmetries. In this Section, we identify the relevant W-symmetries, and show their corresponding group ($\gkpar$) to be an extension of the \emph{ordinary} group ($\gs$) by quasimomentum translations, where $\gs$ corresponds purely to space-time transformations; the inequivalent extensions are classified by the second cohomology group, which we also introduce here. In crystals, $\gs$ would be a magnetic point group\cite{magnetic_groups} for a spinless particle, i.e., $\gs$ comprise the space-time transformations (possibly including time reversal) that preserve at least one real-space point.  It is well-known how $\gs$ may be extended by phase factors to describe half-integer-spin particles, and also by discrete spatial translations to describe nonsymmorphic crystals.\cite{ascher1,ascher2,cohomologyhiller,mermin_fourierspace,rabsonbenji} One lesson learned here is that $\gs$ may be further extended by quasimomentum translations (as represented by the Wilson loop), thus placing real and quasimomentum space on equal footing.  \\

W-symmetries are a special type of constraints on the Wilson loop at high-symmetry momenta ($\kpar$). As exemplified in Eq.\ (\ref{Ttype}) and (\ref{Ptype}), constraints ($\hat{g}$) on a Wilson loop ($\W$) map $\W$ to itself, up to a reversal in orientation: 
\begin{align} \label{groupwilson}
\hat{g} \,\W\, \hat{g}^{-1} = \W^{{\pm 1}},
\end{align} 
where $\W^{\sma{-1}}$ is the inverse of $\W$; all $\hat{g}$ satisfying this equation are defined as elements in the group ($G_{\sma{\kpar}}$) of the Wilson loop. A trivial example of $\hat{g}$ would be the Wilson loop itself; $\hat{g}$ may also represent a space-time transformation, as exemplified by a $2\pi$ real-space rotation ($\bar{E}$).  Particularizing to our context, we let  $k_y {\in} [{-}\pi,\pi)$ parametrize the non-contractible momentum loop, and choose the convention that $\W$ ($\W^{\sma{-1}}$) effects parallel transport in the positive orientation:$+2\pi \vec{y}$ (resp.\ in the reversed orientation:${-}2\pi \vec{y}$), as further elaborated in App.\ \ref{app:notations}.\\

\begin{figure}[H]
\centering
\includegraphics[width=8 cm]{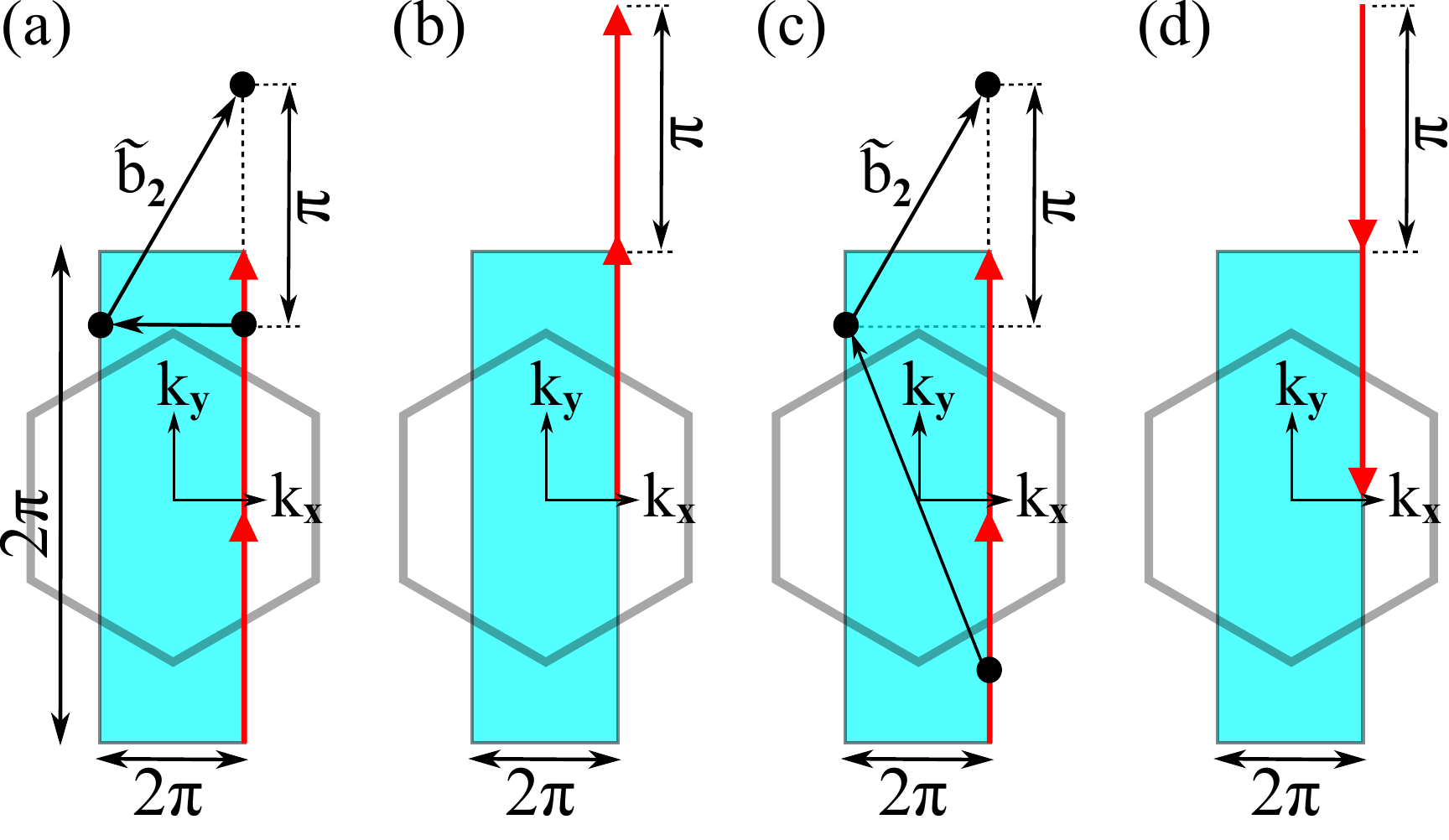}
    \caption{Origin of W-glide and W-time-reversal symmetries. (a-d) are constant-$k_z$ slices of the bulk Brillouin zone. (a) illustrates how the glide ($\bmx$) maps momenta from the glide plane $k_x{=}\pi$. Under $\bmx$, the Wilson loop is mapped from the red vertical arrow in (a) to the red vertical arrow in (b).  (c-d) describe the $k_z{=}0$ plane and illustrate a similar story for the time reversal $T$. } \label{fig:cohomology}
\end{figure}

W-symmetries arise as constraints if a space-time transformation exists that maps: $k_y {\rightarrow} {\pm} k_y {+}\pi$. Our first example of a W-symmetry has been introduced in Sec.\ \ref{sec:rules}, namely that the glide reflection ($\bmx$) maps:  $(k_y,\kpar) {\rightarrow}  (k_y {+}\pi,\kpar)$ for any $\kpar$ along $k_x{=}\pi$ ($\tilx \tilu$), as illustrated in Fig.\ \ref{fig:cohomology}(a). Consequently, the Wilson loop is mapped as
\begin{align} \label{bmxeffect}
\bmx\,\W_{\sma{-\pi}}(\pi,k_z)\,\bmx^{\mo} = \W_{\sma{0}}(\pi,k_z),
\end{align}
where we have indicated the base point of the parameter loop as a subscript of $\W$, i.e., $\W_{\sma{\bar{k}_y}}$  induces parallel transport from $(\bar{k}_y,\pi,k_z)$ to $(\bar{k}_y{+}2\pi,\pi,k_z)$ in the positive orientation. This mapping from $\W_{\sma{-\pi}}$ (vertical arrow in Fig.\ \ref{fig:cohomology}(a)) to $\W_{\sma{0}}$ (arrow in Fig.\ \ref{fig:cohomology}(b)) is also illustrated. As it stands, Eq.\ (\ref{bmxeffect}) is not a constraint as defined in Eq.\ (\ref{groupwilson}). Progress is made by \emph{further} parallel-transporting the occupied space by $-\pi \vec{y}$, such that we return to the initial momentum:  $(k_y,\pi,k_z)$. This motivates the definition of a W-glide symmetry ($\calbmx$) which combines the glide reflection ($\bmx$) with parallel transport across half a reciprocal period -- then by our construction, $\calbmx$ is an element in the group ($\gkpar$) of $\W_{\sma{-\pi}}(\pi,k_z)$. To be precise, let us define the Wilson line $\W_{\sma{-\pi \leftarrow 0}}$ to represent a parallel transport from $(0,\pi,k_z)$ to $(-\pi,\pi,k_z)$, then
\begin{align} \label{Mxwilson}
\calbmx \;\W_{\sma{-\pi}}\;\calbmx^{\sma{-1}} = \W_{\sma{-\pi}}, \;\text{with}\;\;\; \calbmx = \W_{\sma{-\pi \leftarrow 0}}\,\bmx.
\end{align}
The W-glide ($\calbmx$) squares as:
\begin{align} \label{wglidesquare}
\calbmx^2 = \bar{E}\;t(\vec{z})\;{\W}_{\sma{-\pi}}^{\sma{-1}},
\end{align}
which may be understood loosely as follows: the glide component of the W-glide squares as a $2\pi$ rotation ($\bar{E}$) with a lattice translation ($t(\vec{z})$), while the transport component squares as a full-period transport ($\W^{\sma{-1}}$); we defer the detailed derivations of Eq.\ (\ref{bmxeffect})-(\ref{wglidesquare}) to App.\ \ref{app:kxpi}. 
 For a Wilson band with energy $\theta(k_z)$,  Eq.\ (\ref{wglidesquare}) implies the corresponding W-glide eigenvalue depends on the sum of energy and momentum, as in Eq.\ (\ref{relativistic}). Our construction of $\calbmx$ is a quasimomentum-analog of the nonsymmorphic extension of point groups.\cite{ascher1,ascher2,cohomologyhiller,mermin_fourierspace,rabsonbenji} For example, the glide reflection ($\bmx$) combines a reflection with half a \emph{real}-lattice translation -- $\bmx^2$ thus squares to a full lattice translation, which necessitates extending the point group by the group of translations.  Here, we have further combined $\bmx$ with half a \emph{reciprocal}-lattice translation, thus necessitating a further extension by Wilson loops.\\ 



Our second example of a W-symmetry ($\calt$) combines time reversal ($T$) with parallel transport over a half period, and belongs in the groups of $\W(\tilde{X})$ and $\W(\tilde{U})$, which correspond to the two time-reversal-invariant $\kpar$ along $k_x{=}\pi$ (recall Fig.\ \ref{fig:structure}); since both groups are isomorphic, we use a common label: $\gwp$. Under time reversal,
\begin{align}
T: (k_y,\pi,\bar{k}_z) \longrightarrow &\; (-k_y,-\pi,-\bar{k}_z) \lin
\eq (-k_y+\pi,\pi,\bar{k}_z)-\tilde{\boldsymbol{b}}_2-2\bar{k}_z\vec{z},
\end{align}
for $\,\bar{k}_z{\in}\{0,\pi\}$ and $2\bar{k}_z\vec{z}$ a reciprocal vector (possibly zero), as illustrated in Fig.\ \ref{fig:cohomology}(c). Consequently, 
\begin{align} \label{there2}
T \, \W_{\sma{{-}\pi}}\,T^{-1} \kzeq \W_{r,\sma{2\pi}},
\end{align}
where $\W_{r,\sma{2\pi}}$ denotes the reverse-oriented Wilson loop with base point $2\pi$ (see arrow in Fig.\ \ref{fig:cohomology}(d)), and $\kzeq$ indicates that this equality holds for $\kpar \in \{\tilde{X},\tilde{U}\}$. Eq.\ (\ref{there2}) motivates combining $T$ with a half-period transport, such that the combined operation $\calt$ effects
\begin{align} \label{wilsonT}
\calt \, \W_{\sma{{-}\pi}}\,{\calt}^{\,-1}  \kzeq \W_{\sma{{-}\pi}}^{{-1}},\ins{with}\calt \kzeq \W_{\sma{-\pi \leftarrow 0}} \,T.
\end{align}
To complete the Wilsonian algebra, we derive in App.\ \ref{app:kxpi} that
\begin{align} \label{Twilson}
\calt^2 \kzeq \bar{E},\;\;\;\;  \calbmx \;\calt\;\calbmx^{-1}\;\calt^{-1} \kzeq \W^{-1}_{\sma{{-}\pi}}.
\end{align}
This result, together with Eq.\ (\ref{wglidesquare}), may be compared with the ordinary algebra of space-time transformations: 
\begin{align} \label{usualalgebra}
\bmx^2 = \bar{E}\;t(\vec{z}) , \;\;T^2 \kzeq \bar{E},\;\;\bmx \;T \;\bmx^{-1}\;T^{-1} \kzeq I,
\end{align}
as would apply to the surface bands at any time-reversal-invariant $\kpar$. Both algebras are identical modulo factors of $\W$ and its inverse; from hereon, $\W(\kpar)_{\sma{-\pi}}{=}\W$. We emphasize that the \emph{same} edge symmetries are represented differently in the surface Hamiltonion ($H_s$) and in $\W$ -- this difference originates from the out-of-surface translational symmetry ($t(\,\tilde{\boldsymbol{a}}_1\,)$), which is broken for $H_s$ but not for $\W$; recall here that $\tilde{\boldsymbol{a}}_1$ is the out-of-surface Bravais lattice vector drawn in Fig.\ \ref{fig:structure}(b). Where $t(\,\tilde{\boldsymbol{a}}_1\,)$ symmetry is preserved, we can distinguish the bulk wavevectors $k_y$ from $k_y{+}\pi$, and therefore define W-symmetry operators that include the Wilson line $\W_{\sma{-\pi \leftarrow 0}}$. \\

To further describe this difference group-theoretically, let us define $\gs \cong \Z_2\times \Z_2$ as the symmetry group of a spinless particle with glideless-reflection ($M_{\sma{x}}$) and time-reversal ($T$) symmetries:
\begin{align} \label{defineGs}
\gs = \{ \;M_x^a\,T^b \;|\;a,b \in \Z_2 \;\},
\end{align}
with the algebra: 
\begin{align} \label{symmorphicalgebra}
M_x^2 = I,\;\; T^2 = I,\;\; [M_x,T]=0.
\end{align}
The algebra of Eq.\ (\ref{usualalgebra}) describes a well-known, nonsymmorphic extension of $\gs$ for spinful particles;\cite{Lax} we propose that Eq.\ (\ref{wglidesquare}) and (\ref{Twilson}) describe a \emph{further} extension of Eq.\ (\ref{usualalgebra}) by reciprocal translations. That is, $\gwp$ is a nontrivial extension of $\gs$ by $\caln$, where $\caln \cong \Z^2 \times \Z_2$ is an Abelian group generated by $\bar{E}$, $\W$ and $t(\vec{z})$:
\begin{align} \label{defineNgroup}
\caln = \{ \;\bar{E}^a\,t(\vec{z})^b\,\W^c \;|\; a \in \Z_2, b,c \in \Z \; \}.
\end{align}
For an introduction to group extensions and their application to our problem, we refer to the interested reader to App.\ \ref{app:extension}. There exists another extension ($\gwz$, as further elaborated later in this Section) which is inequivalent to $\gwp$, and applies to a different momentum submanifold of our crystal; in  Sec.\ \ref{sec:connect}, we further show that inequivalent extensions lead to different subtopologies for the Wilson bands. \\

From the cohomological perspective, two extensions (of $\gs$ by $\gn$) are equivalent if they correspond to the same element in the second cohomology group $H^2(\gs,\gn)$. The identity element in this group corresponds to a linear representation of $\gs$, which we now define. Let the group element $g_i \in \gs$ be represented by $\hat{g}_i$ in the extension of $\gs$ by $\gn$, and further define $\gij\equiv g_ig_j \in \gs$ by ${\hatgij}$. We insist that $\{\hat{g}_i\}$ satisfy the the associativity condition:
\begin{align}
(\,\hatgi\,\hatgj)\,\hatgk = \hatgi\,(\,\hatgj\,\hatgk).
\end{align}
In a linear representation,
\begin{align}
\hatgi \hatgj = \hatgij \ins{for all} g_i,g_i,\gij \in \gs,
\end{align}
while in a projective representation,
\begin{align} \label{factorsystem}
\hatgi \hatgj = \cij\,\hatgij, \ins{where} \cij \in \caln,
\end{align}
at least one of $\{\cij \}$ (defined as the factor system\cite{SPTandgroupcohomologyXie}) is not trivially identity. Eq.\ (\ref{wglidesquare}) exemplifies Eq.\ (\ref{factorsystem}) for $g_i{=}g_j{=}M_{\sma{x}}$ satisfying $M_{\sma{x}}^2{=}I$, $\hatgi{=}\calbmx$ and $\cij {=}\bar{E}\,t(\vec{z})\,{\W}^{\sma{-1}}$. We say that two representations are equivalent if they are related by the transformation 
\begin{align} \label{gaugetrans}
\hat{g}_i \rightarrow \hat{g}_i'=\di\,\hat{g}_i \ins{with} \di \in \caln.
\end{align}
In either representation, the same constraint is imposed on $\W$ (cf.\ Eq.\ (\ref{groupwilson})):
\begin{align}
\hatgi'\,\W\,\hatgi'^{-1} = \W^{\pm 1} \iff \hat{g}_i\,\W\,\hat{g}_i^{{-1}} = \W^{\pm 1},
\end{align}
since any element of $\caln$ commutes with $\W$. This state of affairs is reminiscent of the $U(1)$ gauge ambiguity in representing symmetries of the Hamiltonian ($\hat{H}$),\cite{weinbergbook1} where if $[\hat{g},\hat{H}]{=}0$, so would $[\hat{g}',\hat{H}]{=}0$ for any $\hat{g}'{=}$exp$[i\phi(g)]\hat{g}$. By this analogy, we also call $\hat{g}_i$ and $\hat{g}_i'$ from Eq.\ (\ref{gaugetrans}) two gauge-equivalent representations of the same element $g_i$, though it should be understood in this paper that the relevant gauge group is $\caln$ and not $U(1)$. To recapitulate, each element in $H^2(\gs,\gn)$ corresponds to an equivalence class of associative representations; in App.\ \ref{app:connectprojrejcoh}, we further connect our theory to group cohomology through the geometrical perspective of coboundaries and cocycles.\\


To exemplify an extension/representation that is inequivalent to $\gwp$, let us consider the group ($\gwz$) of $\W(\tilde{\Gamma})$; $\gwz$ is isomorphic to the group of $\W(\tilde{Z}))$; recall that both $\tilde{\Gamma}$ and $\tilde{Z}$ are time-reversal-invariant $\kpar$ along $k_x{=}0$. $k_x{=}0$ labels a glide line in the 010-surface BZ, which guarantees the $k_x{=}0$ plane (in the bulk BZ) is mapped to itself under the glide $\bmx$; the same could be said for $k_x{=}\pi$. However, unlike $k_x{=}\pi$, $\bmx$ also belongs to the group of any bulk wavevector in the $k_x{=}0$ plane, and therefore
\begin{align}
\bmx\,\W(\tilde{\Gamma})\,\bmx^{\mo} = \W(\tilde{\Gamma})
\end{align}
with $\bmx$ an ordinary space-time symmetry, i.e., unlike $\calbmx$ in Eq.\ (\ref{Mxwilson}), $\bmx$ does not encode parallel transport. Consequently, this element of $\gwz$ satisfies the ordinary algebra in Eq.\ (\ref{usualalgebra}); by an analogous derivation,  the time reversal element in $\gwz$ is also ordinary. It is now apparent why $\gwp$ and $\gwz$ are inequivalent extensions: there exists no gauge transformation, of the form in Eq.\ (\ref{gaugetrans}), that relates their factor systems. For example, the following elements of $\gs$:
\begin{align}
g_1 \equiv \bmx\,T,\;\; g_2 \equiv \bmx^{-1}\,T^{-1},\;\; g_1g_2 \equiv g_{12}=I,
\end{align}
may be represented in $\gwp$ by 
\begin{align}
\hat{g}_1 \equiv \calbmx \calt,\;\; \hat{g}_2 {\equiv}\calbmx^{-1}\calt^{-1}, \;\;\hat{g}_{12} \equiv I,
\end{align} 
such that the second relation in Eq.\ (\ref{Twilson}) translates to
\begin{align} \label{transformsas}
\hat{g}_1 \, \hat{g}_2 = C_{1,2}\,\hat{g}_{12} \ins{with} C_{1,2} \equiv \W^{-1}.
\end{align}
Under the gauge transformation: $\calbmx {\rightarrow} \calbmx'{=} \W^a \calbmx$, $\calt {\rightarrow} \calt'{=}\W^b \calt$, and $a,b \in \Z$, Eq.\ (\ref{transformsas}) transforms as
\begin{align} \label{transform3}
&\calbmx \;\calt\;\calbmx^{-1}\;\calt^{-1} = \W^{-1}  \lin
\rightarrow &\;\calbmx' \;\calt'\;{\calbmx'}^{-1}\;{\calt'}^{-1} = \W^{2a-1},
\end{align}
which ensures that the factor $C_{1,2}$ is always an odd product of $\W$. This must be compared with the analogous algebraic relation in $\gwz$, where with  $\bmx {\rightarrow} \bmx'{=}\W^c \bmx$, $T {\rightarrow} T'{=}\W^d T$, and $c,d \in \Z$,
\begin{align}
\bmx\,T\,\bmx^{-1}\,T^{-1} = I \rightarrow \bmx'\,T'\,{\bmx'}^{-1}\,{T'}^{-1} =\W^{2c};
\end{align}
here, the analogous factor $C_{1,2}$ is always an even product of $\W$ -- there exists no gauge transformation that relates the two factor systems in $\gwp$ and $\gwz$. We say that the factor system of a projective representation can be lifted if, by some choice of gauge, all of $\{C_{i,j}\}$ from Eq.\ (\ref{factorsystem}) may be reduced to the identity element in $\caln$; Eq.\ (\ref{transform3}) demonstrates that $C_{1,2}$ for $\gwp$ can never be transformed to identity. $\gwp$ thus exemplifies an intrinsically projective representation, wherefor its nontrivial factor system can never be lifted. \\





Finally, we remark that this Section does not exhaust all elements in $\gwp$ or $\gwz$; our treatment here minimally conveys their group structures. A complete treatment of $\gwp$ is offered in App.\ \ref{app:kxpi}, where we also derive the above algebraic relations in greater detail.



\section{Discussion and outlook} \label{sec:outlook}



In the topological classification of band insulators, one may sometimes infer the classification purely from the representation theory\cite{AAchen,ChaoxingNonsymm,TCIbyrepresentation} of surface wavefunctions. In our companion work,\cite{Hourglass} we have identified a criterion on the surface group that characterizes all robust surface states which are protected by space-time symmetries.\cite{kane2005B,Classification_Chiu,fu2011,AAchen,ChaoxingNonsymm,TCIbyrepresentation,unpinned,Shiozaki2015,Nonsymm_Shiozaki,moore2010,fang2013,liu2013,magnetic_ti} Our criterion introduces the notion of connectivity within a submanifold ($\calm$) of the surface-Brillouin torus, and generalizes the theory of elementary band representations.\cite{connectivityMichelZak,elementaryenergybands} To restate the criterion briefly, we say there is a $\cald$-fold connectivity within $\calm$ if bands there divide into sets of $\cald$, such that within each set there are enough contact points in $\calm$ to continuously travel through all $\cald$ branches. If $\calm$ is a single wavevector ($\kpar$), $\cald$ coincides with the minimal dimension of the irreducible representation at $\kpar$; $\cald$ generalizes this notion of symmetry-enforced degeneracy where $\calm$ is larger than a wavevector (e.g., a glide line). We are ready to state our criterion: (a) there exist two separated submanifolds $\calm_{\sma{1}}$ and $\calm_{\sma{2}}$, with corresponding $\cald_{\sma{1}}{=}\cald_{\sma{2}}{=} fd$ ($f {\geq} 2$ and $d {\geq} 1$ are integers), and (b) a third submanifold $\calm_{\sma{3}}$ that connects $\calm_{\sma{1}}$ and $\calm_{\sma{2}}$, with corresponding $\cald_{\sma{3}}{=}d$. This surface-centric criterion is technically simple, and has proven to be predictive of the topological classification. However, we also found it is sometimes over-predictive,\cite{Hourglass} in the sense of allowing some surface topologies that are inconsistent with the full set of bulk symmetries.  \\


An alternative and, as far as we know, faithful approach would apply our connectivity criterion\cite{Hourglass} to the Wilson `bands', which properly encode bulk symmetries that are absent on the surface; since Wilson `bands' also live on the surface-Brillouin torus, we could replace the original meaning of surface bands in the above criterion by Wilson `bands'. To determine the possible Wilson `bandstructures', one has to determine how symmetries are represented in the Wilson loop; one lesson learned from classifying $D_{6h}^4$ is that this representation can be projective, requiring an extension of the point group by the Wilson loop itself. Such an extended group forces us to generalize the traditional notion of symmetry as a space-time transformation -- we instead encounter `symmetry' operators that combine both space-time transformations and quasimomentum translations, thus putting real and quasimomentum space on equal footing.  While our case study involved a nonsymmorphic space group, the nonsymmorphicity (i.e., nontrivial extensions by spatial translations) is not a prerequisite for nontrivial quasimomentum extensions, e.g., there are projective mirror planes (e.g., in symmorphic rocksalt structures) where the reflection also relates Bloch waves separated by half a reciprocal period; the implications are left for future study.\\



To restate our finding from a broader perspective, group cohomology specifies how symmetries are represented in the quasimomentum submanifold, which in turn determines the band topology. A case in point is time reversal symmetry ($T$), which may be extended by $2\pi$-spin rotations (which distinguishes half-integer- from integer-spin representations) and also by real-space translations (which distinguishes paramagnetic and antiferromagnetic insulators); only the projective representation ($T^2{=}{-}I$) has a well-known $\Z_2$ topology.\cite{kane2005B,moore2010}  By our cohomological classification of quasimomentum submanifolds through \q{classifymani}, we have provided a unifying framework to classify chiral topological insulators,\cite{Haldane1988}, and topological insulators with robust edge states protected by space-time symmetries.\cite{Classification_Chiu,fu2011,AAchen,ChaoxingNonsymm,unpinned,Shiozaki2015,Nonsymm_Shiozaki,kane2005B,moore2010,fang2013,liu2013,magnetic_ti} Our framework is also useful in classifying some topological insulators without edge states;\cite{AA2014,hughes2011,turner2012} one counter-example that eludes this framework may nevertheless by classified by bent Wilson loops,\cite{berryphaseTCI} rather than the straight Wilson loops of this work. With the recent emergence of Floquet topological phases, an interesting direction would be to consider further extending \q{classifymani} by discrete time translations.\cite{floquet_groups}

\begin{acknowledgments}
We thank Ken Shiozaki and Masatoshi Sato for discussions of their K-theoretic classification, Mykola Dedushenko for discussions on group cohomology, and Chen Fang for suggesting a nonsymmorphic model of the quantum spin Hall insulator. ZW, AA and BAB were supported by NSF CAREER DMR-095242, ONR - N00014-11-1-0635, ARO MURI on topological insulators, grant W911NF-12-1-0461, NSF-MRSEC DMR-1420541, Packard Foundation, Keck grant, “ONR Majorana Fermions” 25812-G0001-10006242-101, and Schmidt fund 23800-E2359-FB625. In finishing stages of this work, AA was further supported by the Yale Prize Fellowship.\\
\end{acknowledgments}


\begin{appendix}

\textbf{
\begin{center}
APPENDIX\\
\end{center}
}

Organization of the Appendix:\\

\noi{\ref{app:tightbinding}} We review how space-time symmetries affect the tight-binding Hamiltonian. Notations are introduced which will be employed in the remaining appendices. \\

\noi{\ref{app:symmPyP}} We derive how symmetries of the Wilson loop are represented, and their constraints on the Wilson bands, as summarized in Tab.\ \ref{rulesgeneric} and \ref{rulestrim}. The first few Sections deal with ordinary symmetry representations along $\tilx \tilg \tilz \tilu$, while the last derives the projective representations along $\tilx \tilu$.\\

\noi{\ref{app:connectivity}} We prove the four-fold connectivity of Hamiltonian bands,  in spin systems with minimally time-reversal and glide-reflection symmetries. This proof is used in the topological classification of Sec.\ \ref{sec:connect}.\\

\noi{\ref{app:wilsonproj}} We introduce group extensions by Wilson loops, as well as rederive the extended algebra in Sec.\ \ref{sec:wilsoniansymm} from a group-theoretic perspective. \\

\section{Review of symmetries in the tight-binding method} \label{app:tightbinding}

We review the effects of spatial symmetries in App.\ \ref{app:rep}, then generalize our discussion to include time-reversal symmetry in App.\ \ref{sec:spacetimetight}.

\begin{widetext}

\subsection{Effect of spatial symmetries on the tight-binding Hamiltonian} \label{app:rep}

Let us denote a spatial transformation by $\pdg{g}_{\bdelta}$, which transforms real-space coordinates as $ \br \rightarrow D_g \br + \bdelta$, where $D_g$ is the orthogonal matrix representation of the point-group transformation $g$ in $\R^d$.  Nonsymmorphic space groups contain symmetry elements where $\bdelta$ is a rational fraction\cite{Lax} of the lattice period; in a symmorphic space group, an origin can be found where $\bdelta=0$ for all symmetry elements. The purpose of this Section is to derive the constraints of $\pdg{g}_{\bdelta}$ on the tight-binding Hamiltonian. First, we clarify how $\pdg{g}_{\bdelta}$ transforms the creation and annihilation operators. We define the creation operator for a \low function\cite{slater1954,goringe1997,lowdin1950} ($\varphi_{\alpha}$) at Bravais lattice vector $\boldsymbol{R}$ as $\dg{c}_{\alpha}(\boldsymbol{R}+\boldsymbol{r_{\alpha}})$. From (\ref{basisvec}), the creation operator for a Bloch-wave-transformed \low orbital $\phi_{\boldsymbol{k},\alpha}$ is
\begin{align} \label{blochbasis}
\dg{c}_{\boldsymbol{k}, \alpha} = \frac{1}{\sqrt{N}} \sum_{\boldsymbol{R}} \,e^{i\boldsymbol{k} \cdot (\boldsymbol{R} + \boldsymbol{r_{\alpha}})}\,\dg{c}_{\alpha}(\boldsymbol{R}+\boldsymbol{r_{\alpha}}); \;\; \alpha =1,\ldots,n_{tot}.
\end{align}
A Bravais lattice (BL) that is symmetric under $\pdg{g}_{\bdelta}$ satisfies two conditions:\\

\noi{i}  for any BL vector $\boldsymbol{R}$, $D_g\boldsymbol{R}$ is also a BL vector: 
\begin{align} \label{pgscond1}
\forall \boldsymbol{R} \in \text{BL}, \;\; D_g\boldsymbol{R} \in \text{BL}.
\end{align}
\noi{ii} If $\pdg{g}_{\bdelta}$ transforms an orbital of type $\alpha$ to another of type $\beta$, then $D_g(\boldsymbol{R}+\boldsymbol{r_{\alpha}})+\bdelta$ must be the spatial coordinate of an orbital of type $\beta$. To restate this formally, we define a matrix $U_{\sma{g\bdelta}}$ such that the creation operators transform as
\begin{align}
\pgdel\;:\;\dg{c}_{\alpha}(\boldsymbol{R}+\boldsymbol{r_{\alpha}}) \; \longrightarrow \; \dg{c}_{\beta}\big(\,D_g\boldsymbol{R}+\boldsymbol{R}^{\sma{g \bdelta}}_{\sma{\beta \alpha}}+\boldsymbol{r_{\beta}}\,\big) \,[U_{\sma{g\bdelta}}]_{\beta \alpha},
\end{align}
with $\boldsymbol{R}^{\sma{g \bdelta}}_{\sma{\beta \alpha}} \equiv D_g\boldsymbol{r_{\alpha}}+\bdelta -\boldsymbol{r_{\beta}}$. Then\begin{align} \label{pgscond2}
[U_{\sma{g\bdelta}}]_{\beta \alpha} \neq 0 \imp  \boldsymbol{R}^{\sma{g \bdelta}}_{\sma{\beta \alpha}} \in \text{BL}. 
\end{align}
Explicitly, the nonzero matrix elements are given by
\begin{align}
[U_{\sma{g\bdelta}}]_{\beta \alpha} = \sum_{s,s'} \int d^d r\; \varphi_{\beta}^*(\br,s')\; [D^{(1/2)}_g]_{s's} \;\varphi_{\alpha}(D_g^{-1}\br,s),
\end{align}
where $\varphi_{\sma{\alpha}}$ is a spinor with spin index $s$, and $D_{\sma{g}}^{\sma{(1/2)}}$ represents $\pdg{g}_{\bdelta}$ in the spinor representation. \\

For fixed $\pdg{g}_{\bdelta},\alpha$ and $\beta$, the mapping $\calt^{\sma{g \bdelta}}_{\sma{\beta \alpha}}: \boldsymbol{R} \rightarrow \boldsymbol{R}^{\sma{g \bdelta}}_{\sma{\beta \alpha}}$ is bijective. Applying (\ref{blochbasis}), (\ref{pgscond1}), (\ref{pgscond2}), the orthogonality of $D_g$ and the bijectivity of $\calt^{\sma{g \bdelta}}_{\sma{\beta \alpha}}$, the Bloch basis vectors transform as
\begin{align} \label{blochspatial}
\pgdel\; : \; \dg{c}_{\boldsymbol{k},\alpha} \; \longrightarrow &\; \frac{1}{\sqrt{N}} \sum_{\boldsymbol{R}} \,e^{i\boldsymbol{k} \cdot (\boldsymbol{R} + \boldsymbol{r_{\alpha}})}\, \dg{c}_{\beta}\big(\,D_g\boldsymbol{R}+\boldsymbol{R}^{\sma{g \bdelta}}_{\sma{\beta \alpha}}+\boldsymbol{r_{\beta}}\,\big) \,[U_{\sma{g\bdelta}}]_{\beta \alpha}\lin
\eq e^{-i (D_g \boldsymbol{k}) \cdot \bdelta } \frac{1}{\sqrt{N}} \sum_{\boldsymbol{R}} \,e^{i[D_g\boldsymbol{k}] \cdot [D_g(\boldsymbol{R} + \boldsymbol{r_{\alpha}})+\bdelta]}\, \dg{c}_{\beta}\big(\,D_g\boldsymbol{R}+\boldsymbol{R}^{\sma{g \bdelta}}_{\sma{\beta \alpha}}+\boldsymbol{r_{\beta}}\,\big) \,[U_{\sma{g\bdelta}}]_{\beta \alpha}\lin
\eq e^{-i (D_g \boldsymbol{k}) \cdot \bdelta } \frac{1}{\sqrt{N}} \sum_{\boldsymbol{R}} \,e^{i[D_g\boldsymbol{k}] \cdot [D_g\boldsymbol{R}+\boldsymbol{R}^{\sma{g \bdelta}}_{\sma{\beta \alpha}}+\boldsymbol{r_{\beta}}]}\, \dg{c}_{\beta}\big(\,D_g\boldsymbol{R}+\boldsymbol{R}^{\sma{g \bdelta}}_{\sma{\beta \alpha}}+\boldsymbol{r_{\beta}}\,\big) \,[U_{\sma{g\bdelta}}]_{\beta \alpha}\lin
\eq e^{-i (D_g \boldsymbol{k}) \cdot \bdelta } \frac{1}{\sqrt{N}} \sum_{\boldsymbol{R}'} \,e^{i[D_g\boldsymbol{k}] \cdot [\boldsymbol{R}'+\boldsymbol{r_{\beta}}]}\, \dg{c}_{\beta}\big(\,\boldsymbol{R}'+\boldsymbol{r_{\beta}}\,\big) \,[U_{\sma{g\bdelta}}]_{\beta \alpha}\lin
\eq  e^{-i (D_g \boldsymbol{k}) \cdot \bdelta } \dg{c}_{D_g\boldsymbol{k},\beta}\,[U_{\sma{g\bdelta}}]_{\beta \alpha}.
\end{align}
\end{widetext}
This motivates a definition of the operator
\begin{align} \label{actbloch}
\pdg{\hat{g}}_{\bdelta}(\bk) \equiv e^{-i (D_g \boldsymbol{k}) \cdot \bdelta }\, U_{\sma{g\bdelta}},
\end{align}
which acts on Bloch wavefunctions ($|u_{n,\bk}\rangle$) as
\begin{align} \label{act}
\pgdel\; : \;\ket{u_{n,\bk}} \;\longrightarrow \; \pdg{\hat{g}}_{\bdelta}(\bk)\,\ket{u_{n,\bk}}.
\end{align}
The operators $\{\pdg{\hat{g}}_{\bdelta}(\bk)\}$ form a representation of the space-group algebra\cite{Lax} in a basis of Bloch-wave-transformed \low orbitals; we call this the \low representation. If the space group is nonsymmorphic, the nontrivial phase factor exp$(-i D_g \bk \cdot \bdelta)$ in $\hatgdel(\bk)$ encodes the effect of the fractional translation, i.e., the momentum-independent matrices $\{U_{\sma{g\bdelta}}\}$ by themselves form a representation of a point group.\\

\begin{figure}[ht]
\centering
\includegraphics[width=8 cm]{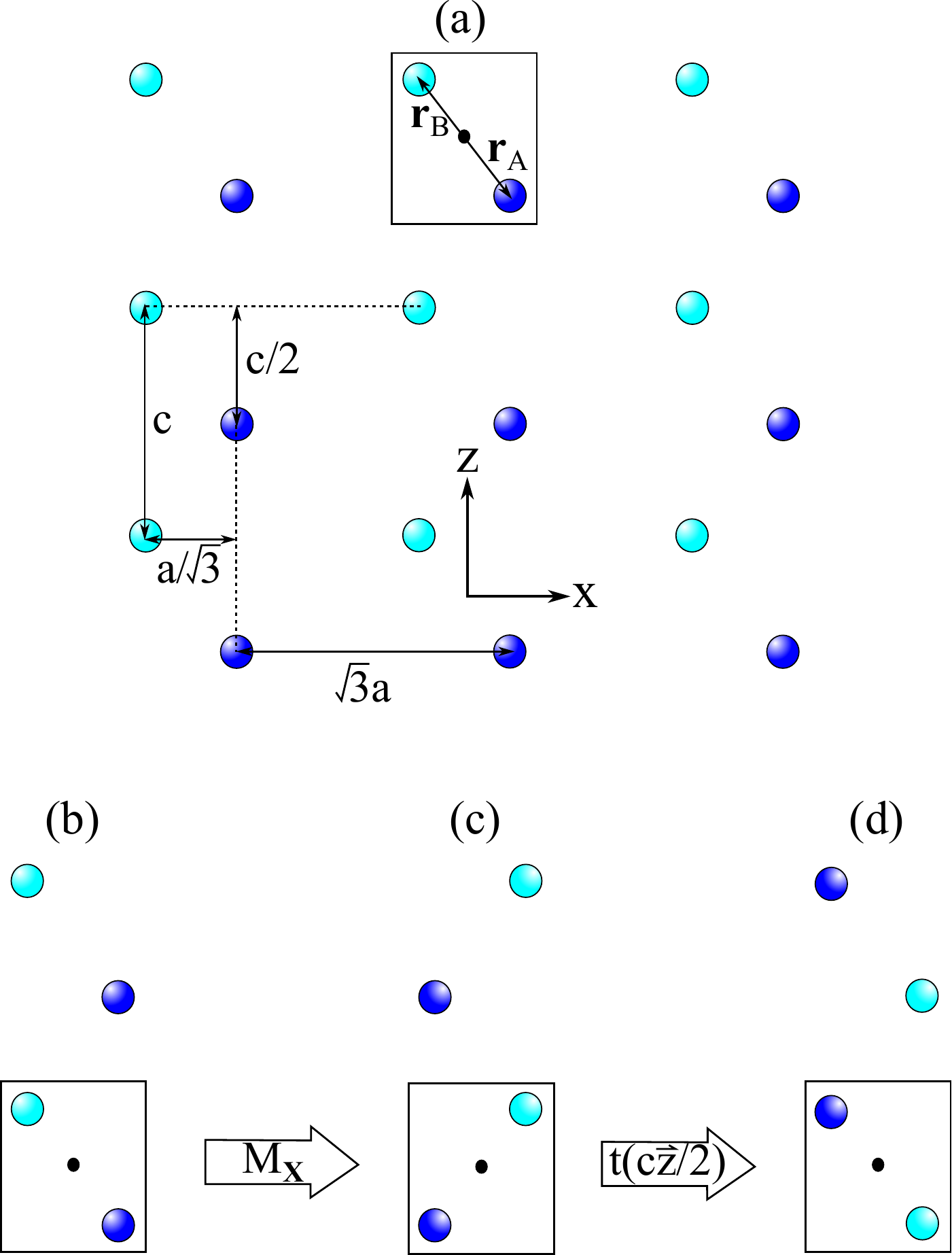}
    \caption{(a) Simple example of a 2D nonsymmorphic crystal. The two sublattices are colored respectively colored dark blue and cyan. (b-d) illustrate the effect of a glide reflection. } \label{fig:nonsymmorphic_ex}
\end{figure}

To exemplify this abstract discussion, we analyze a simple 2D nonsymmorphic crystal in Fig.\ \ref{fig:nonsymmorphic_ex}. As delineated by a square, the unit cell comprises two atoms labelled by subcell coordinates $A$ and $B$, and the spatial origin is chosen at their midpoint, such that $\br_{\sma{A}}=a\vec{x}/\sqrt{3} - c\vec{z}/2 =-\br_{\sma{B}}$, as shown in Fig.\ \ref{fig:nonsymmorphic_ex}(a). The symmetry group ($Pma2$) of this lattice is generated by the elements ${\bmx}$ and $\bmz$, where in the former we first reflect across $\vec{x}$ ($g=M_x$) and then translate by $\bdelta=c\vec{z}/2$. Similarly, $\bmz$ is shorthand for a $z{\rightarrow}{-}z$ reflection followed by a translation by $\bdelta=c\vec{z}/2$. Let us represent these symmetries with spin-doubled $s$ orbitals on each atom. Choosing our basis to diagonalize $S_z$, 
\begin{align}
&\bmx\;:\; \begin{cases} \dg{c}_{\sma{A,S_z}}(\bR+\br_{\sma{A}})  \; \longrightarrow \; -i\dg{c}_{\sma{B,-S_z}}(D_x\bR+\br_{\sma{B}}), \\
\dg{c}_{\sma{B,S_z}}(\bR+\br_{\sma{B}})\; \longrightarrow \;  -i\dg{c}_{\sma{A,-S_z}}(D_x\bR+c\vec{z}+\br_{\sma{A}}), \end{cases}
\end{align}
where $D_x(x,z)^t = (-x,z)^t$, and in the second mapping, we have applied $\bR^{\sma{x,c\vec{z}/2}}_{\sma{AB}} = D_x\br_{\sma{B}}+ c\vec{z}/2 - \br_{\sma{A}} = c\vec{z}$. It is useful to recall here that a reflection is the product of an inversion with a two-fold rotation about the reflection axis: $M_{\sma{j}}=\cali\, C_{\sma{2j}}$ for $j \in \{x,z\}$. Consequently, $\bmx \propto C_{\sma{2x}}$ flips $S_{\sma{z}} \rightarrow -S_{\sma{z}}$. In the basis of Bloch waves,
\begin{align}
&\bmx\;:\;  \dg{c}_{k,\alpha}\; \longrightarrow \;  e^{-ik_zc/2}\,\dg{c}_{D_x\bk,\beta}\,[U_{\bmx}]_{\beta \alpha},
\end{align}
with $U_{\bmx} = -i\,\tx\, \sx.$
Here, we have employed $\tz=+1$ ($-1$) for subcell $A$ ($B$) and $\sz=+1$ for spin up in $\vec{z}$. A similar analysis for the other reflection ($\bmz \propto C_{\sma{2z}} \propto \text{exp}[-iS_z\pi]$) leads to
\begin{align}
\bmz\;:\; \begin{cases} \dg{c}_{\sma{A,S_z}}(\bR+\br_{\sma{A}})  \; \longrightarrow \\
 \myspace \myspace -i\,\text{sign}[S_z]\,\dg{c}_{\sma{A,S_z}}(D_z\bR+c\vec{z}+\br_{\sma{A}}), \\
\dg{c}_{\sma{B,S_z}}(\bR+\br_{\sma{B}})\; \longrightarrow \\
\myspace \myspace -i\,\text{sign}[S_z]\,\dg{c}_{\sma{B,S_z}}(D_z\bR+\br_{\sma{B}}), \end{cases}
\end{align}
with $D_z(x,z)^t = (x,-z)^t$, and in the basis of Bloch-wave-transformed \low orbitals,
\begin{align}
&\bmz\;:\;  \dg{c}_{\bk,\alpha}\; \longrightarrow \;  e^{-ik_zc/2}\,\dg{c}_{D_z\bk,\beta}\,[U_{\bmz}]_{\beta \alpha},  
\end{align}
with $U_{\bmz} = -i\,\sz.$ To recapitulate, we have derived $\{\hatgdel\}$ as 
\begin{align}
&\hat{\bar{M}}_x(\bk) =  -i\,e^{-ik_zc/2}\,\tx\,\sx \ins{and} \lin
&\hat{\bar{M}}_z(\bk) =  -i\,e^{-ik_zc/2}\,\sz, 
\end{align}
which should satisfy the space-group algebra for $Pma2$, namely that
\begin{align}
&\bmx^2 = \bar{E}\,t(c\vec{z}),\;\;\;\; \bmz^2 = \bar{E},\ins{and} \lin
&\bmz\,\bmx = \bar{E}\,t(-c\vec{z})\,\bmx\,\bmz,
\end{align}
where $\bar{E}$ denotes a $2\pi$ rotation and $t(c\vec{z})$ a translation. Indeed, when acting on Bloch waves with momentum $\bk$,
\begin{align}
\hat{\bar{M}}_x&(D_x\bk)\;\hat{\bar{M}}_x(\bk) = -e^{-ik_zc},\;\;\;\; \hat{\bar{M}}_z(D_z\bk)\;\hat{\bar{M}}_z(\bk) = -I,\lin
&\hat{\bar{M}}_{z}(D_x\bk)\,\hat{\bar{M}}_x(\bk) = -e^{-ik_zc}\,\hat{\bar{M}}_x(D_z\bk)\,\hat{\bar{M}}_z(\bk).
\end{align}
Finally, we verify that the momentum-independent matrices $\{U_{\sma{g\bdelta}}\}$ form a representation of the double point group $C_{2v}$, whose algebra is simply
\begin{align} \label{symmorphicalgebratwo}
M_x^2=M_z^2= \bar{E} \ins{and} M_z\,M_x =\bar{E}\,M_x\,M_z.
\end{align}
A simple exercise leads to
\begin{align}
U_{\bmx}^2=U_{\bmz}^2=-I \ins{and} \{U_{\bmx},U_{\bmz}\}=0.
\end{align}
The algebras of $C_{2v}$ and $Pma2$ differ only in the additional elements $t({\pm} c \vec{z})$, which in the \low representation ($\{\hatgdel(\bk)\}$) is accounted for by the phase factors exp$(-ik_zc/2)$.\\

Returning to a general discussion, if the Hamiltonian is symmetric under $\pgdel$:
\begin{align}
\pgdel: \;\; \hat{H} = \sum_{\bk} \dg{c}_{\bk,\alpha}\,H(\bk)_{\ab} \pdg{c}_{\bk,\beta}  \;\;\longrightarrow \;\; \hat{H},
\end{align}
then Eq.\ (\ref{blochspatial}) implies
\begin{align} \label{symmonhk}
\hatgdel(\bk)\,H(\boldsymbol{k})\,\hatgdel(\bk)^{\mo} = H\big(\,D_g\boldsymbol{k}\,\big).
\end{align}
By assumption of an insulating gap, $\pdg{\hat{g}}_{\bdelta}(\bk)|u_{n,\bk}\rangle$ belongs in the occupied-band subspace for any occupied band $|u_{n,\bk}\rangle$. This implies a unitary matrix representation (sometimes called the `sewing matrix') of $\pgdel$ in the occupied-band subspace:
\begin{align} \label{goccupied}
 [\calgdel(D_g\bk+\bG,\bk)]_{mn} = \bra{u_{m,D_g\bk+\bG}} \,V(-\bG)\, \pdg{\hat{g}}_{\bdelta}(\bk) \,\ket{u_{n,\bk}},
\end{align} 
with $m,n=1,\ldots,\noc.$ Here, $\bG$  is any reciprocal vector (including zero), and we have applied Eq.\ (\ref{periodP}) which may be rewritten as:
\begin{align} \label{uuk}
\sum_{n =1}^{\noc} \ket{u_{n,\bk}}\bra{u_{n,\bk}} = V(\bG)\,\sum_{n =1}^{\noc} \ket{u_{n,\bk+\bG}}\bra{u_{n,\bk+\bG}} \,V(\bG)^{\mo}.
\end{align}
To motivate Eq.\ (\ref{goccupied}), we are often interested in high-symmetry $\bk$ which are invariant under $\pdg{g}_{\bdelta}$, i.e., $D_g\bk+\bG=\bk$ for some $\bG$ (possibly zero). At these special momenta, the `sewing matrix' is unitarily equivalent to a diagonal matrix, whose diagonal elements are the $\pdg{g}_{\bdelta}$-eigenvalues of the occupied bands. When we're not at these high-symmetry momenta, we will sometimes use the shorthand: 
\begin{align} \label{shorthand1}
[\calgdel(\bk)]_{mn} \equiv [\calgdel(D_g\bk,\bk)]_{mn} = \bra{u_{m,D_g\bk}} \, \pdg{\hat{g}}_{\bdelta}(\bk) \,\ket{u_{n,\bk}}, 
\end{align}
since the second argument is self-evident. We emphasize that $\hatgdel$ and $\brevegdel$ are different matrix representations of the same symmetry element ($\pgdel$), and moreover the matrix dimensions differ: (i) $\hatgdel$ acts on Bloch-combinations of \low orbitals ($\{\phi_{\sma{\boldsymbol{k}, \alpha}}| \alpha=1,\ldots,n_{tot}\}$) defined in Eq.\ (\ref{basisvec}), while (ii) $\brevegdel$ acts on the occupied eigenfunctions ($\{u_{\sma{n,\bk}}|n=1,\ldots, \noc\}$) of $H(\bk)$. \\

It will also be useful to understand the commutative relation between $\hatgdel(\bk)$ and the diagonal matrix $V(\bG)$ which encodes the spatial embedding; as defined in Eq.\ (\ref{aperiodic}), the diagonal elements are $[V(\bG)]_{\ab} = \delta_{\ab} \text{exp}(i\bG \cdot \br_{\alpha})$. From Eq.\ (\ref{pgscond1}) and (\ref{pgscond2}), 
\begin{align} 
[U_{\sma{g\bdelta}}]_{\alpha \beta} \neq 0 &\imp  D_g^{\mo}\boldsymbol{R}^{\sma{g \bdelta}}_{\sma{\ab}} \in \text{BL} \lin
&\imp  e^{i\bG\cdot ( \boldsymbol{r_{\beta}}+D_{\sma{g}}^{\mo}\bdelta -D_{\sma{g}}^{\mo}\boldsymbol{r_{\alpha}})}=1,
\end{align}
for a reciprocal-lattice (RL) vector $\bG$. Applying this equation in 
\begin{align}
0 \neq [\hatgdel(\bk)\, V(\bG)]_{\ab} \eq e^{-i (D_g \boldsymbol{k})\cdot \bdelta }\, [U_{\sma{g\bdelta}}]_{\ab} \,e^{i\bG \cdot \br_{\beta}} \lin
\eq e^{-i (D_g \boldsymbol{k})\cdot \bdelta }\, [U_{\sma{g\bdelta}}]_{\ab} \,e^{i(D_g \bG) \cdot (\br_{\alpha}-\bdelta)} \lin
\eq e^{-i(D_g \bG) \cdot \bdelta} \,[V(D_g\bG)\,\hatgdel(\bk)]_{\ab}, \notag
\end{align}
we then derive
\begin{align} \label{uv}
\hatgdel(\bk)\, V(\bG) = e^{-i(D_g\bG)\cdot \bdelta}\,V(D_g\bG) \,\hatgdel(\bk),
\end{align}
This equality applies only if the argument of $V$ is a reciprocal vector.

\subsection{Effect of space-time symmetry on the tight-binding Hamiltonian}\label{sec:spacetimetight}

Consider a general space-time transformation $T \pgdel$, where now we include the time-reversal $T$; the following discussion also applies if $\pgdel$ is the trivial transformation.    
\begin{align}
T \pgdel \;: \;\dg{c}_{\alpha}(\boldsymbol{R}+\boldsymbol{r_{\alpha}}) \rightarrow \dg{c}_{\beta}\big(D_g\boldsymbol{R}+\boldsymbol{R}^{\sma{ Tg\bdelta}}_{\sma{\beta \alpha}}+\boldsymbol{r_{\beta}}\,\big) \,[\utg]_{\beta \alpha},\notag
\end{align}
where $\utg$ is the matrix representation of $T \pgdel$ in the \low orbital basis, $\boldsymbol{R}^{\sma{Tg\bdelta}}_{\sma{\beta \alpha}} =D_g\boldsymbol{r_{\alpha}} + \bdelta -\boldsymbol{r_{\beta}}$, 
\begin{align} \label{uitv}
[\utg]_{\beta \alpha} \neq 0 \imp  \boldsymbol{R}^{\sma{Tg\bdelta}}_{\sma{\beta \alpha}} \in \text{BL}, 
\end{align}
and the Bravais-lattice mapping of $\bR$ to $D_g\bR + \boldsymbol{R}^{\sma{Tg\bdelta}}_{\sma{\beta \alpha}}$ is bijective. It follows that the Bloch-wave-transformed \low orbitals transform as
\begin{align}
T \pgdel \;: \;\dg{c}_{\boldsymbol{k},\alpha} \; \longrightarrow \; e^{i D_g \boldsymbol{k} \cdot \bdelta } \dg{c}_{-D_g\boldsymbol{k},\beta}\,[\utg]_{\beta \alpha}.
\end{align}
This motivates the following definition for the \low representation of $T \pgdel$:
\begin{align} \label{Tgdrep}
{\hat{T}}_{\sma{g\bdelta}}(\bk) \equiv e^{i (D_g \boldsymbol{k}) \cdot \bdelta } \,\utg\,K,
\end{align}
where $K$ implements complex conjugation, such that a symmetric Hamiltonian ($T \pgdel : \hat{H} \rightarrow \hat{H})$ satisfies
\begin{align} \label{sptiHam}
\hatTg(\bk)\,H(\bk)\,\hatTg(\bk)^{\mo}= H\big({-}D_g\bk\,\big).
\end{align}
For a simple illustration, we return to the lattice of Fig.\ \ref{fig:nonsymmorphic_ex}, where time-reversal symmetry is represented by $\hat{T}(\bk)=-i\sy K$ in a basis where $\sz=+1$ corresponds to spin up in $\vec{z}$. Observe that time reversal commutes with any spatial transformation:
\begin{align}
\text{for} \;\; j \in \{x,z\}, \;\;\;\;\hat{T}(D_j\bk)\;\hat{\bar{M}}_{\sma{j}}(\bk)=\hat{\bar{M}}_{\sma{j}}(-\bk)\;\hat{T}(\bk). \notag
\end{align}
If the Hamiltonian is gapped, there exists an antiunitary representation of $T\pgdel$ in the occupied-band subspace:
\begin{align} \label{sew}
&[\caltgdel(\bG-D_g\bk,\bk)]_{mn} \lin
\eq \bra{u_{m,\bG-D_g\bk}} \,V(-\bG)\, {\hat{T}}_{\sma{g\bdelta}}(\bk) \,\ket{u_{n,\bk}},
\end{align} 
where $m,n=1,\ldots,\noc,$ $\bG$ is any reciprocal vector and we have applied Eq.\ (\ref{uuk}). Once again, we introduce the shorthand: 
\begin{align} \label{shorthand2}
[\caltgdel(\bk)]_{mn} &\equiv\; [\caltgdel(-D_g\bk,\bk)]_{mn} \lin
\eq \bra{u_{m,-D_g\bk}} \, {\hat{T}}_{\sma{g\bdelta}}(\bk) \,\ket{u_{n,\bk}}.
\end{align}
 Eq.\ (\ref{uitv}) and (\ref{pgscond1}) further imply that
\begin{align} 
[U_{\sma{Tg\bdelta}}]_{\alpha \beta} \neq 0 &\imp  D_g^{\mo}\boldsymbol{R}^{\sma{g \bdelta}}_{\sma{\ab}} \in \text{BL} \lin
&\imp  e^{i\bG\cdot ( \boldsymbol{r_{\beta}}+D_{\sma{g}}^{\mo}\bdelta -D_{\sma{g}}^{\mo}\boldsymbol{r_{\alpha}})}=1,
\end{align}
which when applied to
\begin{align}
0 &\neq\; [\hatTg(\bk)\, V(\bG)\, K]_{\ab} \lin
\eq e^{i (D_g \boldsymbol{k})\cdot \bdelta }\, [U_{\sma{Tg\bdelta}}]_{\ab} \,e^{-i\bG \cdot \br_{\beta}} \lin
\eq e^{i (D_g \boldsymbol{k})\cdot \bdelta }\, [U_{T\sma{g\bdelta}}]_{\ab} \,e^{-i(D_g \bG) \cdot (\br_{\alpha}-\bdelta)} \lin
\eq e^{+i(D_g \bG) \cdot \bdelta} \,[V(-D_g\bG)\,\hatTg(\bk)\,K]_{\ab},
\end{align}
leads finally to
\begin{align} \label{UU}
\hatTg(\bk)\,V(\bG) = e^{i(D_g\bG)\cdot \bdelta}\,V(-D_g\bG)\,\hatTg(\bk).
\end{align}

\section{Symmetries of the Wilson loop} \label{app:symmPyP}

The goal of this Appendix is to derive how symmetries of the Wilson loop are represented, and their implications for the `rules of the curves', as summarized in Tab.\ \ref{rulesgeneric} and \ref{rulestrim}. After introducing the notations and basic analytic properties of Wilson loops in App.\ \ref{app:notations}, we consider in App.\ \ref{sec:010} the effect of spatial symmetries, with particular emphasis on glide symmetry. We then generalize our discussion to space-time symmetries in Sec.\ \ref{sec:spacetime}. These first sections apply only to symmetries of the Wilson loops along $\tilx \tilg$, $\tilg \tilz$ and $\tilz \tilu$. These symmetry representations are shown to be ordinary, i.e., they do not encode quasimomentum transport; their well-known algebra includes:
\begin{align} \label{ordinaryalgebra}
&\bmx^2 = \bar{E}\;t(c\vec{z}) ,\;\;\;\;(T\bmz)^2=I,\lin
&  \bmx\,T\,\bmz = \bar{E} \,t(c\vec{z})\,T\,\bmz\,\bmx,\;\;\;\;T^2=(\cali T)^2 = \bar{E},\lin
& \bmx \, \cali T  = t(c\vec{z})\,\cali T\,\bmx,\;\;\;\;  \bmx \;T = T \;\bmx.
\end{align}
In App.\ \ref{app:kxpi}, we move on to derive the projective representations which apply along $\tilx \tilu$.


\subsection{Notations and analytic properties of the Wilson loop} \label{app:notations}

Consider the parallel transport of occupied bands along the non-contractible loops of Sec.\ \ref{subsec:reviewwilson}.
In the \low-orbital basis, such transport is represented by the Wilson-loop operator:\cite{AA2014}
\begin{align} \label{pt}
\hat{W}(\kpar) = V(2\pi \vec{y})\prod_{k_y}^{\pi \leftarrow -\pi} P(k_y,\kpar).
\end{align}
Recall here our unconventional ordering: $\bk = (k_y,k_x,k_z) = (k_y,\kpar)$. We have discretized the momentum as $k_y = 2\pi m/ N_y$ for integer $m=1,\ldots,N_y$, and $(\pi \leftarrow -\pi)$ indicates that the product of projections is path-ordered. The role of the path-ordered product is to map a state in the occupied subspace ($\calh(-\pi,\kpar)$) at $(-\pi,\kpar)$ to one ($|{\tilde{u}}\rangle$) in the occupied subspace at $(\pi,\kpar)$; the effect of $V(2\pi \vec{y})$ is to subsequently map $|{\tilde{u}}\rangle$ back to $\calh(-\pi,\kpar)$, thus closing the parameter loop; cf. Eq.\ (\ref{uuk}). Equivalently stated, we may represent this same parallel transport in the basis of $\noc$ occupied bands:
\begin{align} \label{con3}
[\W(\kpar)]_{ij} = \bra{u_{i,(-\pi,\kpar)}}\,\hat{W}(\kpar)\,\ket{u_{j,(-\pi,\kpar)}}.
\end{align}
While $\W$ depends on the choice of gauge for $|u_{j,{-}\pi,\kpar}\rangle$, its eigenspectrum does not. Indeed, under the gauge transformation
\begin{align}
\ket{u_{j,(-\pi,\kpar)}} &\rightarrow \sum_{i=1}^{\noc} \ket{u_{i,(-\pi,\kpar)}} S_{ij}, \;\text{with}\; S \in U(\noc),  \lin
&\W \rightarrow \dg{S}\W S = S^{-1}\W S.
\end{align}
The eigenspectrum is also independent of the base point of the loop;\cite{AA2014} our choice of $(-\pi,\kpar)$ as the base point merely renders certain symmetries transparent. In the limit of large $N_{\sma{y}}$, $\W$ becomes unitary and its full eigenspectrum comprises the unimodular eigenvalues of $\hat{W}$, which we label by exp$[i\theta_{\sma{n,\kpar}}]$ with $n=1,\ldots, \noc$. Denoting the eigenvalues of $\pper\hat{y}\pper$ as ${y_{\sma{n,\kpar}}}$, the two spectra are related as $y_{\sma{n,\kpar}} = \theta_{\sma{n,\kpar}}/2\pi$ modulo one.\cite{AA2014}  \\

On occasion, we will also need the reverse-oriented Wilson loop ($\hat{W}_r$), which transports a state from base point $(\pi,\kpar) \rightarrow (-\pi,\kpar)$:
\begin{align} \label{reverseorient}
\hat{W}_r( \kpar) = V\big({{-} }2\pi \vec{y}\,\big) \prod_{k_y}^{{-} \pi \leftarrow \pi} P\big(\,k_y, \kpar\,\big).
\end{align}
In the occupied-band basis, Wilson loops of opposite orientations are mutual inverses:
\begin{align} \label{inversewilson}
[\W(\kpar)^{-1}]_{ij} = [\W(\kpar)]^*_{ji}= \bra{u_{i,(\pi,\kpar)}}\,\hat{W}_r(\kpar)\,\ket{u_{j,(\pi,\kpar)}},
\end{align}
with the gauge choice
\begin{align} \label{reverse}
\ket{u_{j,(\pi,\kpar)}} = V(-2\pi \vec{y})\,\ket{u_{j,(-\pi,\kpar)}} \ins{for} j=1,\ldots, \noc.
\end{align}
The second equality in Eq.\ (\ref{inversewilson}) follows from
\begin{align}
\W^*_{ji} \eq \bra{u_{j,-\pi}}\,V(2\pi\vec{y})\prod_{k_y}^{\pi \leftarrow -\pi} P(k_y) \,\ket{u_{i,-\pi}}^* \lin
\eq \bra{u_{i,-\pi}}\,\prod_{k_y}^{{-} \pi \leftarrow \pi} P(k_y) \,V(-2\pi \vec{y})\,\ket{u_{j,-\pi}} \lin
\eq \bra{u_{i,\pi}}\,V(-2\pi \vec{y})\,\prod_{k_y}^{{-} \pi \leftarrow \pi} P(k_y) \,\ket{u_{j,\pi}} \lin
\eq \bra{u_{i,\pi}}\,\hat{W}_r \,\ket{u_{j,\pi}},
\end{align}
where we have dropped the constant argument $\kpar$ for notational simplicity.


\subsection{Effect of spatial symmetries of the 010 surface} \label{sec:010}

Let us describe the effect of symmetry on the spectrum of $\W$. First, we consider a generic spatial symmetry $\pgdel$, which transforms real-space coordinates as $ \br \rightarrow D_g \br + \bdelta$. From Eq.\ (\ref{symmonhk}), we obtain the constraints on the projections as
\begin{align} \label{gdelconstrainsP}
\hatgdel(\bk)\,P(\boldsymbol{k})\,\hatgdel(\bk)^{\mo} = P\big(\,D_g\boldsymbol{k}\,\big).
\end{align}
The constraints on $\W$ arise only from a subset of the symmetries that either (i) map one loop parametrized by $\kpar$ to another loop at a different $\kpar$, or (ii) map a loop to itself at a high-symmetry $\kpar$. We say that two loops are mapped to each other even if the mapping reverses the loop orientation (i.e., $\W \rightarrow \W^{\sma{-1}}$; cf. Eq.\ (\ref{reverseorient}) and (\ref{inversewilson})), or translates the base point of the loop. For illustration, we consider spatial symmetries of the 010 surface which necessarily preserve the spatial $y$-coordinate; for these symmetries we add an additional subscript to $\pgdel \equiv \pgdelpar$, $D_g \equiv \Dpar$ and $\bdelta \equiv \bdeltapar$, such that $\Dpar \vec{y} = \vec{y}$ and $\bdeltapar \cdot \vec{y}=0$. We now demonstrate that the Wilson loop is constrained as
\begin{align} \label{spacewilson}
\brevegdelpar({-}\pi,\kpar)\, \W(\kpar)\,\brevegdelpar({-}\pi,\kpar)^{\mo} = \W\big(\,\Dpar \kpar\,\big),
\end{align}
where the argument of $\brevegdelpar$ is the base point of $\W(\kpar)$.\\

\begin{widetext}

\noindent \emph{Proof:} we note that the \low representation of $\pgdel$ only depends on momentum through a multiplicative phase factor: exp$(-i(D_g\bk)\cdot \bdelta)$, and for $\pgdelpar$ that this same phase factor is independent of $k_y$, hence the \low representation of $\pgdel$ may be written $\hatgdelpar(\bk) \equiv \hatgdelpar(\kpar)$. Eq.\ (\ref{gdelconstrainsP}) then particularizes to 
\begin{align} \label{proj010}
\hatgdelpar(\kpar)\,P(k_y,\kpar)\,\hatgdelpar(\kpar)^{-1} = P\big(\,k_y,\Dpar\kpar\,\big),
\end{align}
and Eq.\ (\ref{uv}) to
\begin{align} \label{uvsimplie}
\hatgdelpar(\kpar)\, V(2\pi \vec{y}) = e^{-i2\pi \vec{y}\cdot \bdeltapar}\,V(2\pi\vec{y}) \,\hatgdelpar(\kpar) = V(2\pi\vec{y}) \,\hatgdelpar(\kpar).
\end{align}
Applying Eq.\ (\ref{pt}), (\ref{proj010}) and (\ref{uvsimplie}), 
\begin{align} 
\hatgdelpar(\kpar)\,\hat{W}(\kpar)\,\hatgdelpar(\kpar)^{-1} = \hat{W}\big(\,\Dpar \kpar\,\big).
\end{align}
Into this equation, we then insert complete sets of states $(\hat{I}(\bk))$:
\begin{align} \label{whatever}
\hat{I}(-\pi,\Dpar \kpar)\,\hatgdelpar(\kpar)\,\hat{I}(-\pi,\kpar)\,\hat{W}(\kpar)\,\hat{I}(-\pi,\kpar)\,\hatgdelpar(\kpar)^{-1}\hat{I}(-\pi,\Dpar \kpar) = \hat{I}(-\pi,\Dpar \kpar) \,\hat{W}\big(\,\Dpar \kpar\,\big)\,\hat{I}(-\pi,\Dpar \kpar).
\end{align}
where $\hat{I}(\bk)$ is resolved by the energy eigenbasis at $\bk$: 
\begin{align}
\hat{I}(\bk) = \sum_{n =1}^{n_{\sma{tot}}} \ket{u_{n,\bk}}\bra{u_{n,\bk}}.
\end{align}
Since all symmetry representations are block-diagonal with respect to occupied and empty subspaces, Eq.\ (\ref{whatever}) is equivalent to 
\begin{align} \notag
P(-\pi,\Dpar \kpar)\,\hatgdelpar(\kpar)\,P(-\pi,\kpar)\,\hat{W}(\kpar)\,P(-\pi,\kpar)\,\hatgdelpar(\kpar)^{-1}P(-\pi,\Dpar \kpar)
= P(-\pi,\Dpar \kpar) \,\hat{W}\big(\,\Dpar \kpar\,\big)\,P(-\pi,\Dpar \kpar).
\end{align}
Finally, we apply Eq.\ (\ref{goccupied}) and (\ref{con3}) to obtain Eq.\ (\ref{spacewilson}), as desired. $\blacksquare$\\

\end{widetext}

Let us exemplify this discussion with the glide reflection which transforms spatial coordinates as $(x,y,z) \rightarrow (-x,y,z+1/2)$. This symmetry acts on Bloch waves as $\hat{\bar{M}}_x(\bk)=e^{-ik_z/2}\ubmx$ (from Eq.\ (\ref{actbloch})), with $\ubmx^2=-I$ representing a $2\pi$ rotation. Eq.\ (\ref{spacewilson}) assumes the form
\begin{align} \label{bmxw}
\brevebmx({-}\pi,\kpar)\, \W(k_x,k_z)\,\brevebmx({-}\pi,\kpar)^{\mo} = \W\big({-}k_x,k_z\,\big).
\end{align}
Since $\bmx$ transforms momentum as $\bk \rightarrow (k_y,-k_x,k_z)$, it belongs in the little group of any wavevector with $k_x=0$. Indeed, $[\brevebmx({-}\pi,0,k_z),\W(0,k_z)]=0$, and each Wilson band may be labelled by an eigenvalue of $\brevebmx$, which again falls into either branch of ${\pm} i \,\text{exp}({-}ik_z/2)$, as we now show: 
\begin{align}
& [\brevebmx(\bk_1)]^2_{mn} \lin
\eq e^{-ik_z}\,\sum_{a=1}^{\noc} \bra{u_{m,\bk_1}} \, \ubmx \,\ket{u_{a,\bk_1}}\bra{u_{a,\bk_1}} \, \ubmx \,\ket{u_{n,\bk_1}} \lin
\eq e^{-ik_z}\,\sum_{a=1}^{n_{tot}} \bra{u_{m,\bk_1}} \, \ubmx \,\ket{u_{a,\bk_1}}\bra{u_{a,\bk_1}} \, \ubmx \,\ket{u_{n,\bk_1}} \lin
\eq  e^{-ik_z}\,\bra{u_{m,\bk_1}} \, (\ubmx)^2  \,\ket{u_{n,\bk_1}} =-e^{-ik_z}\,\delta_{mn}.
\end{align}
Here, $\bk_1 \equiv (-\pi,0,k_z)$; in the second equality, we denote $n_{tot}$ as the total number of bands, and applied that the symmetry representations are block-diagonal with respect to the occupied and empty subspaces (i.e., $\langle u_{a,\bk_1} | \,\ubmx\,| u_{n,\bk_1} \rangle =0$ if the bra (ket) state is occupied (empty)); the completeness relation was used in the third, and $(\ubmx)^2=-I$ represents a $2\pi$ rotation.\\

\subsection{Effect of space-time symmetries} \label{sec:spacetime}

Suppose our Hamiltonian is symmetric under a space-time transformation $T\pgdel$, where $\pgdel$ is any of the following: (a) a symmorphic spatial transformation (i.e., the fractional translation $\bdelta=0$) that is not necessarily a symmetry of the 010 surface, and (b) a nonsymmorphic transformation that is a symmetry of the 010 surface. In the group we study (space group $D_{6h}^4$ with time-reversal symmetry), if one considers the subset of space-time symmetries which map the Wilson loop to itself (recall what `to itself' means in App.\ \ref{sec:010}), elements of this subset are either of (a)- or (b)-type. \\

Since the \low representation (recall $\hatTg(\bk)$ in Eq.\ (\ref{Tgdrep})) of $T\pgdel$ depends on momentum only through the phase factor exp$[\,{i (D_g \boldsymbol{k}) \cdot \bdelta }\,]$, we deduce that $\hatTg(\bk)$ is independent of $k_y$. In case (a), this follows trivially from $\bdelta=0$, while in case (b) we apply that $D_g\vec{y}=\vec{y}$ and $\bdelta \cdot \vec{y}=0$. Consequently, we will write $\hatTg(\bk) \equiv \hatTg(\kpar)$ with possible $\kpar$-dependence through a phase factor. \\

Following Eq.\ (\ref{sptiHam}), the occupied-band projection is constrained as
\begin{align}
\hatTg(\kpar)\,P(\bk)\,\hatTg(\kpar)^{-1}= P\big({{-} }D_g\bk\,\big).
\end{align}
This, in combination with  Eq.\ (\ref{UU}) and (\ref{pt}), implies 
\begin{align} \label{wilsonsptime}
&\hatTg(\kpar)\,\hat{W}(\kpar)\,\hatTg(\kpar)^{\mo} \lin
\eq \hatTg(\kpar)\,V(2\pi \vec{y})\,\prod_{k_y}^{\pi \leftarrow -\pi} P(\bk)\,\hatTg(\kpar)^{\mo} \lin
\eq \hatTg(\kpar)\,V(2\pi \vec{y})\,\hatTg(\kpar)^{\mo}\,\prod_{k_y}^{\pi \leftarrow -\pi} P(-D_g\bk) \lin
\eq e^{iD_g(2\pi \vec{y})\cdot \bdelta}\,V\big({{-} }D_g(2\pi \vec{y})\,\big) \prod_{k_y}^{\pi \leftarrow {-} \pi} P\big({{-} }D_g\bk\,\big).
\end{align}
In the next few subsections, we particularize to a few examples of $T\pgdel$ that are relevant to the topology of our space group.

\subsubsection{Effect of space-time inversion symmetry} \label{app:spacetimeinversion}

The  space-time inversion symmetry ($T\cali$) maps $(x,y,z,t) \rightarrow -(x,y,z,t)$. Let us show how this results in the eigenvalues of $\W(\kpar)$ forming complex-conjugate pairs at each $\kpar$. If we interpret the phase ($\theta$) of each eigenvalue as an `energy', then the spectrum has a $\theta \rightarrow -\theta$ symmetry; this may be likened to a particle-hole symmetry that unconventionally preserves the momentum coordinate ($\kpar$).\\

\noindent \emph{Proof:} Inserting $D_g=-I$ and $\bdelta=0$ in  Eq.\ (\ref{wilsonsptime}),
\begin{align} \label{uitW}
\hat{T}_{\sma{\cali}}(\kpar)\,\hat{W}(\kpar)\,\hat{T}_{\sma{\cali}}(\kpar)^{\mo} = \hat{W}(\kpar).
\end{align}
Then by inserting in Eq.\ (\ref{uitW}) a complete set of states at momentum $\bk = ({-} \pi,\kpar)$, and applying the definitions (\ref{con3}) and (\ref{sew}), 
\begin{align}
\breveti(-\pi,\kpar)\,\W(\kpar)\,\breveti(-\pi,\kpar)^{\mo} = \W(\kpar).
\end{align}
Thus if an eigensolution exists with eigenvalue exp$[i\theta(\kpar)]$,
there exists a partner solution with the complex-conjugate eigenvalue exp$[{-} i\theta(\kpar)]$. These two solutions are always mutually orthogonal, even in cases where the eigenvalues are real. The orthogonality follows from $\breveti^2=-I$, as we now show:
\begin{align}
\big[\breveti(\bk)^2]_{mn} \eq \sum_{a=1}^{\noc} \bra{u_{m,\bk}} \, \hat{T}_{\sma{\cali}}(\bk) \,\ket{u_{a,\bk}}\bra{u_{a,\bk}} \, \hat{T}_{\sma{\cali}}(\bk) \,\ket{u_{n,\bk}} \lin
\eq \sum_{a=1}^{n_{tot}} \bra{u_{m,\bk}} \, \hat{T}_{\sma{\cali}}(\bk) \,\ket{u_{a,\bk}}\bra{u_{a,\bk}} \, \hat{T}_{\sma{\cali}}(\bk) \,\ket{u_{n,\bk}} \lin
\eq \bra{u_{m,\bk}} \, \big[\,\hat{T}_{\sma{\cali}}(\bk)\,\big]^2  \,\ket{u_{n,\bk}} \lin
\eq \bra{u_{m,\bk}} \, \uit\,\uit^*  \,\ket{u_{n,\bk}} \lin
\eq -\braket{u_{m,\bk}}{u_{n,\bk}} = -\delta_{mn}.
\end{align}
In the second equality, we denote $n_{tot}$ as the total number of bands, and applied that the symmetry representations are block-diagonal with respect to the occupied and empty subspaces; the completeness relation was used in the third equality, and $\uit\,\uit^*=-I$ follows because $(T\cali)^2$ is a $2\pi$ rotation.

\subsubsection{Effect of time reversal with a spatial glide-reflection} \label{sec:tmx}



The symmetry $T\bmx$ maps space-time coordinates as $(x,y,z,t) \rightarrow (-x,y,z+c/2,-t)$, i.e.,
\begin{align} \label{Tbmxspace}
D_{\sma{M_x}}=\text{diag}[-1,1,1] \ins{and} \bdelta=c\vec{z}/2.
\end{align}
The momentum coordinates are mapped as 
\begin{align} \notag
(k_y,k_x,k_z)^t \rightarrow -D_{\sma{M_x}}(k_y,k_x,k_z)^t=(-k_y,k_x,-k_z)^t,
\end{align}
which implies that a Wilson loop at fixed $(k_x,k_z)$ is mapped to a loop at $(k_x,-k_z)$, with a reversal in orientation since $k_y \rightarrow -k_y$. In more detail, we insert Eq.\ (\ref{Tbmxspace}) into  Eq.\ (\ref{wilsonsptime}),
\begin{align} \notag
\hat{T}_{\sma{\bmx}}(k_x,{k}_z)\,\hat{W}(k_x,{k}_z)\,\hat{T}_{\sma{\bmx}}(k_x,{k}_z)^{\mo} = \hat{W}_r(k_x,-{k}_z),
\end{align}
where the reversed Wilson-loop operator ($\hat{W}_r$) is defined in Eq.\ (\ref{reverseorient}). An equivalent expression in the occupied-band basis is 
\begin{align} \label{yperiod}
&\brevetbmx({-} \pi,k_x,{k}_z)\,\W(k_x,{k}_z)\,\brevetbmx({-} \pi,k_x,{k}_z)^{\mo}\lin
\eq {\W(k_x,-{k}_z)}^{\mo},
\end{align}
where the inverse Wilson loop is defined in Eq.\ (\ref{inversewilson}); it is worth clarifying that $\brevetbmx$ particularizes Eq.\ (\ref{shorthand2}) as
\begin{align} \label{gaugeiny}
&\brevetbmx({-} \pi,k_x,{k}_z)_{mn} \lin
\eq \bra{u_{m,(\pi,k_x,-k_z)}}\,\hattbmx(k_x,k_z)\,\ket{u_{n,(-\pi,k_x,k_z)}},\ins{with} \lin
& \ket{u_{m,(\pi,k_x,-k_z)}} = V(-2\pi\vec{y})\, \ket{u_{m,(-\pi,k_x,-k_z)}}.
\end{align}
We now focus on $k_z=\bar{k}_z$ satisfying $\bar{k}_z=-\bar{k}_z$ modulo $2\pi$, such that Eq.\ (\ref{yperiod}) particularizes to 
\begin{align} \label{yuuu}
&\brevetbmx({-} \pi,k_x,\bar{k}_z)\,\W(k_x,\bar{k}_z)\,\brevetbmx({-} \pi,k_x,\bar{k}_z)^{\mo}\lin
\eq {\W(k_x,-\bar{k}_z)}^{-1} ={\W(k_x,\bar{k}_z)}^{-1},
\end{align}
in the gauge 
\begin{align} \label{gaugeinz}
\ket{u_{j,(\pi,k_x,-\bar{k}_z)}} = V(2\bar{k}_z \vec{z})\,\ket{u_{j,(\pi,k_x,\bar{k}_z)} },
\end{align}
for $j \in \{1,2,\ldots, \noc\}$, and $2\bar{k}_z\vec{z}$ a reciprocal vector (possibly zero). Eq.\ (\ref{yuuu}) shows that the symmetry maps the Wilson loop to itself, with a reversal of orientation.\\

\begin{widetext}

Let us prove that
\begin{align}
\brevetbmx({-} \pi,k_x,\bar{k}_z)^2 = \begin{cases} +I, & \bar{k}_z =0 \\ -I, & \bar{k}_z = \pi, \end{cases}
\end{align}
from which we may deduce a Kramers-like degeneracy in the spectrum of $\W(k_x,k_z=\pi)$ but not in $\W(k_x,k_z=0)$. Employing the shorthand
\begin{align}
\bkone = (-\pi,k_x,\bar{k}_z), \myspace \bktwo^t = -D_{M_x}\bkone^t =(\pi,k_x,-\bar{k}_z)^t,
\end{align} 
and the gauge conditions assumed in Eq.\ (\ref{gaugeiny}) and (\ref{gaugeinz}),
\begin{align}
\big[\, \brevetbmx(\bkone)^2\,\big]_{mn} \eq \sum_{a=1}^{\noc} \bra{u_{m,\bktwo}} \, \hat{T}_{\sma{\bmx}}(\bkone) \,\ket{u_{a,\bkone}}\bra{u_{a,\bktwo}} \, \hat{T}_{\sma{\bmx}}(\bkone) \,\ket{u_{n,\bkone}}  \lin
 \eq \sum_{a=1}^{\noc} \bra{u_{m,\bkone}} \, V(2\pi \vec{y} {-}2\bar{k}_z \vec{z}) \,\hat{T}_{\sma{\bmx}}(\bkone) \,\ket{u_{a,\bkone}}\bra{u_{a,\bkone}} \,V(2\pi \vec{y} {-}2\bar{k}_z \vec{z}) \, \hat{T}_{\sma{\bmx}}(\bkone) \,\ket{u_{n,\bkone}} \lin
\eq \sum_{a=1}^{n_{tot}} \bra{u_{m,\bkone}} \, V(\,2\pi\vec{y}-2\bar{k}_z \vec{z}\,)\,\hat{T}_{\sma{\bmx}}(\bkone) \,\ket{u_{a,\bkone}}\bra{u_{a,\bkone}}\,V(\,2\pi\vec{y}-2\bar{k}_z \vec{z}\,) \, \hat{T}_{\sma{\bmx}}(\bkone) \,\ket{u_{n,\bkone}} \lin
\eq \bra{u_{m,\bkone}} \, V(\,2\pi\vec{y}-2\bar{k}_z \vec{z}\,)\,\hat{T}_{\sma{\bmx}}(\bkone) \,V(\,2\pi\vec{y}-2\bar{k}_z \vec{z}\,) \, \hat{T}_{\sma{\bmx}}(\bkone) \,\ket{u_{n,\bkone}} \lin
\eq e^{-i\bar{k}_z} \bra{u_{m,\bkone}} \, V(\,2\pi\vec{y}-2\bar{k}_z \vec{z}\,)\,V(-2\pi\vec{y}+2\bar{k}_z \vec{z}\,)\,\hat{T}_{\sma{\bmx}}(\bkone) \, \hat{T}_{\sma{\bmx}}(\bkone) \,\ket{u_{n,\bkone}} \lin
\eq e^{-i\bar{k}_z} \bra{u_{m,\bkone}} \, U_{\sma{T\bmx}} \,  U_{\sma{T\bmx}}^* \,\ket{u_{n,\bkone}} =e^{-i\bar{k}_z} \braket{u_{m,\bkone}}{u_{n,\bkone}} = e^{-i\bar{k}_z}\,\delta_{mn}.
\end{align}
In the second equality, we applied that the symmetry representations are block-diagonal with respect to the occupied and empty subspaces; the completeness relation was used in the third equality, Eq.\ (\ref{UU}) in the fourth equality, and  $U_{\sma{T\bmx}} \,  U_{\sma{T\bmx}}^*=+I$ represents the point-group relation that $(TM_x)^2$ is just the identity transformation; cf. our discussion in App.\ \ref{app:rep}.\\

\end{widetext}

\subsubsection{Effect of time-reversal symmetry}

Let us particularize the discussion in Sec.\ \ref{sec:spacetimetight} by letting $\pgdel$ in $T\pgdel$ be the trivial transformation. In the \low representation, $\hat{T}=U_TK$, where $U_TU_T^*=-I$ corresponds to a $2\pi$ rotation of a half-integer spin. We obtain from Eq.\ (\ref{wilsonsptime}) that
\begin{align} \label{gaugeinv}
&\hat{T} \,\hat{W}(\kpar)\,\hat{T}^{\mo} = V\big({{-} }2\pi \vec{y}\,\big) \prod_{k_y}^{\pi \leftarrow {-} \pi} P\big({{-} }\bk\,\big) \lin
\eq V\big({{-} }2\pi \vec{y}\,\big) \prod_{k_y}^{{-} \pi \leftarrow \pi} P\big(\,k_y,{-} \kpar\,\big) = \hat{W}_r({-} \kpar),
\end{align}
where in the last equality we identify the reverse-oriented Wilson loop defined in Eq.\ (\ref{reverseorient}). Equivalently, in the occupied-band basis,
\begin{align} \label{gaugedep}
\breve{T}({-} \pi,\kpar)\,\W(\kpar)\,\breve{T}({-} \pi,\kpar)^{\mo}={\W({-} \kpar)}^{\mo},
\end{align}
with the inverse Wilson loop defined in Eq.\ (\ref{inversewilson}). Time reversal thus maps exp$[{i\theta_{\kpar}}] \rightarrow \text{exp}[{i\theta_{-\kpar}}]$. Following an exercise similar to the previous section (Sec.\ \ref{sec:tmx}), one may derive a Kramers degeneracy where $\kpar=-\kpar$ (up to a reciprocal vector).

\subsection{Extended group algebra of the W-symmetries along $\tilx \tilu$} \label{app:kxpi}

Sec.\ \ref{sec:wilsoniansymm} introduced the notion of W-symmetries and should be read in advance of this Section. Our aim is to derive the algebra of the group ($\gkpar$) of $\W(\pi,k_z)$, which we introduced in Sec.\ \ref{sec:wilsoniansymm}. $k_z=0$ and $\pi$ mark the time-reversal invariant $\kpar$ (namely, $\tilx$ and $\tilu$). Here, $\gx \equiv G_{\sma{\pi,0}} \cong G_{\sma{\tilde{U}}} \equiv G_{\sma{\pi,\pi}}$ has the elements: $2\pi$ rotation ($\bar{E}$), the lattice translation $t(\vec{z})$, the Wilson loop ($\W$), and analogs of time reversal ($\calt$), spatial inversion ($\cali$) and glide reflection ($\calbmx$) that additionally encode parallel transport; the latter three are referred to as W-symmetries. In addition to deriving the algebraic relations in Eq.\ (\ref{Mxwilson}) and (\ref{Twilson}), we also show here that: \\

\noi{a} The combination of time reversal, spatial glide and parallel transport is an element $\caltbmz$ with the algebra:
\begin{align} \label{TMzwilson}
\caltbmz\;\W\;\caltbmz^{-1} = \W^{\mo},& \ins{with} \caltbmz^2 = I  \lin
\ins{and} \calbmx\;\caltbmz \eq  \bar{E}\;t(\vec{z})\; \caltbmz \;\calbmx\;\W.
\end{align}

\noi{b} The space-time inversion symmetry acts in the ordinary manner: 
\begin{align} \label{TIwilson}
&\calti \;\W\;\calti^{\mo} = \W, \ins{with}\lin
&\calti^2= \bar{E},\;\;\;\;\calbmx \,\calti = t(\vec{z})\, \calti \,\calbmx.
\end{align}

\noindent The algebra that we derive here extends the ordinary algebra of space-time transformatons, which we showed in Eq.\ (\ref{ordinaryalgebra}). For time-reversal-variant $\kpar$ along the same glide line, $\gkpar$ ($k_z \notin \{0,\pi\}$) is a subgroup of $\gx \equiv G_{\sma{\pi,0}}$, and is instead generated by $\bar{E}, t(\bR_{\shortparallel}), \W, \calbmx, \caltbmz$ and $\calti$. Therefore, Eq.\ (\ref{Mxwilson}), (\ref{TMzwilson}) and (\ref{TIwilson}) (but not Eq.\ (\ref{Twilson})) would apply to $\gkpar$ ($k_z \notin \{0,\pi\}$). Eq.\ (\ref{Mxwilson}), (\ref{TMzwilson}), (\ref{TIwilson}) and (\ref{Twilson}) are respectively derived in App.\ \ref{app:wilsonreflection}, \ref{app:TMzwilson}, \ref{app:TIwilson} and \ref{app:wilsontime}. \\ 

One motivation for deriving the Wilsonian algebra is that it determines the possible topologies of the Wilson bands along $\tilx \tilu$. This determination is through `rules of the curves' that we summarize in two tables: Tab.\ \ref{rulesgeneric} is derived from the algebra of $\gkpar$ and applies to any $\kpar$ along $\tilx \tilu$; Tab.\ \ref{rulestrim} is derived from $\gx$ and applies only to $\tilx$ and $\tilu$.

\subsubsection{Wilsonian glide-reflection symmetry} \label{app:wilsonreflection}

Consider the glide reflection $\bmx$, which transforms spatial coordinates as $(x,y,z)\rightarrow (-x,y,z+1/2)$. In App.\ \ref{sec:010}, we have described how $\bmx$ constrains Wilson loops in the $k_x=0$ plane, where $\bmx$ belongs in the little group of each wavevector, and therefore the projections on this plane separates into two mirror representations. This is no longer true for the $k_x=\pi$ plane, since $\bmx$ maps between two momenta which are separated by half a reciprocal period, i.e., $\bmx: \; (k_y,\pi,k_z) \longrightarrow (k_y,-\pi,k_z) = (k_y+\pi,\pi,k_z)-\tilde{\boldsymbol{b}}_2$, as illustrated in Fig.\ \ref{fig:wherewilson}(b). Consequently,  the Wilson-loop operator is symmetric under a combination of a glide reflection with parallel transport over half a reciprocal period, as we now prove.\\

\begin{figure}[H]
\centering
\includegraphics[width=5.2 cm]{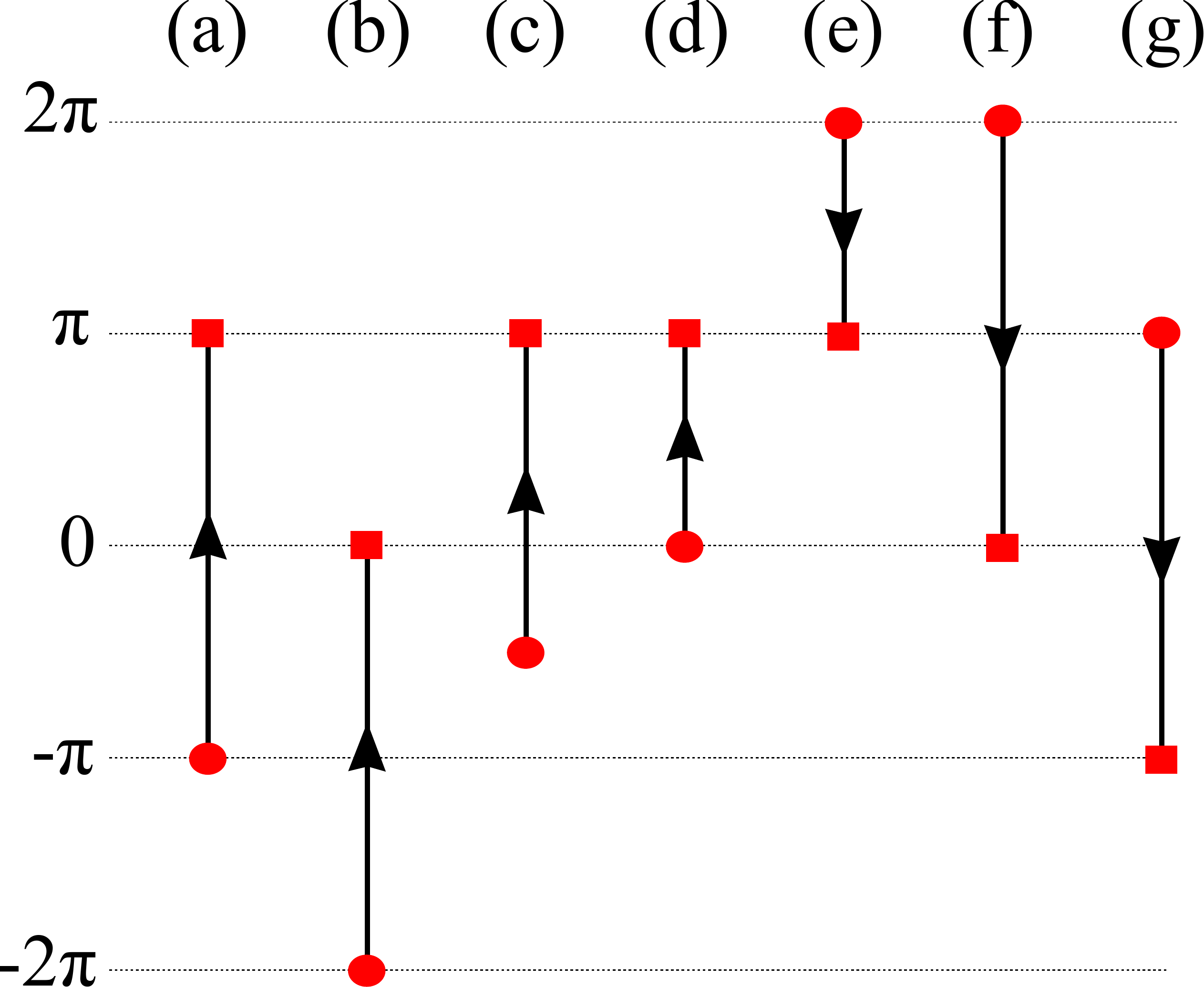}
    \caption{To clarify our notations for Wilson loops and lines, we draw several examples, all of which occur at fixed $\kpar$ and variable $k_y$ (vertical axis). The arrow indicates the orientation for parallel transport. (a) The Wilson loop at base point $-\pi$ (encircled) is labelled by $\W_{-\pi} \equiv \W_{\pi \leftarrow -\pi}$. (b) $\W_{-2\pi} \equiv \W_{0 \leftarrow -2\pi}$. (c) $\W_{\pi \leftarrow -\pi/2}$. (d) $\W_{\pi \leftarrow 0}$. (e) $\W_{\pi \leftarrow 2\pi}$. $\W_{r,\bar{k}_y}$ denotes a Wilson loop with base point $\bar{k}_y$ and oriented in the direction of decreasing $k_y$, e.g.,  $\W_{r,2\pi}$ in (f) and $\W_{r,\pi}$ in (g).} \label{fig:wilsonlines}
\end{figure}

\noindent {\textbf{Proof}} A crucial observation is that the glide symmetry relates projections at $(k_y,\pi,k_z)$ and $(k_y+\pi,\pi,k_z)$. That is, by particularizing Eq.\ (\ref{symmonhk}) and (\ref{uuk}), we obtain the symmetry constraint on the projections:
\begin{align} \label{glideconstr}
V(\tilde{\boldsymbol{b}}_2)^{\mo}\,\ubmx\,P(k_y,\pi,k_z)\,\ubmx^{\mo}\,V(\tilde{\boldsymbol{b}}_2) = P(k_y+\pi,\pi,k_z).
\end{align}
This relation allows us to define a unitary `sewing matrix' between states at $(0,\kpar)$ and $(-\pi,\kpar)$: 
\begin{align} \label{brevesew}
 &[\brevebmx((0,\kpar),(-\pi,\kpar))]_{mn} \lin
\eq e^{-ik_z/2}\,\bra{u_{m,(0,\kpar)}}\,V({-}\tilde{\boldsymbol{b}}_2)\,\ubmx\,\ket{u_{n,(-\pi,\kpar)}},
\end{align}
which particularizes Eq.\ (\ref{goccupied}) with $\bG=\tilde{\boldsymbol{b}}_2$. \\

Presently, it becomes useful to distinguish the base point of a Wilson loop, and also to define Wilson lines which do not close into a loop. The remaining discussion in this Section occurs at fixed $\kpar=(\pi,k_z )$ (for any $k_z \in [-\pi,\pi)$) and variable $k_y$; subsequently, we will suppress the label $\kpar$, e.g.,
\begin{align} \label{shorthand44}
&\ket{u_{m,k_y,\kpar}} \equiv \ket{u_{m,k_y}};\lin
&\brevebmx((0,\kpar),(-\pi,\kpar)) \equiv \brevebmx(0,-\pi).
\end{align}
We denote the base point ($\bar{k}_y$) of a Wilson loop by the subscript in $\W_{\sma{\bar{k}_y}}$ and choose an orientation of increasing $k_y$, i.e., we would parallel transport from $\bar{k}_y\rightarrow \bar{k}_y+2\pi$, as exemplified by Fig.\ \ref{fig:wilsonlines}(a-b). We denote a Wilson line between two distinct momenta by 
\begin{align}
[\W_{\sma{k_{2}\leftarrow k_{1}}}]_{mn} = \bra{u_{m,k_2}}\,\prod_{k_y}^{k_2 \leftarrow k_1} P(k_y)\,\ket{u_{n,k_1}} ,
\end{align}
where $\prod_{k_y}^{k_2 \leftarrow k_1} P(k_y)$ denotes a path-ordered product of projections sandwiched by $P(k_1)$ (rightmost) and $P(k_2)$ (leftmost); while $k_1$ and $k_2$ are more generally mod $2\pi$ variables due to the periodicity of momentum space, we always choose a branch that includes both $k_1$ and $k_2$ in our definition of $\W_{\sma{k_{2}\leftarrow k_{1}}}$, such that if $k_2>k_1$ we parallel transport in the direction of increasing $k_y$ (e.g., Fig.\ \ref{fig:wilsonlines}(c-d)), and vice versa (e.g., Fig.\ \ref{fig:wilsonlines}(e)). Therefore, this Wilson line reverts to the familiar Wilson loop if $k_2-k_1=2\pi$, i.e., $\W_{\pi \leftarrow -\pi} = \W_{-\pi}$ of Eq.\ (\ref{con3}) with the gauge condition of Eq.\ (\ref{reverse}). Since parallel transport is unitary within the occupied subspace,\cite{AA2014} 
\begin{align} \label{inversew}
\dg{\W_{\sma{k_{2}\leftarrow k_{1}}}} = \W_{\sma{k_{1}\leftarrow k_{2}}}  \imp \W_{\sma{k_{2}\leftarrow k_{1}}}\W_{\sma{k_{1}\leftarrow k_{2}}}=I.
\end{align}
With these definitions in hand, we return to the proof.\\ 

\begin{widetext}

To proceed, we first show that the glide symmetry translates the Wilson loop by half a reciprocal period in $\vec{y}$:
\begin{align} \label{wul}
&\brevebmx(0,-\pi) \, \W_{\sma{{-}\pi}}\,\brevebmx(0,-\pi)^{-1} = \W_{\sma{0}}.
\end{align}
This is proven by applying Eq.\ (\ref{glideconstr}) and (\ref{uv}) to
\begin{align}
&V(\tilde{\boldsymbol{b}}_2)^{\mo}\,\ubmx\,\hat{W}\,\ubmx^{\mo}\,V(\tilde{\boldsymbol{b}}_2) = V(\tilde{\boldsymbol{b}}_2)^{\mo}\,\ubmx\,V(2\pi\vec{y})\prod_{k_y}^{\pi \leftarrow -\pi}P(k_y)\,\ubmx^{\mo}\,V(\tilde{\boldsymbol{b}}_2) \lin
\eq V(\tilde{\boldsymbol{b}}_2)^{\mo}\,\ubmx\,V(2\pi\vec{y})\,\ubmx^{\mo}\,V(\tilde{\boldsymbol{b}}_2)\,\prod_{k_y}^{\pi \leftarrow -\pi}P(k_y+\pi) = V(2\pi\vec{y})\,\prod_{k_y}^{2\pi \leftarrow 0}P(k_y), 
\end{align}
which then implies Eq.\ (\ref{wul}) in the occupied-band basis; cf.\ Eq.\ (\ref{shorthand44}) and (\ref{brevesew}). 
\begin{figure}[H]
\centering
\includegraphics[width=7 cm]{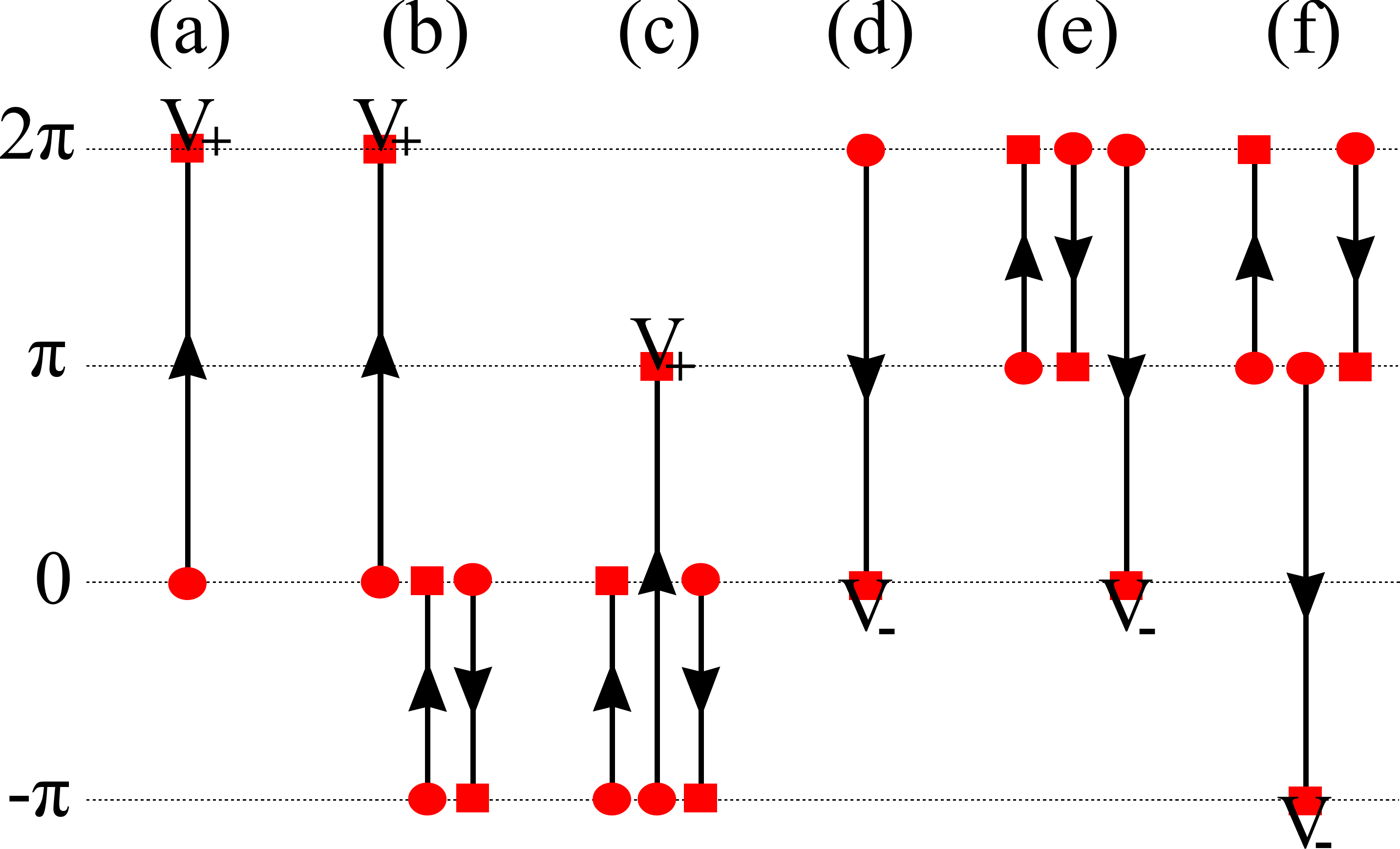}
    \caption{(a-c) Pictorial representation of Eq.\ (\ref{W0}), with solid line indicating a product of projections that is path-ordered according to the arrow on the same line, and $V_+$ indicating an insertion of the spatial-embedding matrix $V(2\pi\vec{y})$. (d-f) represent Eq.\ (\ref{wilsonreversegauge}), with $V_+$ indicating an insertion of $V(-2\pi\vec{y})$.} \label{fig:wilsonidentity}
\end{figure}
We now apply the following identity, 
\begin{align} \label{W0}
[\W_{\sma{0}}]_{mn} \eq \bra{u_{m,0}}\,V(2\pi \vec{y})\, \prod_{k_y}^{2\pi \leftarrow 0} P(k_y)\,\ket{u_{n,0}} = \bra{u_{m,0}}\,V(2\pi \vec{y})\, \prod_{k_y}^{2\pi \leftarrow 0}\,\prod_{r_y}^{0 \leftarrow -\pi} P(r_y)\,\prod_{q_y}^{-\pi \leftarrow 0} P(q_y)\,\ket{u_{n,0}} \lin
\eq  \bra{u_{m,0}}\,\prod_{k_y}^{0 \leftarrow -\pi} P(k_y)\,V(2\pi \vec{y})\,\prod_{r_y}^{\pi \leftarrow -\pi} P(r_y) \,\prod_{q_y}^{-\pi \leftarrow 0} P(q_y)\,\ket{u_{n,0}} = [\W_{\sma{0 \leftarrow -\pi}}\,\W_{\sma{-\pi}}\,\W_{\sma{-\pi \leftarrow 0}}]_{mn},
\end{align}
which is pictorially represented in Fig.\ \ref{fig:wilsonidentity}: (a) represents $\W_{\sma{0}}$, (b) an intermediate step in Eq.\ (\ref{W0}), and (c) the final result. The second equality in Eq.\ (\ref{W0}) follows from Eq.\ (\ref{inversew}) and the third from Eq.\ (\ref{periodP}). Combining Eq.\ (\ref{wul}) and (\ref{W0}), 
\begin{align} \label{commute}
[\calbmx,\; \W_{\sma{{-}\pi}}]=0, \ins{with} \calbmx = \W_{\sma{-\pi \leftarrow 0}}\,\brevebmx(0,-\pi).
\end{align}
To interpret this result, the Wilson-loop operator commutes with a combination of glide reflection (encoded in $\brevebmx$) with parallel transport over half a reciprocal period (encoded in $\W_{\sma{-\pi \leftarrow 0}}$); we call any such `symmetry' that combines a space-time transformation with parallel transport a Wilsonian symmetry, or just a W-symmetry. $\blacksquare$\\

\end{widetext}

It is worth mentioning that $\calbmx$ and $\W_{\sma{{-}\pi}}$ always commute as in Eq.\ (\ref{commute}), though the specific representations of $\calbmx$ and $\W_{\sma{{-}\pi}}$ depend on the gauge of $|u_{\sma{n,-\pi}}\rangle$. That is, in a different gauge labelled by a tilde header:
\begin{align}
\ket{\tilde{u}_{n,-\pi}} \equiv \sum_{m=1}^{\noc} \ket{{u}_{m,-\pi}}\, S_{mn}; \myspace S \in U(\noc),
\end{align}
we would represent the W-reflection and the Wilson loop by
\begin{align}
{\cal \tilde{\bar{M}}}_x =S^{-1}\,\calbmx \,S \ins{and} \tilde{\W}_{\sma{{-}\pi}} =S^{-1}\,\W_{\sma{{-}\pi}}\,S,
\end{align}
which would still commute.\\

Eq.\ (\ref{commute}) implies we can simultaneously diagonalize both Wilson-loop and W-symmetry operators; for each simultaneous eigenstate, the two eigenvalues are related in the following manner: if exp$[{i\theta}]$ is the Wilson-loop eigenvalue, then the W-symmetry eigenvalue falls into either branch of $\lambda_x(k_z{+}\theta){\equiv}{\pm} i \text{exp}[{{-}i(k_z{+}\theta)/2}]$, as claimed in Eq.\ (\ref{relativistic}) and as we proceed to prove.\\

\begin{widetext}

\noindent \textbf{Proof} Up to a $k_z$-dependent phase factor, this W-symmetry operator squares to the reverse-oriented Wilson loop:
\begin{align}
[\;\calbmx^2\;]_{mn} \eq [\;\W_{\sma{-\pi \leftarrow 0}}\,\brevebmx(0,-\pi)\,\W_{\sma{-\pi \leftarrow 0}}\,\brevebmx(0,-\pi)\;]_{mn} \lin
\eq e^{-ik_z}\bra{u_{m,-\pi}}\,\prod_{k_y}^{-\pi \leftarrow 0} P(k_y)\,V(\tilde{\boldsymbol{b}}_2)^{\mo}\,\ubmx\,\prod_{q_y}^{-\pi \leftarrow 0} P(q_y)\,V(\tilde{\boldsymbol{b}}_2)^{\mo}\,\ubmx\,\ket{u_{n,-\pi}}\lin
\eq e^{-ik_z}\bra{u_{m,-\pi}}\,\prod_{k_y}^{-\pi \leftarrow 0} P(k_y)\,\prod_{q_y}^{-\pi \leftarrow 0} P(q_y+\pi)\,V(\tilde{\boldsymbol{b}}_2)^{\mo}\,\ubmx\,V(\tilde{\boldsymbol{b}}_2)^{\mo}\,\ubmx\,\ket{u_{n,-\pi}}\lin
\eq e^{-ik_z}\bra{u_{m,-\pi}}\,\prod_{k_y}^{-\pi \leftarrow \pi} P(k_y)\,V(\tilde{\boldsymbol{b}}_2)^{\mo}\,V(\tilde{\boldsymbol{b}}_2-2\pi \vec{y})\,(\ubmx)^2\,\ket{u_{n,-\pi}}\lin
\eq -e^{-ik_z}\bra{u_{m,-\pi}}\,\prod_{k_y}^{-\pi \leftarrow \pi} P(k_y)\,V(-2\pi \vec{y})\,\ket{u_{n,-\pi}}\lin
\eq -e^{-ik_z}\bra{u_{m,\pi}}\,V(-2\pi\vec{y})\,\prod_{k_y}^{-\pi \leftarrow \pi} P(k_y)\,\ket{u_{n,\pi}}\lin
\eq -e^{-ik_z}\;[\;({\W}_{-\pi})^{-1}\;]_{mn}.
\end{align}
Here, $\ubmx^2=-I$ represents a $2\pi$ rotation, and in the fourth equality, we made use of
\begin{align} \label{uvpart}
\ubmx\, V(-\tilde{\boldsymbol{b}}_2) = \text{exp}[\;{iD_{\sma{\bmx}} \tilde{\boldsymbol{b}}_2 \cdot \tfrac{\vec{z}}{2}}\;]\,V(-D_{\sma{\bmx}}\tilde{\boldsymbol{b}}_2) \,\ubmx = V(\tilde{\boldsymbol{b}}_2-2\pi \vec{y}) \,\ubmx,
\end{align}
which follows from Eq.\ (\ref{uv}). $\blacksquare$\\

\end{widetext}

The situation is analogous to that of the 010-surface bands along both glide lines ($\tilg \tilz$ and $\tilx \tilu$), where on each line there are also two branches for the glide-mirror eigenvalues, namely $\lambda_z(k_z){\equiv} {\pm} i \text{exp}({{-}ik_z/2})$. A curious difference is that the W-symmetry eigenvalue of a Wilson band (only along $\tilx \tilu$) also depends on the energy ($\theta$) through Eq.\ (\ref{relativistic}). \\

We remark on how a W-reflection symmetry more generally arises. While our nonsymmorphic case study is a momentum plane with W-glide symmetry, the nonsymmorphicity is not a prerequisite for W-symmetries. Indeed, certain momentum planes in symmorphic crystals (e.g., rocksalt structures) exhibit a \emph{glideless} W-reflection symmetry. The rocksalt structure and our case study are each characterized by a momentum plane (precisely, a torus within a plane) that is: (a) orthogonal to the reflection axis, and (b) is reflected not directly to itself, but to itself translated by half a reciprocal period ($\bG/2$), with $\bG$ lying parallel to the same plane; in our nonsymmorphic case study, the plane is $k_x{=}\pi$ and $\bG{=}2\pi \vec{y}$.

\subsubsection{Wilsonian $T\bmz$-symmetry} \label{app:TMzwilson}

$T\bmz$ transforms space-time coordinates as $(x,y,z,t) \rightarrow (x,y,-z+1/2,-t)$, and momentum coordinates as $(k_y,\pi,k_z) \longrightarrow (-k_y,-\pi,k_z) = (\pi-k_y,\pi,k_z)-\tilde{\boldsymbol{b}}_2$, as illustrated in Fig.\ \ref{fig:wherewilson}(c). From Eq.\ (\ref{Tgdrep}) and (\ref{sptiHam}), we obtain its action on Bloch waves:
\begin{align}
\hat{T}_{\sma{\bmz}}(\bk) = e^{-ik_z/2}\;U_{\sma{T\bmz}}\;K,
\end{align}
and the constraint on the occupied-band projections:
\begin{align} \label{consttmx}
V(\tilde{\boldsymbol{b}}_2)^{\mo}\, \hat{T}_{\sma{\bmz}}\,P(k_y,\pi,k_z) \,\hat{T}_{\sma{\bmz}}^{\mo}\,V(\tilde{\boldsymbol{b}}_2) = P(\pi-k_y,\pi,{k}_z),
\end{align}
following similar steps that were used to derive Eq.\ (\ref{glideconstr}). This relation allows us to define a unitary `sewing matrix' between states at $(2\pi,\kpar)$ and $(-\pi,\kpar)$: 
\begin{align} \label{brevesew}
 &[\brevetbmz((2\pi,\kpar),(-\pi,\kpar))]_{mn} \lin
\eq \bra{u_{m,(2\pi,\kpar)}}\,V(\tilde{\boldsymbol{b}}_2)^{\mo}\,\hat{T}_{\sma{\bmz}}\,\ket{u_{n,(-\pi,\kpar)}},
\end{align}
which particularizes Eq.\ (\ref{sew}) with $\bG=-\tilde{\boldsymbol{b}}_2$. Henceforth suppressing the $\kpar$ labels, and by similar manipulations that were used to derive Eq.\ (\ref{wul}), we are led to
\begin{align} \label{there}
\brevetbmz(2\pi,{-}\pi) \, \W_{\sma{{-}\pi}}\,\brevetbmz(2\pi,{-}\pi)^{-1} = \W_{r,\sma{2\pi}},
\end{align}
and $\W_{r,\sma{2\pi}}$ denotes the reverse-oriented Wilson loop with base point $2\pi$, as illustrated in Fig.\ \ref{fig:wilsonlines}(f).\\

\begin{widetext}

 We will use the following two identities: as pictorially represented in Fig.\ \ref{fig:wilsonidentity}(d-f), the first identity 
\begin{align} \label{wilsonreversegauge}
\text{(i)}\myspace \W_{\sma{r,2\pi}} = \W_{\sma{2\pi \leftarrow \pi}}\,\W_{\sma{r,\pi}}\,\W_{\sma{\pi \leftarrow 2\pi}},
\end{align} 
follows from a generalization of Eq.\ (\ref{W0}), and
\begin{align} \label{iden2}
&\text{(ii)}\myspace [\W_{\sma{\pi \leftarrow 2\pi}} \,\brevetbmz(2\pi,{-}\pi)]_{mn} \lin
\eq \bra{u_{m,\pi}}\,\prod_{k_y}^{\pi \leftarrow 2\pi}P(k_y)\,V(\tilde{\boldsymbol{b}}_2)^{\mo}\,\hat{T}_{\sma{\bmz}}\,\ket{u_{n,-\pi}} = \bra{u_{m,-\pi}}\,\prod_{k_y}^{-\pi \leftarrow 0}P(k_y)\,V(2\pi \vec{y}-\tilde{\boldsymbol{b}}_2 )\,\hat{T}_{\sma{\bmz}}\,\ket{u_{n,-\pi}} \lin
\eq \sum_{a=1}^{\noc} \bra{u_{m,-\pi}}\,\prod_{k_y}^{-\pi \leftarrow 0}P(k_y)\,\ket{u_{a,0}}\bra{u_{a,0}}\,V(2\pi \vec{y}-\tilde{\boldsymbol{b}}_2 )\,\hat{T}_{\sma{\bmz}}\,\ket{u_{n,-\pi}} \equiv \sum_{a=1}^{\noc}[\W_{\sma{-\pi \leftarrow 0}}]_{ma} \, [\brevetbmz(0,{-}\pi)]_{an}.
\end{align}
Inserting (i) and (ii) into Eq.\ (\ref{there}), we arrive at
\begin{align}
\caltbmz \, \W_{\sma{{-}\pi}}\,{\caltbmz}^{-1}  = [\W_{\sma{{-}\pi}}]^{-1},\ins{with}\caltbmz = \W_{\sma{-\pi \leftarrow 0}} \,\brevetbmz(0,{-}\pi) .
\end{align}
The Wilson-loop operator is thus W-symmetric under $\caltbmz$, which combines a space-time transformation (encoded in $\brevetbmz$) with parallel transport over half a reciprocal period (encoded in $\W_{\sma{-\pi \leftarrow 0}}$). This constraint does not produce any degeneracy, since (i) $\caltbmz^2=+I$ and furthermore (ii) the eigenvalues of $\calbmx$ (cf. Eq.\ (\ref{relativistic})) are preserved under $\caltbmz$. The proof of (i) follows as    
\begin{align}
[\caltbmz^2]_{mn} \eq \bra{u_{\sma{m,-\pi}}}\,\prod_{k_y}^{-\pi \leftarrow 0} P(k_y)\,V(2\pi \vec{y}-\tilde{\boldsymbol{b}}_2 )\,\hat{T}_{\sma{\bmz}}\,\prod_{q_y}^{-\pi \leftarrow 0} P(q_y)\,V(2\pi \vec{y}-\tilde{\boldsymbol{b}}_2 )\,\hat{T}_{\sma{\bmz}}\,\ket{u_{\sma{n,-\pi}}} \lin
\eq \bra{u_{\sma{m,-\pi}}}\,\prod_{k_y}^{-\pi \leftarrow 0} P(k_y)\,\prod_{q_y}^{-\pi \leftarrow 0} P(-\pi-q_y)\,V(2\pi \vec{y}-\tilde{\boldsymbol{b}}_2 )\,\hat{T}_{\sma{\bmz}}\,V(2\pi \vec{y}-\tilde{\boldsymbol{b}}_2 )\,\hat{T}_{\sma{\bmz}}\,\ket{u_{\sma{n,-\pi}}} \lin
\eq \bra{u_{\sma{m,-\pi}}}\,\prod_{k_y}^{-\pi \leftarrow 0} P(k_y)\,\prod_{q_y}^{0 \leftarrow -\pi} P(q_y)\,V(2\pi \vec{y}-\tilde{\boldsymbol{b}}_2 )\,V(-2\pi \vec{y}+\tilde{\boldsymbol{b}}_2 )\,(\hat{T}_{\sma{\bmz}})^2\,\ket{u_{\sma{n,-\pi}}} \lin
\eq \bra{u_{\sma{m,-\pi}}}\,\prod_{k_y}^{-\pi \leftarrow 0} P(k_y)\,\prod_{q_y}^{0 \leftarrow -\pi} P(q_y)\,\ket{u_{\sma{n,-\pi}}}=  [\W_{\sma{-\pi \leftarrow 0}}\,\W_{\sma{0 \leftarrow -\pi}}\,]_{mn} = \delta_{mn}, 
\end{align}
where in the second equality we used Eq.\ (\ref{periodP}) and (\ref{consttmx}), in the third Eq.\ (\ref{UU}), and in the fourth $(\hat{T}_{\sma{\bmz}})^2=\utbmz \,\utbmz^*=I$. To prove (ii), we first demonstrate that $\ubmx$ and $\utbmz\,K$ anticommute, which follows from $\ubmx$ and $\utbmz\,K$ forming a \emph{symmorphic} representation of $M_x$ and $TM_z$ (where $M_j$ are glideless reflections; see discussion of Eq.\ (\ref{symmorphicalgebratwo})). Indeed, $M_xM_z= \bar{E}M_zM_x$  in the half-integer-spin representation, and time reversal commutes with any spatial transformation, leading to $\{\ubmx,\utbmz\,K\}=0$. This anticommutation is applied in the fourth equality of 
\begin{align}
&[\calbmx \;\caltbmz]_{mn}\lin
 \eq e^{-ik_z}\;\bra{u_{\sma{m,-\pi}}} \,\prod_{k_y}^{-\pi \leftarrow 0} P(k_y)  \,V(-\tilde{\boldsymbol{b}}_2)\,\ubmx \,\prod_{q_y}^{-\pi \leftarrow 0} P(q_y)\,V(2\pi \vec{y}-\tilde{\boldsymbol{b}}_2 )\,\utbmz\,K  \,\ket{u_{\sma{n,-\pi}}} \lin
 \eq e^{-ik_z}\;\bra{u_{\sma{m,-\pi}}} \,\prod_{k_y}^{-\pi \leftarrow 0} P(k_y)  \,\prod_{q_y}^{-\pi \leftarrow 0} P(q_y+\pi)\,V(-\tilde{\boldsymbol{b}}_2)\,\ubmx \,V(2\pi \vec{y}-\tilde{\boldsymbol{b}}_2 )\,\utbmz\,K  \,\ket{u_{\sma{n,-\pi}}} \lin
 \eq e^{-ik_z}\;\bra{u_{\sma{m,-\pi}}} \,\prod_{k_y}^{-\pi \leftarrow 0} P(k_y)  \,\prod_{q_y}^{0 \leftarrow \pi} P(q_y)\,V(-\tilde{\boldsymbol{b}}_2) \,V(\tilde{\boldsymbol{b}}_2 )\,\ubmx\,\utbmz\,K  \,\ket{u_{\sma{n,-\pi}}} \lin
\eq -e^{-ik_z}\;\bra{u_{\sma{m,-\pi}}} \,\prod_{k_y}^{-\pi \leftarrow 0} P(k_y)   \,\prod_{q_y}^{0 \leftarrow \pi} P(q_y)\,\utbmz\,K\,\ubmx \,\ket{u_{\sma{n,-\pi}}} \lin
\eq -e^{-ik_z}\;\bra{u_{\sma{m,-\pi}}} \,\prod_{k_y}^{-\pi \leftarrow 0} P(k_y)   \,\prod_{q_y}^{0 \leftarrow \pi} P(q_y)\,V(2\pi \vec{y}-\tilde{\boldsymbol{b}}_2 )\,V(-2\pi \vec{y}+\tilde{\boldsymbol{b}}_2 )\,(\,\hat{T}_{\sma{\bmz}}e^{-ik_z/2}\,)\,\ubmx \,\ket{u_{\sma{n,-\pi}}} \lin
\eq -e^{-ik_z}\;\bra{u_{\sma{m,-\pi}}} \,\prod_{k_y}^{-\pi \leftarrow 0} P(k_y)   \,\prod_{q_y}^{0 \leftarrow \pi} P(q_y)\,V(2\pi \vec{y}-\tilde{\boldsymbol{b}}_2 )\,\hat{T}_{\sma{\bmz}}\,V(2\pi \vec{y})\,e^{-ik_z/2}\,V(-\tilde{\boldsymbol{b}}_2 )\,\ubmx \,\ket{u_{\sma{n,-\pi}}} \lin
\eq -e^{-ik_z}\;\bra{u_{\sma{m,-\pi}}} \,\prod_{k_y}^{-\pi \leftarrow 0} P(k_y)   \,V(2\pi \vec{y}-\tilde{\boldsymbol{b}}_2 )\,\hat{T}_{\sma{\bmz}}\,V(2\pi \vec{y})\,\prod_{q_y}^{0 \leftarrow \pi} P(\pi-q_y)\,e^{-ik_z/2}\,V(-\tilde{\boldsymbol{b}}_2 )\,\ubmx \,\ket{u_{\sma{n,-\pi}}} \lin
\eq -e^{-ik_z}\;\bra{u_{\sma{m,-\pi}}} \,\prod_{k_y}^{-\pi \leftarrow 0} P(k_y)  \,V(2\pi \vec{y}-\tilde{\boldsymbol{b}}_2 )\,\hat{T}_{\sma{\bmz}} \,V(2\pi\vec{y})\,\prod_{q_y}^{\pi \leftarrow -\pi} P(q_y)\,\prod_{l_y}^{-\pi \leftarrow 0} P(l_y)\,  e^{-ik_z/2}\,V(-\tilde{\boldsymbol{b}}_2)\,\ubmx \,\ket{u_{\sma{n,-\pi}}} \lin
\eq -e^{-ik_z}\,[\caltbmz\;\W_{\sma{{-}\pi}}\;\calbmx]_{mn} = -e^{-ik_z}\,[\caltbmz\;\calbmx\;\W_{\sma{{-}\pi}}]_{mn},
\end{align}
where we have also applied Eq.\ (\ref{glideconstr}) and Eq.\ (\ref{uv}) in multiple instances. Then define simultaneous eigenstates of $\W_{\sma{{-}\pi}}$ and $\calbmx$ such that
\begin{align} \label{defineeigenstate}
\W_{\sma{{-}\pi}} \,\ket{e^{i\theta},\lambda_x;k_z} =  e^{i\theta}\,\ket{e^{i\theta},\lambda_x;k_z} \ins{and} \calbmx \,\ket{e^{i\theta},\lambda_x;k_z} =  \lambda_x\,\ket{e^{i\theta},\lambda_x;k_z}.
\end{align}
Once again, all operators and eigenvalues here depend on $k_z$. Finally,
\begin{align} \notag
& \calbmx \,\caltbmz\,\ket{e^{i\theta},\lambda_x;k_z} = -e^{-ik_z}\,\caltbmz\,\calbmx\,\W_{\sma{{-}\pi}}\,\ket{e^{i\theta},\lambda_x;k_z} = -e^{-i(k_z+\theta)}\,\lambda_x^*\,\caltbmz\,\ket{e^{i\theta},\lambda_x;k_z} =  \lambda_x\,\caltbmz\,\ket{e^{i\theta},\lambda_x;k_z} .
\end{align}
In the last equality, we applied $\lambda_x^2 =-\text{exp}[{-i(\theta+{k}_z)}]$ from Eq.\ (\ref{relativistic}).

\subsubsection{Effect of space-time inversion symmetry} \label{app:TIwilson}

The first two relations of Eq.\ (\ref{TIwilson}) may be carried over from App.\ \ref{app:spacetimeinversion}, if we identify $\calti \equiv \breveti$. What remains is to show: $\calbmx \,\calti {=} t(\vec{z})\, \calti \,\calbmx.$ 
\begin{align} 
[\calbmx \,\calti]_{mn} \eq e^{-i{k}_z/2}\;\bra{u_{\sma{m,-\pi}}} \,\prod_{k_y}^{-\pi \leftarrow 0} P(k_y)  \,V(-\tilde{\boldsymbol{b}}_2)\,\ubmx \,\hat{T}_{\sma{\cali}}  \,\ket{u_{\sma{n,-\pi}}} \lin
\eq e^{-ik_z}\;\bra{u_{\sma{m,-\pi}}} \,\hat{T}_{\sma{\cali}}\,e^{-ik_z/2}\,\prod_{k_y}^{-\pi \leftarrow 0} P(k_y)  \,V(-\tilde{\boldsymbol{b}}_2)\,\ubmx   \,\ket{u_{\sma{n,-\pi}}}  = e^{-ik_z}\,[\calti \,\calbmx]_{mn}.
 \end{align}
Recalling our definitions in Eq.\ (\ref{defineeigenstate}), we now show that $|e^{i\theta},\lambda_x;k_z \rangle $ and $\calti\,|e^{i\theta},\lambda_x;k_z \rangle $ belong in opposite mirror branches. To be precise, since $\lambda_x(k_z{+}\theta){\equiv}{\pm} i $exp$[{-}i(k_z{+}\theta)/2]$ is both momentum- and energy-dependent, and $T\cali$ maps $\theta\rightarrow -\theta, k_z \rightarrow k_z$, we would show that two space-time-inverted partners have $\calbmx$-eigenvalues $\lambda_x(k_z{+}\theta)$ and $-\lambda_x(k_z{-}\theta)$:
\begin{align}
\calbmx \,\calti\,\ket{e^{i\theta},\lambda_x;k_z} = e^{-ik_z}\,\lambda_x^*\,\calti\,\ket{e^{i\theta},\lambda_x;k_z} = \mp i \,e^{-i(k_z-\theta)/2}\,\calti\,\ket{e^{i\theta},\lambda_x;k_z}.
\end{align}

\subsubsection{Wilsonian time-reversal operator} \label{app:wilsontime}

In the remaining discussion, we particularize to two Wilson loops at fixed $\kpar = (\pi,\bar{k}_z)$, with $\bar{k}_z=0$ or $\pi$ only. Under time reversal,  $(k_y,\pi,\bar{k}_z) \longrightarrow (-k_y,-\pi,-\bar{k}_z) = (\pi-k_y,\pi,\bar{k}_z)-\tilde{\boldsymbol{b}}_2-2\bar{k}_z\vec{z}$, as illustrated in Fig.\ \ref{fig:wherewilson}(d). This implies that the occupied-band projections along these lines are constrained as
\begin{align}
V(\tilde{\boldsymbol{b}}_2+2\bar{k}_z \vec{z})^{\mo}\, \hat{T}\,P(k_y,\pi,\bar{k}_z) \,\hat{T}^{\mo}\,V(\tilde{\boldsymbol{b}}_2+2\bar{k}_z \vec{z}) = P(\pi-k_y,\pi,\bar{k}_z),
\end{align}
where $\hat{T}=U_T K$ is the antiunitary representation of time reversal. This relation allows us to define a unitary `sewing matrix' between states at $(2\pi,\kpar)$ and $(-\pi,\kpar)$: 
\begin{align} 
 [\breve{T}((2\pi,\kpar),(-\pi,\kpar))]_{mn} = \bra{u_{m,(2\pi,\kpar)}}\,V(-\tilde{\boldsymbol{b}}_2-2\bar{k}_z\vec{z})\,\hat{T}\,\ket{u_{n,(-\pi,\kpar)}},
\end{align}
which particularizes Eq.\ (\ref{sew}) with $\bG=-\tilde{\boldsymbol{b}}_2-2\bar{k}_z\vec{z}$. Henceforth suppressing the $\kpar$ labels, we are led to
\begin{align} \label{there2}
\breve{T}(2\pi,{-}\pi) \, \W_{\sma{{-}\pi}}\,\breve{T}(2\pi,{-}\pi)^{-1} = \W_{r,\sma{2\pi}} \ins{where} [\breve{T}(2\pi,{-}\pi)]_{mn} = \bra{u_{m,2\pi}}\,V(-\tilde{\boldsymbol{b}}_2-2\bar{k}_z \vec{z})\,\hat{T}\,\ket{u_{n,-\pi}},
\end{align}
and $\W_{r,\sma{2\pi}}$ denotes the reverse-oriented Wilson loop with base point $2\pi$, as drawn in Fig.\ \ref{fig:wilsonlines}(f). Combining this result with Eq.\ (\ref{wilsonreversegauge}) and the identity 
\begin{align}
 [\W_{\sma{\pi \leftarrow 2\pi}} \,\breve{T}(2\pi,-\pi)]_{mn} \eq  \bra{u_{m,\pi}}\,\prod_{k_y}^{\pi \leftarrow 2\pi}P(k_y)\,V(-\tilde{\boldsymbol{b}}_2-2\bar{k}_z \vec{z})\,\hat{T}\,\ket{u_{n,-\pi}}  \lin
\eq  \bra{u_{m,-\pi}}\,V(2\pi \vec{y})\,\prod_{k_y}^{\pi \leftarrow 2\pi}P(k_y)\,V(-\tilde{\boldsymbol{b}}_2-2\bar{k}_z \vec{z})\,\hat{T}\,\ket{u_{n,-\pi}}  \lin
\eq  \bra{u_{m,-\pi}}\,\prod_{k_y}^{-\pi \leftarrow 0}P(k_y)\,\,V(2\pi \vec{y}-\tilde{\boldsymbol{b}}_2-2\bar{k}_z \vec{z})\,\hat{T}\,\ket{u_{n,-\pi}} \lin
\equiv&\; [\W_{\sma{-\pi \leftarrow 0}}\,\breve{T}(0,-\pi)]_{mn},
\end{align}
we arrive at
\begin{align} \label{timereversewilson}
\calt \, \W_{\sma{{-}\pi}}\,{\calt}^{\,-1}  = [\W_{\sma{{-}\pi}}]^{-1},\ins{with}\calt = \W_{\sma{-\pi \leftarrow 0}} \,\breve{T}(0,-\pi) .
\end{align}
The Wilson-loop operator is thus W-symmetric under $\calt$, which combines time reversal (encoded in $\breve{T}$) with parallel transport over half a reciprocal period (encoded in $\W_{\sma{-\pi \leftarrow 0}}$). While many properties of the ordinary time reversal are well-known (e.g., Kramers degeneracy, the commutivity of time reversal with spatial transformations), it is not a priori obvious that these properties are applicable to the W-symmetry ($\calt$). We will find that $\calt$ indeed enforces a Kramers degeneracy in the $\W$-spectrum, but it only commutes with the W-glide ($\calbmx$) modulo a Wilson loop. The Kramers degeneracy follows from (a) $\calt$ relating two eigenstates of $\W_{\sma{{-}\pi}}$ with the same eigenvalue, as follows from Eq.\ (\ref{timereversewilson}), and (b) $\calt^2=-I$, as we now show:   
\begin{align}
[\calt^2]_{mn} 
\eq \bra{u_{\sma{m,-\pi}}}\,\prod_{k_y}^{-\pi \leftarrow 0} P(k_y)\,V(2\pi \vec{y}-\tilde{\boldsymbol{b}}_2-2\bar{k}_z \vec{z})\,\hat{T}\,\prod_{q_y}^{-\pi \leftarrow 0} P(q_y)\,V(2\pi \vec{y}-\tilde{\boldsymbol{b}}_2-2\bar{k}_z \vec{z})\,\hat{T}\,\ket{u_{\sma{n,-\pi}}} \lin
\eq \bra{u_{\sma{m,-\pi}}}\,\prod_{k_y}^{-\pi \leftarrow 0} P(k_y)\,\prod_{q_y}^{0 \leftarrow -\pi} P(q_y)\,V(2\pi \vec{y}-\tilde{\boldsymbol{b}}_2-2\bar{k}_z \vec{z})\,V(-2\pi \vec{y}+\tilde{\boldsymbol{b}}_2+2\bar{k}_z \vec{z})\,(\hat{T})^2\,\ket{u_{\sma{n,-\pi}}} \lin
\eq -\bra{u_{\sma{m,-\pi}}}\,\prod_{k_y}^{-\pi \leftarrow 0} P(k_y)\,\prod_{q_y}^{0 \leftarrow -\pi} P(q_y)\,\ket{u_{\sma{n,-\pi}}}= - [\W_{\sma{-\pi \leftarrow 0}}\,\W_{\sma{0 \leftarrow -\pi}}\,]_{mn} = - \delta_{mn}, 
\end{align}
where $\hat{T}^2=-I$ represents a $2\pi$ rotation. We further investigate if Kramers partners share identical or opposite eigenvalues under $\calbmx$. We find at $k_z=0$ that  $\calt: \;\lambda_x \longrightarrow -\lambda_x$, while at $k_z=\pi$, $\calt: \;\lambda_x \longrightarrow \lambda_x$, as we now prove. \\

\noindent \textbf{Proof} Applying $[\hat{T},\ubmx]=0$,
\begin{align}
[\calbmx \;\calt]_{mn} \eq e^{-i\bar{k}_z/2}\;\bra{u_{\sma{m,-\pi}}} \,\prod_{k_y}^{-\pi \leftarrow 0} P(k_y)  \,V(-\tilde{\boldsymbol{b}}_2)\,\ubmx \,\prod_{q_y}^{-\pi \leftarrow 0} P(q_y)\,V(2\pi \vec{y}-\tilde{\boldsymbol{b}}_2-2\bar{k}_z \vec{z})\,\hat{T}  \,\ket{u_{\sma{n,-\pi}}} \lin
\eq e^{-i\bar{k}_z/2}\;\bra{u_{\sma{m,-\pi}}} \,\prod_{k_y}^{-\pi \leftarrow 0} P(k_y)   \,\prod_{q_y}^{0 \leftarrow \pi} P(q_y)\,V(-\tilde{\boldsymbol{b}}_2)\, e^{i\bar{k}_z}\,V(\tilde{\boldsymbol{b}}_2-2\bar{k}_z \vec{z})\,\hat{T} \,\ubmx \,\ket{u_{\sma{n,-\pi}}} \lin
\eq \bra{u_{\sma{m,-\pi}}} \,\prod_{k_y}^{-\pi \leftarrow 0} P(k_y)  \,V(2\pi \vec{y}-\tilde{\boldsymbol{b}}_2-2\bar{k}_z \vec{z})\,\hat{T} \,V(2\pi\vec{y})\,\prod_{q_y}^{\pi \leftarrow -\pi} P(q_y)\,\prod_{l_y}^{-\pi \leftarrow 0} P(l_y)\, e^{-i\bar{k}_z/2}\, V(-\tilde{\boldsymbol{b}}_2)\,\ubmx \,\ket{u_{\sma{n,-\pi}}} \lin
\eq [\calt\;\W_{\sma{{-}\pi}}\;\calbmx]_{mn} = [\calt\;\calbmx\;\W_{\sma{{-}\pi}}]_{mn}.
\end{align}
This confirms our previous claim that $\calt$ commutes with $\calbmx$ modulo a Wilson loop, unlike the algebra of ordinary space-time symmetries. Recalling Eq.\ (\ref{defineeigenstate}),
\begin{align}
\calbmx \,\calt \,\ket{e^{i\theta},\lambda_x;\bar{k}_z} = \calt\,\calbmx\,\W_{\sma{{-}\pi}}\,\ket{e^{i\theta},\lambda_x;\bar{k}_z} = e^{-i\theta}\,\lambda_x^*\,\calt\,\ket{e^{i\theta},\lambda_x;\bar{k}_z} =  - e^{i\bar{k}_z}\,\lambda_x\,\calt\,\ket{e^{i\theta},\lambda_x;\bar{k}_z} .
\end{align}
In the last equality, we applied $\lambda_x(\theta+k_z)^2 =-\text{exp}[{-i(\theta+{k}_z)}]$. $\blacksquare$\\

An analog of this result occurs for the surface bands, where the eigenvalues of $\bmx$ are imaginary (real) at $k_z=0$ (resp.\ $\pi$), and time-reversal pairs up complex-conjugate eigenvalues. \\

\end{widetext}

\section{Connectivity of bulk Hamiltonian bands in spin systems with glide and time-reversal symmetries}\label{app:connectivity}

\begin{figure}[H]
\centering
\includegraphics[width=5 cm]{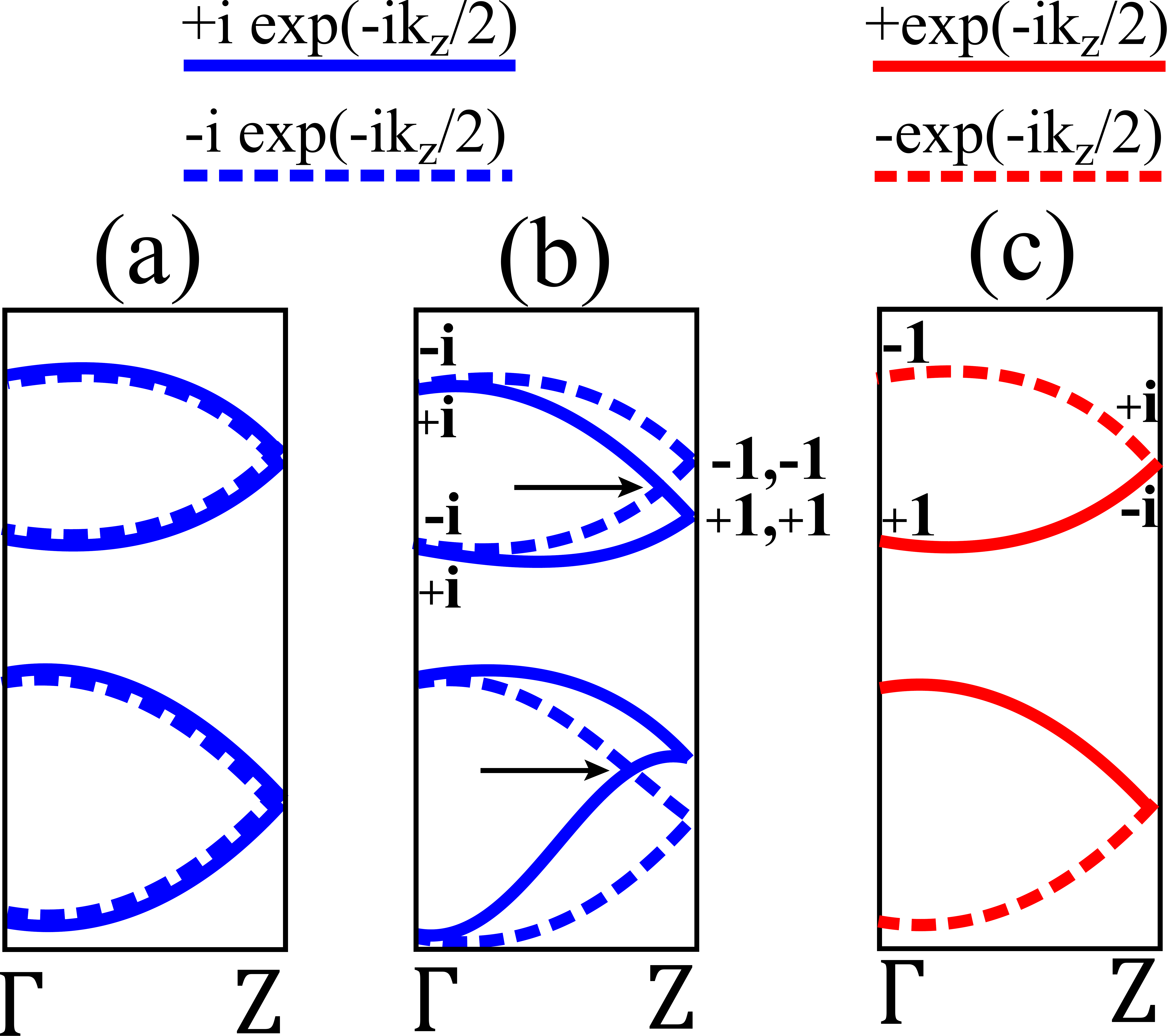}
    \caption{Bulk bandstructures with glide and time-reversal symmetries, for systems with spin (a-b) and without (c). $\Gamma \equiv (0,0,k_z=0)$ and $Z \equiv (0,0,k_z=\pi)$ are high-symmetry points that are selected because the fractional translation (in the glide) is parallel to $\vec{z}$. (a) either has no spin-orbit coupling, or has spin-orbit coupling with an additional spatial-inversion symmetry. (b) has spin-orbit coupling but breaks spatial-inversion symmetry. The crossings between orthogonal mirror branches (indicated by arrows) are movable along $\Gamma Z$ but unremovable so long as glide and time-reversal symmetries are preserved. The glide eigenvalues are indicated at $\Gamma$ and $Z$ for one of the two hourglasses. (c) could apply to an intrinsically spinless system (e.g., bosonic cold atoms and photonic crystals), and also to an effectively spinless system (e.g., a single-spin species in an electronic system without spin-orbit coupling).} \label{fig:connectivity}
\end{figure}

In spin systems with \emph{minimally} time-reversal ($T$) and glide-reflection ($\bmx$) symmetries, we prove that bulk Hamiltonian bands divide into quadruplet sets of hourglasses, along the momentum circle parametrized by $(0,0,k_z)$. Each quadruplet is connected, in the sense that there are enough contact points to continuously travel through all four branches. With the addition of other spatial symmetries in our space group, we further describe how degeneracies \emph{within} each hourglass may be further enhanced. Our proof of connectivity generalizes a previous proof\cite{connectivityMichelZak} for integer-spin representations of nonsymmorphic space groups. \\

The outline of our proof: we first consider a spin system with vanishing spin-orbit coupling, such that it has a spin $SU(2)$ symmetry. In this limit, we prove that bands divide into doubly-degenerate quadruplets with a four-fold intersection at $(0,0,\pi)$, as illustrated in Fig.\ \ref{fig:connectivity}(a). Then by introducing spin-orbit coupling and assuming no other spatial symmetries, we show that each quadruplet splits into a connected hourglass (Fig.\ \ref{fig:connectivity}(b)).\\

With vanishing spin-orbit coupling, the system is additionally symmetric under the spin flip $(F_x)$ that rotates spin by $\pi$ about $\vec{x}$. The double group ($G$) relations include
\begin{align}
&T^2=F_x^2=\bar{E},\myspace \bmx^2 = \bar{E}\,t(c\vec{z}), \lin
& [T,F_x]=[T,\bmx]=[F_x,\bmx]=0,
\end{align}
with $\bar{E}$ a $2\pi$ rotation and $t$ a lattice translation. It follows from this relations that we can define two operators that act like time reversal and glide reflection in a spinless system:
\begin{align} \label{spinlessalgebra}
&T_x \equiv T\, F_x,\;\;\bar{m}_x \equiv \bmx \,F_x; \lin
 &[T_x,\bar{m}_x]=0,\;\;T_x^2 = I,\;\;\bar{m}_x^2 = t(c\vec{z}).
\end{align}
These operations preserve the spin component in $\vec{z}$ -- the group ($\tilde{G}$) of a single spin species (aligned in ${\pm} \vec{z}$) is generated by $T_x$, $\bar{m}_x$ and the lattice translations. It is known from Ref.\ \onlinecite{connectivityMichelZak} that the elementary band representation\cite{elementaryenergybands} of $\tilde{G}$ is two-dimensional, i.e., single-spin bands divide into sets of two which cannot be decomposed as direct sums, and in each set there are enough contact points (to be found anywhere in the Brillouin zone) to continuously travel through both branches; this latter property they call `connectivity'. We reproduce here their proof of connectivity by monodromy:\\

\noindent Let us consider the single-spin Bloch representation ($\tilde{D}_{k_z}$) of $\tilde{G}$ along the circle $(0,0,k_z)$; $k_z \in [0,2\pi)$. For $k_z \neq \pi$, there are two irreducible representations (labelled by $\rho$) which are each one-dimensional: 
\begin{align} 
&\tilde{D}^{\rho}_{k_z}\big(\,t(c\vec{z})\,\big) = e^{-ik_z} \lin
&\imp\; \tilde{D}^{\rho}_{k_z}(\bar{m}_x) = e^{-i(k_z+2\pi \rho)/2}, \;\; k_z \neq \pi,\;\; \rho \in \{0,1\},\notag
\end{align}
as follows from Eq.\ (\ref{spinlessalgebra}). Making one full turn in this momentum circle ($k_z \rightarrow k_z+2\pi$) effectively permutes the representations as $\rho \rightarrow \rho+1$. This implies that if we follow continuously an energy function of one of the two branches, we would evolve to the next branch after making one circle, and finally return to the starting point after making two circles -- both branches form a connected graph. The contact point between the two branches is determined by the time-reversal symmetry ($T_x$), which pinches together complex-conjugate representations of $\bar{m}_x$ at $k_z{=}\pi$; on the other hand, real representations at $k_z{=}0$ are not degenerate, as illustrated in Fig.\ \ref{fig:connectivity}(c). We applied here that $[T_x,\bar{m}_x]{=}0$, and the eigenvalues of $\bar{m}_x$ are imaginary (real) at $k_z{=}\pi$ (resp.\ $0$). $\blacksquare$\\

Due to the spin $SU(2)$ symmetry, we double all irreducible representations of $\tilde{G}$ to obtain  representations of $G$, i.e., in the absence of spin-orbit coupling but accounting for both spin species, bands are everywhere spin-degenerate, and especially four-fold degenerate at $k_z{=}\pi$. For illustration, Fig.\ \ref{fig:connectivity}(a) may be interpreted as a spin-doubled copy of Fig.\ \ref{fig:connectivity}(c). Recall that the eigenvalues of $\bmx$ fall into two branches labelled by $\eta{=} {\pm} 1$ in $\lambda_x(k_z) {=} \eta\, i\text{ exp}(-ik_z/2)$. For each spin-degenerate doublet, the two spin species fall into opposite branches of $\bmx$, as distinguished by solid and dashed curves in Fig.\ \ref{fig:connectivity}(a). This result follows from continuity to $k_z{=}0$, where Kramers partners have opposite, \emph{imaginary} eigenvalues under $\bmx$. \\

Without additional spatial symmetries, the effect of spin-orbit coupling is to split the spin degeneracy for generic $k_z$, while preserving the Kramers degeneracy at $k_z{=}0$ and $\pi$ -- the final result is the hourglass illustrated in Fig.\ \ref{fig:connectivity}(b). To demonstrate this, we note at $k_z{=}\pi$ that each four-dimensional subspace (without the coupling) splits into two Kramers subspace, where Kramers partners have identical, real eigenvalues under $\bmx$; pictorially, two solid curves emerge from the one of the two Kramers subspaces, and for the other subspace both curves are dashed, as shown in Fig.\ \ref{fig:connectivity}(b). Furthermore, we know from the previous paragraph that each Kramers pair at $k_z{=}0$ combines a solid and dashed curve. These constraints may be interpreted as curve boundary conditions at $0$ and $\pi$, which impose a solid-dashed crossing between the boundaries, as indicated by arrows in Fig.\ \ref{fig:connectivity}(b). There is then an `unavoidable degeneracy'\cite{elementaryenergybands}  which can move along the half-circle, but cannot be removed. This contact point, in addition to the unmovable Kramers degeneracies at $0$ and $\pi$, guarantee that each quadruplet is connected. \\

Let us consider how other spatial symmetries (beyond $\bmx$) may enhance degeneracies within each hourglass. For illustration, we consider the spatial inversion ($\cali$) symmetry, which applies to the space group of KHg$X$. Since $T\cali$ belongs in the group of every bulk wavevector, the spin degeneracy along $(0,0,k_z)$ does not split, and $k_z {=}\pi$ remains a point of four-fold degeneracy, as illustrated in Fig.\ \ref{fig:connectivity}(a).\\

One final remark is that the notion of `connectivity' over the entire Brillouin zone, as originally formulated in Ref.\ \onlinecite{connectivityMichelZak}, can fruitfully be particularized to `connectivity of a submanifold', which we introduced in our companion work\cite{Hourglass} as a criterion for topological surface bands; our notion differs from the original formulation in that contact points must be found only within the submanifold in question, rather than the entire Brillouin zone.



\section{Projective representations, group extensions and group cohomology} \la{app:wilsonproj}

Wilsonian symmetries describe the extension of a point group by quasimomentum translations; such extensions are also called projective representations. In this Appendix, we elaborate on the connection between projective representations and group extensions in App.\ \ref{app:extension}, then proceed in App.\ \ref{app:connectprojrejcoh} to describe projective representations from the more abstract perspective of cochains, which emphasizes the connection with group cohomology. Finally in App.\ \ref{app:simplecalc}, we exemplify a simple calculation of the second group cohomology. 

\subsection{Connection between projective representations and group extensions} \label{app:extension}

In Sec.\ \ref{sec:wilsoniansymm}, we introduced three groups: \\

\noi{i} As defined in Eq.\ (\ref{defineGs}), $\gs \cong \Z_2 \times \Z_2$ is a symmorphic, spinless group generated by time reversal ($T$) and a glideless reflection ($M_x$), with an algebra summarized in Eq.\ (\ref{symmorphicalgebra}). \\

\noi{ii} The group ($\gwp$) of the Wilson loop is generated by $2\pi$-rotation ($\bar{E}$), a lattice translation ($t(c\vec{z})$), the Wilson loop ($\W$), and analogs of time reversal ($\calt$) and glide reflection ($\calbmx$) additionally encode parallel transport. With regard to our KHg$X$ material class, this Appendix does not exhaust all elements in $\gwp$ or $\gs$; our treatment here minimally conveys their group structures. A more complete treatment of the material class has been described in App.\ \ref{app:kxpi} for the different purpose of topological classification.\\

\noi{iii} As defined in Eq.\ (\ref{defineNgroup}), $\caln \cong \Z^2 \times \Z_2$ is an Abelian group generated by $2\pi$ rotations ($\bar{E}$),  momentum translations ($\W$) and real-space translations ($t(\vec{z})$). $\gs$ induces an automorphism on $\caln$, which we proceed to define. Letting the group element $g_i \in \gs$ be represented in $\gwp$ by $\hat{g}_i$, we say that $g_i$ induces the automorphism $a {\rightarrow} \sigma_i(a)$, where
\begin{align} \label{automorphism}
& \sigma_i(\,\bar{E}^a\,t(\vec{z})^b\,\W^c\,) \equiv \hatgi\,\bar{E}^a\,t(\vec{z})^b\,\W^c  \,\hatgi^{-1} \lin
\eq \bar{E}^a\,t(\vec{z})^{\kappa(\hatgi) b}\,\W^{\gamma(\hatgi) c}\;\;\text{with}\;\; \kappa(\hatgi),\gamma(\hatgi) \in \{\pm 1\}.
\end{align}
$\gamma(\hatgi){=}{+}1$ if $\hatgi$ preserves the Wilson loop (e.g., $\hatgi{=}\calbmx$ in Eq.\ (\ref{Mxwilson})), and $\gamma(\hatgi){=}{-}1$ if $\hatgi$ is orientation-reversing (e.g., $\hatgi{=}\calt$ in Eq.\ (\ref{wilsonT})). Similarly, $\kappa(\hatgi){=}{+}1$ (${-}1$) if $\hatgi$ preserves (inverts) the spatial translation $t(\vec{z})$; in our example, $\kappa(\calbmx){=}\kappa(\calt){=}{+}1$. Equivalently, we may say that $\gn$ is a normal subgroup of $\gwp$. \\

\noindent To show that $\gwp$ is an extension of $\gs$ by $\gn$, it is sufficient to demonstrate that $\gs$ is isomorphic to the factor group $\gwp/\gn$.\cite{tinkhambook} This factor group has as elements the left cosets $g \gn$ with $g \in \gwp$; by the normality of $\gn$, $g\gn {=} \gn g$. This isomorphism respectively maps the elements $\gn, \calt \gn, \calbmx \gn$ (in $\gwp/\gn$) to identity $(I), T, M_x$ (in $\gs$); this is a group isomorphism in the sense that their multiplication rules are identical. For example, $M_x^2{=}I$ is isomorphic to:
\begin{align}
&(\calbmx \,\gn)^2 = \calbmx\, (\gn\,\calbmx)\, \gn = \calbmx^2 \gn^2 \lin
\eq \bar{E}\,t(c\vec{z})\,\W^{\sma{-1}}\,\gn  = \gn,
\end{align}
where we applied that $\gn$ is a normal subgroup of $\gwp$, and the third equality relies on Eq.\ (\ref{wglidesquare}). Two extensions are equivalent if there exists a group isomorphism between them; the second group cohomology ($H^2(\gs,\gn)$, as further elaborated in App.\ \ref{app:connectprojrejcoh}) classifies the isomorphism classes of all extensions of $\gs$ by $\gn$, of which $\gwp$ is one example. We have also described an inequivalent extension ($\gwz$) toward the end of Sec.\ \ref{sec:wilsoniansymm}, which is nontrivially extended in $t(\vec{z})$ (i.e., we are dealing with a glide instead of a glideless reflection symmetry) and also in $\bar{E}$ (i.e., this is a half-integer-spin representation), but not in $\W$. More generally, we could have either an integer-spin or a half-integer-spin representation, with either a glide or glideless reflection symmetry, and we could be describing a reflection plane (glide/glideless) in which the reflection either preserves every wavevector, or translates each wavevector by half a reciprocal period.

\subsection{Connection between projective representations and the second cohomology group} \label{app:connectprojrejcoh}

For a group $G_{\circ}$, we define a $G_{\circ}$-module (denoted $\caln$) as an abelian group on which $G_{\circ}$ acts compatibly with the multiplication operation in $\caln$.\cite{cohomologysharifi} In our application, $G_{\circ}$ of Eq.\ (\ref{defineGs}) is the point group of a spinless particle, and $\caln$ of Eq.\ (\ref{defineNgroup}) is the group generated by real- and quasimomentum-space translations, as well as $2\pi$ rotations. Let the $i$'th element ($g_i$) of $G_{\circ}$ act on $a\in \caln$ by the automorphism: $a \rightarrow \sigma_i(a) \in \caln$; we say this action is compatible if 
\begin{align}
\sigma_i(\,ab\,) = \sigma_i(\,a\,)\sigma_i(\,b\,) \ins{for every} a,b \in \caln.
\end{align}
We showed in Eq.\ (\ref{automorphism}) that $g_i$ acts on $a$ by conjugation, i.e., $\sigma_i(a) = \hat{g}_ia\hat{g}_i^{\mo}$, which guarantees that the action is compatible. \\

To every factor ($\cij \in \caln$) of a projective representation, defined again by
\begin{align} \label{reminder}
\hatgi \,\hatgj = \cij\, \hatgij,
\end{align}
where $g_{ij} \equiv g_ig_j$ and $\hat{g}_i$ is the representation of $g_i$, there corresponds a 2-cochain ($\nu_2$):
\begin{align} \label{factorcochain}
\cij = \nu_2(I,g_i,g_{ij}) \in \caln,
\end{align}
with the first argument in $\nu_2$ set as the identity element in $G_{\circ}$. $\nu_2$ more generally is a map $: G_{\circ}^{3} \rightarrow \caln$ -- besides informing of the factor system through Eq.\ (\ref{factorcochain}), it also encodes how each factor transforms under $G_{\circ}$, through
\begin{align} \label{encodessymm}
&\nu_2(g_ig_0,g_ig_1,g_ig_2) \equiv \hatgi\,\nu_2(g_0,g_1,g_2)\,\hatgi^{-1} \lin
&\equiv\; \sigma_i\big(\,\nu_2(g_0,g_1,g_2)\,\big) \in \caln.
\end{align}
Presently, we would review group cohomology from the perspective of cochains before establishing its connection with projective representations. Our review closely follows that of Ref.\ \onlinecite{SPTandgroupcohomologyXie} which described only $U(1)$ modules; our review demonstrates that the structure of cochains exists for more general modules (e.g., $\caln$). Moreover, we are motivated by possible generalizations of our ideas to higher-than-two cohomology groups, though presently we do not know if such exist. We adopt the convention of Ref.\ \onlinecite{SPTandgroupcohomologyXie} in defining cochains and the coboundary operator, which they have shown to be equivalent to standard\cite{cohomologysharifi} definitions; the advantage gained is a more compact definition of the coboundary operator. Generalizing Eq.\ (\ref{encodessymm}), an $n$-cochain is a map $\nu_n:G_{\circ}^{n+1} \rightarrow \caln$ satisfying
\begin{align} \label{ncochain}
\nu_n(g_ig_0,g_ig_1,\ldots, g_ig_{n}) \equiv  \sigma_i\big(\,\nu_n(g_0,g_1,\ldots,g_{n})\,\big) \in \caln.
\end{align}
Given any $\nu_{n-1}$, we can construct a special type of $n$-cochain, which we call an $n$-coboundary, by applying the coboundary operator $d_{n-1}$, defined as 
\begin{align} \label{definecoboundary}
&[d_{n-1}\nu_{n-1}](g_0,g_1,\ldots,g_{n})\lin
\eq \prod_{j=0}^n \nu_{n-1}(g_0,g_1,\ldots,g_{j-1},g_{j+1},\ldots,g_n)^{(-1)^j},\lin
&e.g.,\;\;[d_1\nu_1](g_0,g_1,g_2) \lin
&\myspace \myspace = \nu_1(g_1,g_2)\nu_1(g_0,g_2)^{-1}\nu_1(g_0,g_1).
\end{align} 
Formally, let $\calc^n(\caln)=\{\nu_n\}$ be the space of all $n$-cochains, and the space of all $n$-coboundaries is an abelian subgroup of $\calc^n$ defined by 
\begin{align}
\calb^n =\{ \nu_n| \nu_n = d_{n-1}\nu_{n-1},\; \nu_{n-1} \in \calc^{n-1} \}.
\end{align}

\begin{widetext}

Applying $d_n$ to an $n$-coboundary always gives the identity element in $\caln$:
\begin{align} \label{cobhasno}
 [d_nd_{n-1}\nu_{n-1}](g_0,g_1,\ldots,g_{n+1}) = \prod_{j=0}^{n+1} [d_{n-1}\nu_{n-1}](g_0,\ldots,g_{j-1},g_{j+1},\ldots,g_{n+1})^{(-1)^j} =I.
\end{align}
Succintly, a coboundary has no coboundary. Though this conclusion is well-known, it might interest the reader how this result is derived with our nonstandard definition of $d_n$. First express Eq.\ (\ref{cobhasno}) as a product of $\nu_{n-1}^{\pm 1}$ by inserting Eq.\ (\ref{definecoboundary}), e.g.,
\begin{align} \label{egcobcob}
&[d_2d_1\nu_1](g_0,g_1,g_2,g_3) = \left\{ \nu_1(g_2,g_3)\nu_1(g_1,g_3)^{-1}\nu_1(g_1,g_2)\right\}\left\{ \nu_1(g_2,g_3)\nu_1(g_0,g_3)^{-1}\nu_1(g_0,g_2)\right\}^{-1}\lin
& \myspace \myspace \times \left\{ \nu_1(g_1,g_3)\nu_1(g_0,g_3)^{-1}\nu_1(g_0,g_1)\right\}\left\{\nu_1(g_1,g_2)\nu_1(g_0,g_2)^{-1}\nu_1(g_0,g_1)  \right\}^{-1} =I.
\end{align}
Each of $\nu_{n-1}^{\pm 1}$ has $n$ distinct arguments drawn from the set $\{g_0,g_1,\ldots,g_{n+1}\}$ of $n+2$ elements. Equivalently, we may label $\nu_{n-1}^{\pm 1}$ by the two elements ($g_i$ and $g_j$; $i,j \in \{0,1,\ldots,n+1\}$; $i< j$) which have been deleted from this cardinality-$(n+2)$ set; there are always two such $\nu_{n-1}^{\pm 1}$ arising from two different ways to delete $\{g_i,g_j\}$:  
 either (a) $g_i$ was deleted by $d_{n}$ and $g_j$ by $d_{n-1}$, or (b) vice versa. For example, $\{g_0,g_1\}$ corresponds to 
\begin{align}
[d_2d_1\nu_1](g_0,g_1,g_2,g_3) \propto \left\{ \nu_1(g_2,g_3)\ldots\right\}\left\{ \nu_1(g_2,g_3)\ldots\right\}^{-1}\left\{ \ldots \right\}\left\{ \ldots  \right\}^{-1},
\end{align}
where (a) $\nu_1(g_2,g_3)$ originates from $d_2$ deleting $g_0$ and $d_1$ deleting $g_1$, while (b) $\nu_1(g_2,g_3)^{-1}$ arises from $d_2$ deleting $g_1$ and $d_1$ deleting $g_0$. These two factors, from (a) and (b), multiply to identity, as is more generally true for any  $\{g_i,g_j\}$ (recall $i<j$) and $d_nd_{n-1}\nu_{n-1}$, since 
\begin{align}
&[d_nd_{n-1}\nu_{n-1}](g_0,\ldots,g_{n+1}) \lin
&\propto \; \nu_{n-1}(\ldots, g_{i-1},g_{i+1},\ldots,g_{j-1},g_{j+1},\ldots)^{(-1)^i(-1)^{j-1}}\;\nu_{n-1}(\ldots, g_{i-1},g_{i+1},\ldots,g_{j-1},g_{j+1},\ldots)^{(-1)^j(-1)^i}=I,
\end{align}
where $\nu_{n-1}^{(-1)^j(-1)^i}$ originates from $d_n$ deleting $g_j$, and $\nu_{n-1}^{(-1)^i(-1)^{j-1}}$ from $d_n$ deleting $g_i$. We have thus shown that  $d_nd_{n-1}\nu_{n-1}=I$.\\

An $n$-coboundary is an example of an $n$-cocyle, which is more generally defined as any $n$-cochain with a trivial coboundary. The space of all $n$-cocycles is an abelian subgroup of $\calc^n$ defined as 
\begin{align}
\calz^n = \{ \nu_n| d_n\nu_n=I,\;\nu_n \in \calc^n \}.
\end{align}
The $n$'th cohomology group is defined by the quotient group
\begin{align} \label{definencohgroup}
H^n(G_{\circ},\caln) = \frac{\calz^n(G_{\circ},\caln)}{\calb^n(G_{\circ},\caln)};
\end{align}
its elements are equivalence classes of $n$-cocyles, in which we identify any two $n$-cocycles that differ by an $n$-coboundary. The rest of this Section establishes how $H^2(G_{\circ},\caln)$ classifies the different projective representations of $G_{\circ}$, as extended by $\caln$. Indeed, we have already identified the factor system of a projective representation with a $2$-cochain through Eq.\ (\ref{factorcochain}), and the associativity condition on the factor system will shortly be derived as
\begin{align}
C^{-1}_{ij,k}\,C^{-1}_{i,j}\,{\sigma}_i(\pdg{C}_{j,k})\,\pdg{C}_{i,jk} =I,
\end{align}
which translates to a constraint that the $2$-cochain is a $2$-cocyle. Indeed, from inserting Eq.\ (\ref{reminder}-\ref{encodessymm}) into $\hatgone (\hatgtwo \hatgthree)=(\hatgone \hatgtwo) \hatgthree$, find that
\begin{align}
& \pdg{\sigma}_1(\pdg{C}_{2,3})\,\pdg{C}_{1,23}\,\hat{g}_{123} = C_{12,3}\,C_{1,2}\,\hat{g}_{123},\lin
\imp &I=\pdg{\sigma}_1(\pdg{C}_{2,3})\,C^{-1}_{12,3}\,\pdg{C}_{1,23}\,C^{-1}_{1,2} \lin
 \imp & I=\nu_2(g_1,g_{12},g_{123})\,\nu_2(I,g_{12},g_{123})^{-1}\,\nu_2(I,g_1,g_{123})\,\nu_2(I,g_1,g_{12})^{-1} \lin
 \imp & I=\hat{g}_0\,I\,\hat{g}_0^{-1} = \hat{g}_0\,\nu_2(g_1,g_{12},g_{123})\,\nu_2(I,g_{12},g_{123})^{-1}\,\nu_2(I,g_1,g_{123})\,\nu_2(I,g_1,g_{12})^{-1} \,\hat{g}_0^{-1}\lin
\imp & I=\nu_2(g_{01},g_{012},g_{0123})\,\nu_2(g_0,g_{012},g_{0123})^{-1}\,\nu_2(g_0,g_{01},g_{0123})\,\nu_2(g_0,g_{01},g_{012})^{-1} \lin   
\imp &I=\nu_2(g_1,g_{2},g_{3})\nu_2(g_0,g_{2},g_{3})^{-1}\nu_2(g_0,g_1,g_{3})\nu_2(g_0,g_1,g_{2})^{-1} \equiv [d_2\nu_2](g_0,g_1,g_2,g_3),
\end{align}
where in the last ${\imp}$ we relabelled $g_{01\ldots k} \rightarrow g_k$ and $g_0 \rightarrow g_0$. Furthermore, we recall from Eq.\ (\ref{gaugetrans}) that two projective representations are equivalent if they are related by the gauge transformation $\hat{g}_i \rightarrow \hat{g}_i'=\di\,\hat{g}_i$ with $\di \in \caln$. This may be reexpressed as
\begin{align} \label{gaugetransonecc}
\hatgi \rightarrow \hatgi' = \nu_1(I,g_i)^{-1}\,\hatgi,
\end{align}
by relabelling $D_i \equiv \nu_1(I,g_i)^{\mo} \in \caln$. The motivation for calling $\hatgi$ and $\hatgi'$ gauge-equivalent is that both representations induce the same automorphism on the abelian group $\caln$, i.e.,
\begin{align} \label{equivauto}
\hatgi \,a\,\hatgi^{-1} \equiv \pdg{\sigma}_i(a) \ins{for any} a \in \caln  \imp \hatgi' \,a\,\hatgi'^{-1} \equiv \pdg{\sigma}_i(a).
\end{align}
Let us demonstrate that this gauge-equivalence condition may be expressed as an equivalence of 2-cochains modulo 1-coboundaries. By inserting Eq.\ (\ref{gaugetransonecc}) and (\ref{equivauto}) into Eq.\ (\ref{reminder}) with $i=1$ and $j=2$:
\begin{align}
& \hatgone \,\hatgtwo = C_{1,2}\,\hatgonetwo \imp (\nu_1(I,g_1)\,\hatgone') \,(\nu_1(I,g_2)\,\hatgtwo') = C_{1,2}\,\nu_1(I,g_{12})\,\hatgonetwo' \equiv \nu_2(I,g_1,g_{12})\,\nu_1(I,g_{12})\,\hatgonetwo'\lin
\imp & \nu_1(I,g_1)\,\pdg{\sigma}_1(\nu_1(I,g_2))\,\hatgone'\,\hatgtwo' = \nu_2(I,g_1,g_{12})\,\nu_1(I,g_{12})\,\hatgonetwo' \lin
\imp & \hatgone'\,\hatgtwo' = \nu_2(I,g_1,g_{12})\,\nu_1(I,g_1)^{-1}\,\nu_1(I,g_{12})\,\nu_1(g_1,g_{12})^{-1}\,\hatgonetwo' \equiv  \nu_2(I,g_1,g_{12})'\,\hatgonetwo'.
\end{align}
To reiterate, the $\nu_2'$ and $\nu_2$ are two gauge-equivalent 2-cochains differing only by multiplication with a 1-coboundary:
\begin{align}
&\nu_2(1,g_1,g_{12}) = \nu_2(1,g_1,g_{12})' \,\nu_1(I,g_1)\,\nu_1(I,g_{12})^{-1}\,\nu_1(g_1,g_{12}) \lin
 \imp &  \hat{g}_0\,\nu_2(1,g_1,g_{12})\,\hat{g}_0^{-1} =\nu_2(g_0,g_{01},g_{012}) = \nu_2(g_0,g_{01},g_{012})'\,\nu_1(g_0,g_{01})\,\nu_1(g_0,g_{012})^{-1}\,\nu_1(g_{01},g_{012})   \lin
 \imp & \nu_2(g_0,g_{1},g_{2}) = \nu_2(g_0,g_{1},g_{2})' \,\nu_1(g_{1},g_{2})\,\nu_1(g_0,g_{2})^{-1} \,\nu_1(g_0,g_{1})\equiv  \nu_2(g_0,g_{1},g_{2})'\,[d_1\nu_1](g_0,g_1,g_2),
\end{align}
where in the last ${\imp}$ we relabelled $g_{01\ldots k} \rightarrow g_k$ and $g_0 \rightarrow g_0$. We have thus demonstrated that different equivalence classes of projective representations correspond to equivalence classes of 2-cocycles, where equivalence is defined modulo 1-coboundaries, i.e., different projective representations are elements of the second cohomology group (recall $H^2(G_{\circ},\caln)$ from Eq.\ (\ref{definencohgroup})).\\

\end{widetext}

\subsection{A simple example} \label{app:simplecalc}

For a simple example of $H^2$, consider a reduced problem where we extend $\gs\cong \Z_2\times \Z_2$ (as generated by $M_{\sma{x}}$ and $T$)  by the group of Wilson loops
\begin{align}
N = \{\W^n \;|\;n \in \Z \} \cong \Z,
\end{align}
which differs from $\caln$ in lacking the generators $t(\vec{z})$ and $\bar{E}$; nontrivial extensions by $t(\vec{z})$ and $\bar{E}$ respectively describe nonsymmorphic and half-integer-spin representations, and are already well-known.\cite{Lax} Here, we focus on extensions purely by momentum translations. $\gs$ acts on $N$ as
\begin{align}  \label{autorules}
T \W T^{-1} = \W^{\mo} \ins{and} M_x\W M_x^{-1} =\W.
\end{align}
Let us follow the procedure outlined in Ref.\ \onlinecite{essin2013} to determine the possible extensions of $\gs$. First we collect all nonequivalent products of generators that multiply to identity according to the multiplication rules of $\gs$: from Eq.\ (\ref{symmorphicalgebra}), these would be
\begin{align}
M_x^2 = I,\;\; T^2 =I \ins{and} M_x\,T\,M_x^{-1}\,T^{-1} = I.
\end{align}
A projective representation is obtained by replacing $I$ on the right-hand side with an element in the $\gs$-module $N$:
\begin{align} \notag
M_x^2 = \W^a,\;\; T^2 =\W^b\ins{and} M_x\,T\,M_x^{-1}\,T^{-1} = \W^c.
\end{align}
That $a,b,c$ are integers does not imply $\Z^3$ inequivalent extensions; rather, we will see that not all three integers are independent, some integers are only gauge-invariant modulo two, and moreover one of them vanishes so that the representation is associative. Indeed, from the associativity of $T^3$,
\begin{align} \label{associattt}
&T\W^b = T(TT)=(TT)T = \W^b T =  T\W^{-b} \lin
&\imp \W^{2b}=I.
\end{align}
Lacking spatial-inversion symmetry, the eigenvalues of $\W$ are generically not quantized to any special value, and the only integral solution to $\W^{2b}{=}I$ is $b{=}0$; we comment on the effect of spatial-inversion symmetry at the end of this example.  Similarly, $a{=}c$ follows from 
\begin{align} \notag
\W^aT = M_x^2T = \W^{2c}TM_x^2 = \W^{2c}T\W^a = \W^{2c-a}T.
\end{align}
Moreover, we will clarify that only the parity of $a$ labels the inequivalent classes. This follows from $M_x$ and $M_x'=\W^{n_x}M_x$  ($n_x \in \Z$) inducing the same automorphism on $N$:
\begin{align}
M_x\W M_x^{-1}= \W \iff M_x'\W {M_x'}^{-1} = \W.
\end{align}
We say that $M_x$ and $M_x'$ are gauge-equivalent representations; consequently, only the parity of the exponent ($a$) of $\W$ in $M_x^2=\W^a$ is gauge-invariant, as seen from
\begin{align} \label{transforma}
&M_x^2=\W^a \imp  (\W^{-n_x}M_x')^2=\W^a \lin
&\imp\; M_x'^2= W^{a+2n_x} \equiv \W^{a'}.
\end{align}
One may verify that the relation $a{=}c$ is gauge-invariant, since if $a {\rightarrow} a'{=} a{+}2n_x$ (as we have just shown), likewise $c {\rightarrow} c'{=}c{+}2n_x$ (as we will now show). To determine the gauge-transformed $c'$, we consider two gauge-equivalent representations of time reversal related by $T'{=}\W^{n_T}T$ with $n_T {\in}\Z$. By application of Eq.\ (\ref{autorules}), we derive
\begin{align}
\W^c \eq  M_x\,T\,M_x^{-1}\,T^{-1} \lin
\eq (\W^{-n_x}\,M'_x)\,(\W^{-n_T}\,T')\,({M_x'}^{-1}\,\W^{n_x})\,({T'}^{-1}\,\W^{n_T}), \lin 
\imp &\W^{-n_x-n_T-n_x+n_T}\,  M'_x\,T'\,{M'_x}^{-1}\,{T'}^{-1} = \W^c \lin
\imp & M'_x\,T'\,{M'_x}^{-1}\,{T'}^{-1} = \W^{c+2n_x} \equiv \W^{c'},
\end{align}
as desired. We conclude that there are only two elements of $H_2(\gs,N)$: (i) the first is gauge-equivalent to $a{=}0$: 
\begin{align} \label{split}
M_x^2 = I,\;\; T^2 = I, \;\; [T,M_x]=0,
\end{align}
as expected from the algebra of $\gs$; (ii) the second element of $H_2(\gs,N)$ is gauge-equivalent to $a{=}{-}1$:
\begin{align} \label{unsplit}
M_x^2 = \W^{\mo},\;\; T^2 = I, \;\; T\,M_x=\W^{\mo}\,M_x\,T.
\end{align}
The first extension is split (i.e., it is isomorphic to a semi-direct product of $\gs$ with $N$), and corresponds to the identity element of $H_2(\gs,N) \cong \Z_2$. Multiplication of two elements corresponds to multiplying the factor systems, e.g., the two non-split elements multiply as
\begin{align}
M_x^2 = \W^{\sma{-2}},\;\; T^2 = I, \;\; T\,M_x=\W^{\sma{-2}}\,M_x\,T,
\end{align}
which is gauge-equivalent to Eq.\ (\ref{split}) by the transformation of Eq.\ (\ref{transforma}) with $n_x{=}1$.\\

To realize the nontrivial algebra in Eq.\ (\ref{unsplit}), we need that $M_x$ is a Wilsonian symmetry, i.e., it describes not purely a spatial reflection, but also induces parallel transport. As elaborated in Sec.\ \ref{sec:wilsoniansymm}, this Wilsonian symmetry is realized in mirror planes where any wavevector is mapped to itself by a combination of spatial reflection and quasimomentum translation across half a reciprocal period. To clarify a possible confusion, Sec.\ \ref{sec:wilsoniansymm} described a nonsymmorphic, half-integer-spin representation of a space group where $\bmx^2$ is a product of a spatial translation ($t(\vec{z})$) and a $2\pi$ rotation ($\bar{E}$), as is relevant to the KHg$X$ material class\cite{Hourglass}; this Appendix describes a symmorphic, integer-spin case study where $M_x^2=I$. Indeed, we have rederived Eq.\ (\ref{wglidesquare}) and (\ref{Twilson}) in Sec.\ \ref{sec:wilsoniansymm}, modulo factors of $\bar{E}$ and $t(\vec{z})$. Despite this difference, all Wilsonian reflections, whether glide or glideless, have the same physical origin: some crystal structures host mirror planes (of glide-type for KHg$X$, but glideless in this Appendix) where the group of any wavevector includes the product of spatial glide/reflection with a fractional recriprocal translation.   \\

Finally, we address a different example where $\gs$ includes a spatial-inversion ($\cali$) symmetry. We have shown in Ref.\ \onlinecite{AA2014} that a subset of the $\W$-eigenvalues may be quantized to ${\pm} 1$ depending on the $\cali$-eigenvalues of the occupied bands. Indeed, if we focus only on this quantized eigenvalue-subset, we might conclude that $\W^{2b}{=}I$ (whose derivation in Eq.\ (\ref{associattt}) carries through in the presence of $\cali$ symmetry) could be solved for any $b {\in}\Z$. However, our perspective is that Wilson-loop extensions classify different momentum submanifolds in the Brillouin zone; this classification should therefore be independent of specific $\cali$-representations of the occupied bands. Thus assuming that a finite subset of $\W$-eigenvalues are generically not quantized, we conclude that $b{=}0$ even with $\cali$ symmetry.



\end{appendix}

\bibliography{bib_29mar}

\end{document}